\documentclass[11pt,a4paper]{article}
\pdfoutput=1
\usepackage{jcappub}
\usepackage{ifthen}
\usepackage{epsfig}
\usepackage{booktabs} 
\usepackage{natbib}
\usepackage{bbold}

\newcommand{\beqra}{\begin{flalign}}
\newcommand{\eeqra}{\end{flalign}}
\newcommand{\beq}{\begin{equation}}
\newcommand{\eeq}{\end{equation}}
\usepackage{ccicons} 
\usepackage{listings}
\usepackage{color}   
\usepackage{braket}  
\usepackage{hyperref}
\usepackage{dsfont}  
\usepackage{leftidx} 
\usepackage{bm}
\allowdisplaybreaks

\title{Dark matter signals at neutrino telescopes in effective theories}
\author[a]{Riccardo Catena}
\affiliation[a]{Institut f\"ur Theoretische Physik, Friedrich-Hund-Platz 1, 37077 G\"ottingen, Germany}
\emailAdd{riccardo.catena@theorie.physik.uni-goettingen.de}

\abstract{
We constrain the effective theory of one-body dark matter-nucleon interactions using neutrino telescope observations. We derive exclusion limits on the 28 coupling constants of the theory, exploring interaction operators previously considered in dark matter direct detection only, and using new nuclear response functions recently derived through nuclear structure calculations. 
We determine for what interactions neutrino telescopes are superior to current direct detection experiments, and 
show that Hydrogen is not the most important element in the exclusion limit calculation for the majority of the spin-dependent operators.}


\begin{document}
\maketitle

\section{Introduction}
The clustering of dark matter in large spheroidal halos hosting galaxies and other visible astrophysical structures has inspired complementary approaches to reveal the nature of dark matter~\cite{Kuhlen:2012ft}. 
The indirect detection of energetic neutrinos from dark matter annihilation in the Sun is a prime example of strategies designed for discovering dark matter in astroparticle physics~\cite{Silk:1985ax}. 

Dark matter particles from the Milky Way dark matter halo might be captured and annihilate in the Sun, if they loose energy while crossing and scattering in the solar medium~\cite{Gould:1987ir}. Energetic neutrinos from dark matter annihilation in the Sun are expected in a variety of models for weakly interacting dark matter~\cite{Jungman:1995df,Bergstrom:2000pn,Bertone:2004pz,Catena:2013pka}, and could potentially be observed on Earth at neutrino telescopes~\cite{Tanaka:2011uf,Aartsen:2012kia}.

The predicted flux of dark matter-induced neutrinos from the Sun depends on how dark matter is distributed in the solar neighborhood, on the primary dark matter annihilation channel and total annihilation rate, and on the cross-section for dark matter-nucleus scattering in the Sun~\cite{Wikstrom:2009kw,Blennow:2007tw}.

In this work we focus on the cross-section dependence of the expected dark matter-induced neutrino flux, in a comprehensive analysis of a variety of dark matter-nucleon interactions. 

We calculate the flux of neutrino-induced upward muons from dark matter annihilation in the Sun at neutrino telescopes in the general effective theory of one-body dark matter-nucleon interactions. We then compare our predictions with current neutrino telescope observations, deriving exclusion limits on the isoscalar and isovector coupling constants of the theory. We also compare our limits from neutrino telescopes with current limits from dark matter direct detection experiments, establishing when the former are superior to the latter. For the elements in the Sun, we consider new nuclear response functions which recently appeared in the literature~\cite{Catena:2015uha}, and analyze the contribution of different isotopes to the final exclusion limits in detail. 

Compared to previous analyses, here we explore a larger set of dark matter-nucleon interactions, including several momentum or velocity dependent operators previously considered in dark matter direct detection only. Most importantly, for the dark matter-nucleon interactions studied here, and for the 16 most abundant isotopes in the Sun, we consider new nuclear response functions recently computed through numerical nuclear structure calculations in~\cite{Catena:2015uha}. The use of improved nuclear response functions strengthen our conclusions, which do not rely on the approximate exponential form factors often used in the literature in the analysis of neutrino telescope data.   

Effective theories for dark matter-nucleon interactions were explored in the context of dark matter direct detection in~\cite{Chang:2009yt,Fan:2010gt,Fornengo:2011sz,Fitzpatrick:2012ix,Fitzpatrick:2012ib,Menendez:2012tm,Cirigliano:2012pq,Anand:2013yka,DelNobile:2013sia,Klos:2013rwa,Peter:2013aha,Hill:2013hoa,Catena:2014uqa,Catena:2014hla,Catena:2014epa,Gluscevic:2014vga,Panci:2014gga,Vietze:2014vsa,Barello:2014uda,Schneck:2015eqa}. Neutrino telescope observations were interpreted in terms of spin and momentum/velocity dependent dark matter-nucleon interactions in~\cite{Guo:2013ypa}, and more recently in~\cite{Liang:2013dsa}. Both analyses focus on a subset of operators studied here, and consider either dark matter scattering from Hydrogen only, or simple Helm form factors. The effective field theory interpretation of searches for dark matter annihilation in the Sun in~\cite{Blumenthal:2014cwa} considers 4 operators, and constant spin-dependent cross-sections. Recently, advances in the theory of dark matter heat transport in the Sun within models with momentum or velocity dependent scattering cross-sections, and in the context of Helioseismology were done in~\cite{Vincent:2013lua,Lopes:2014aoa,Vincent:2014jia}.

The paper is organized as follows. In Sec.~\ref{sec:eft} we review the general effective theory of one-body dark matter-nucleon interactions. We compare the predictions of this theory to physical observables introduced in Sec.~\ref{sec:signals} as explained in Sec.~\ref{sec:analysis}, where we also interpret our findings.  We conclude in Sec.~\ref{sec:conclusions}. Finally, we describe our LUX and COUPP analysis in Appendix~\ref{sec:dd}, and list key equations in Appendix~\ref{sec:appDM}. 

\section{Theoretical framework}
\label{sec:eft}
In this section we review the general effective theory of one-body dark matter-nucleon interactions mediated by a heavy spin-1 or spin-0 particle. We compare the predictions of this theory with current neutrino telescope observations in Sec.~\ref{sec:signals}.

\subsection{Effective theory of dark matter-nucleon interactions}
\label{sec:dmnu}
We start with an analysis of the amplitude $\mathcal{M}$ for non-relativistic elastic scattering of a dark matter particle of mass $m_\chi$ and initial (final) momentum ${\bf{p}}$ (${\bf{p}}'$) from a single nucleon of mass $m_N$ and initial (final) momentum ${\bf{k}}$ (${\bf{k}}'$).
The amplitude $\mathcal{M}$ is restricted by Galilean invariance and momentum conservation. Momentum conservation implies that only three out of the four momenta ${\bf{p}}$, ${\bf{p}}'$, ${\bf{k}}$ and ${\bf{k}}'$ are independent. A possible choice of independent momenta is ${\bf{p}}$, ${\bf{k}}$ and ${\bf{q}}={\bf{k}}-{\bf{k}}'$, where ${\bf{q}}$ is the momentum transfer. Galilean invariance constrains $\mathcal{M}$ to be a function of the initial relative velocity ${\bf{v}}={\bf{p}}/m_\chi-{\bf{k}}/m_N$, rather than of the momenta ${\bf{p}}$ and ${\bf{k}}$ separately. Therefore, the scattering amplitude $\mathcal{M}$ can in general be written as a function of ${\bf{q}}$ and ${\bf{v}}$, and of the nucleon and dark matter particle spins, ${\bf{S}}_N$ and ${\bf{S}}_\chi$, respectively. 

The non-relativistic quantum mechanical Hamiltonian density that underlies the amplitude $\mathcal{M}$ must have the form~\cite{Fitzpatrick:2012ix}
\begin{eqnarray}
{\bf\hat{\mathcal{H}}}({\bf{r}})&=& \sum_{\tau=0,1} \sum_{k} c_k^{\tau} \,\hat{\mathcal{O}}_{k}({\bf{r}}) \, t^{\tau} \,,
\label{eq:Hc0c1}
\end{eqnarray}
where ${\mathbf{r}}$ is the dark matter-nucleon relative distance, $t^{0}= \mathbb{1}$ is the identity in isospin space, and $t^{1}= \tau_3$ is the third Pauli matrix.
The quantum mechanical operators $\hat{\mathcal{O}}_{k}$ in Eq.~(\ref{eq:Hc0c1}) are constructed from the momentum transfer operator ${\hat{\bf{q}}}$, the transverse relative velocity operator $\hat{\bf{v}}^{\perp}$, and the nucleon and dark matter particle spin operators ${\hat{\bf{S}}}_N$ and ${\hat{\bf{S}}}_\chi$, respectively. If we demand that the operators $\hat{\mathcal{O}}_{k}$ are at most linear in $\hat{\bf{v}}^{\perp}$, $\hat{\bf{S}}_N$ and $\hat{\bf{S}}_\chi$, then only 14 independent operators appear in the linear combination~(\ref{eq:Hc0c1})\footnote{Following~\cite{Anand:2013yka}, we neglect the operator $\hat{\mathcal{O}}_{2}=({\hat{\bf{v}}}^{\perp})^2$, since it is quadratic in ${\hat{\bf{v}}}^{\perp}$, and it cannot be a leading operator in effective theories.}. They are listed in Tab.~\ref{tab:operators}. The operators in Tab.~\ref{tab:operators} define the general effective theory of dark matter-nucleon interactions mediated by a heavy spin-1 or spin-0 particle.
The isoscalar and isovecotor coupling constants in~(\ref{eq:Hc0c1}), $c_k^{0}$ and $c_k^{1}$ respectively, are related to the coupling constants for protons ($c^{p}_k$) and neutrons ($c^{n}_k$) as follows: $c^{p}_k=(c^{0}_k+c^{1}_k)/2$, and $c^{n}_k=(c^{0}_k-c^{1}_k)/2$. These constants have dimension mass to the power $-2$.
\begin{table}[t]
    \centering
    \begin{tabular}{ll}
    \toprule
        $\hat{\mathcal{O}}_1 = \mathbb{1}_{\chi N}$ & $\hat{\mathcal{O}}_9 = i{\hat{\bf{S}}}_\chi\cdot\left(\hat{{\bf{S}}}_N\times\frac{{\hat{\bf{q}}}}{m_N}\right)$  \\
        $\hat{\mathcal{O}}_3 = i\hat{{\bf{S}}}_N\cdot\left(\frac{{\hat{\bf{q}}}}{m_N}\times{\hat{\bf{v}}}^{\perp}\right)$ \hspace{2 cm} &   $\hat{\mathcal{O}}_{10} = i\hat{{\bf{S}}}_N\cdot\frac{{\hat{\bf{q}}}}{m_N}$   \\
        $\hat{\mathcal{O}}_4 = \hat{{\bf{S}}}_{\chi}\cdot \hat{{\bf{S}}}_{N}$ &   $\hat{\mathcal{O}}_{11} = i{\hat{\bf{S}}}_\chi\cdot\frac{{\hat{\bf{q}}}}{m_N}$   \\                                                                             
        $\hat{\mathcal{O}}_5 = i{\hat{\bf{S}}}_\chi\cdot\left(\frac{{\hat{\bf{q}}}}{m_N}\times{\hat{\bf{v}}}^{\perp}\right)$ &  $\hat{\mathcal{O}}_{12} = \hat{{\bf{S}}}_{\chi}\cdot \left(\hat{{\bf{S}}}_{N} \times{\hat{\bf{v}}}^{\perp} \right)$ \\                                                                                                                 
        $\hat{\mathcal{O}}_6 = \left({\hat{\bf{S}}}_\chi\cdot\frac{{\hat{\bf{q}}}}{m_N}\right) \left(\hat{{\bf{S}}}_N\cdot\frac{\hat{{\bf{q}}}}{m_N}\right)$ &  $\hat{\mathcal{O}}_{13} =i \left(\hat{{\bf{S}}}_{\chi}\cdot {\hat{\bf{v}}}^{\perp}\right)\left(\hat{{\bf{S}}}_{N}\cdot \frac{{\hat{\bf{q}}}}{m_N}\right)$ \\   
        $\hat{\mathcal{O}}_7 = \hat{{\bf{S}}}_{N}\cdot {\hat{\bf{v}}}^{\perp}$ &  $\hat{\mathcal{O}}_{14} = i\left(\hat{{\bf{S}}}_{\chi}\cdot \frac{{\hat{\bf{q}}}}{m_N}\right)\left(\hat{{\bf{S}}}_{N}\cdot {\hat{\bf{v}}}^{\perp}\right)$  \\
        $\hat{\mathcal{O}}_8 = \hat{{\bf{S}}}_{\chi}\cdot {\hat{\bf{v}}}^{\perp}$  & $\hat{\mathcal{O}}_{15} = -\left(\hat{{\bf{S}}}_{\chi}\cdot \frac{{\hat{\bf{q}}}}{m_N}\right)\left[ \left(\hat{{\bf{S}}}_{N}\times {\hat{\bf{v}}}^{\perp} \right) \cdot \frac{{\hat{\bf{q}}}}{m_N}\right] $ \\                                                                               
    \bottomrule
    \end{tabular}
    \caption{Complete set of non-relativistic quantum mechanical operators that are at most linear in the transverse relative velocity operator ${\bf{v}}^{\perp}$, and in nucleon and dark matter particle spin operators, $\hat{\bf{S}}_N$ and $\hat{\bf{S}}_\chi$, respectively. Introducing the nucleon mass, $m_N$, in the equations above all operators have the same mass dimension.} 
    \label{tab:operators}
\end{table}

Assuming one-body dark matter-nucleon interactions, the most general Hamiltonian density for elastic dark matter-nucleus interactions is given by
\begin{equation}
\hat{\mathcal{H}}_{\rm T}({\bf{r}})= \sum_{i=1}^{A}  \sum_{\tau=0,1} \sum_{k} c_k^{\tau}\hat{\mathcal{O}}_{k}^{(i)}({\bf{r}}) \, t^{\tau}_{(i)} \,,
\label{eq:H_I}
\end{equation}
which is the sum of $A$ identical terms. The index $i$ in Eq.~(\ref{eq:H_I}) identifies the $i$th-nucleon in the target nucleus, $A$ is the mass number, and ${\bf{r}}$ is the relative distance from the nucleus centre of mass to the dark matter particle. 

The Hamiltonian density $\hat{\mathcal{H}}_{\rm T}$ depends on the nuclear charges and currents, and admits the following coordinate space representation~\cite{Fitzpatrick:2012ix} 
\begin{eqnarray}
\hat{\mathcal{H}}_{\rm T}({\bf{r}}) &=& \sum_{\tau=0,1} \Bigg\{
\sum_{i=1}^A  \hat{l}_0^{\tau}~ \delta({\bf{r}}-{\bf{r}}_i)
+ \sum_{i=1}^A \hat{l}_{0A}^{\tau}~ \frac{1}{2m_N} \Bigg[i \overleftarrow{\nabla}_{\bf{r}} \cdot  \vec{\sigma}_i\delta({\bf{r}}-{\bf{r}}_i) -i \delta({\bf{r}}-{\bf{r}}_i) \
\vec{\sigma}_i  \cdot  \overrightarrow{\nabla}_{\bf{r}} \Bigg]  \nonumber \\
 &+& \sum_{i=1}^A  {\hat{\bf{l}}}_5^{\tau} \cdot \vec{\sigma}_i \delta({\bf{r}}-{\bf{r}}_i) +   \sum_{i=1}^A {\hat{\bf{l}}}_M^{\tau} \cdot \frac{1}{2 m_N} \Bigg[i \overleftarrow{\nabla}_{\bf{r}}\delta({\bf{r}}-{\bf{r}}_i) -i \delta({\bf{r}}-{\bf{r}}_i)\overrightarrow{\nabla}_{\bf{r}} \Bigg]  \nonumber \\
&+& \sum_{i=1}^A {\hat{\bf{l}}}_E^{\tau} \cdot \frac{1}{2m_N} \Bigg[ \overleftarrow{\nabla}_{\bf{r}} \times \vec{\sigma}_i \delta({\bf{r}}-{\bf{r}}_i) +\delta({\bf{r}}-{\bf{r}}_i)\
  \vec{\sigma}_i \times \overrightarrow{\nabla}_{\bf{r}} \Bigg] \Bigg\} t^{\tau}_{(i)}
\label{eq:Hx}
\end{eqnarray}
where $\vec{\sigma}_i/2$ and ${\bf{r}}_i$ represent the the spin operator, and the position in the nucleus centre of mass frame of the $i$th-nucleon in the target nucleus, respectively. In Eq.~(\ref{eq:Hx}) we use the definitions
\begin{eqnarray}
\label{eq:ls}
\hat{l}_0^\tau &=& c_1^\tau + i  \left( {{\hat{\bf{q}}} \over m_N}  \times {\hat{\bf{v}}}_{T}^\perp \right) \cdot  {\hat{\bf{S}}}_\chi  ~c_5^\tau
+ {\hat{\bf{v}}}_{T}^\perp \cdot {\hat{\bf{S}}}_\chi  ~c_8^\tau + i {{\hat{\bf{q}}} \over m_N} \cdot {\hat{\bf{S}}}_\chi ~c_{11}^\tau \nonumber \\
\hat{l}_{0A}^{\tau} &=& -{1 \over 2}  \left[ c_7 ^\tau  +i {{\hat{\bf{q}}} \over m_N} \cdot {\hat{\bf{S}}}_\chi~ c_{14}^\tau \right] \nonumber \\
{\hat{\bf{l}}}_5^{\tau}&=& {1 \over 2} \left[ i {{\hat{\bf{q}}} \over m_N} \times {\hat{\bf{v}}}_{T}^\perp~ c_3^\tau + {\hat{\bf{S}}}_\chi ~c_4^\tau
+  {{\hat{\bf{q}}} \over m_N}~{{\hat{\bf{q}}} \over m_N} \cdot {\hat{\bf{S}}}_\chi ~c_6^\tau
+   {\hat{\bf{v}}}_{T}^\perp ~c_7^\tau + i {{\hat{\bf{q}}} \over m_N} \times {\hat{\bf{S}}}_\chi ~c_9^\tau + i {{\hat{\bf{q}}} \over m_N}~c_{10}^\tau \right. \nonumber \\
 && \left.  +  {\hat{\bf{v}}}_{T}^\perp \times {\hat{\bf{S}}}_\chi ~c_{12}^\tau
+i  {{\hat{\bf{q}}} \over m_N} {\hat{\bf{v}}}_{T}^\perp \cdot {\hat{\bf{S}}}_\chi ~c_{13}^\tau+i {\hat{\bf{v}}}_{T}^\perp {{\hat{\bf{q}}} \over m_N} \cdot {\hat{\bf{S}}}_\chi ~ c_{14}^\tau+{{\hat{\bf{q}}} \over\
 m_N} \times {\hat{\bf{v}}}_{T}^\perp~ {{\hat{\bf{q}}} \over m_N} \cdot {\hat{\bf{S}}}_\chi ~ c_{15}^\tau  \right]\nonumber \\
{\hat{\bf{l}}}_M^{\tau} &=&   i {{\hat{\bf{q}}} \over m_N}  \times {\hat{\bf{S}}}_\chi ~c_5^\tau - {\hat{\bf{S}}}_\chi ~c_8^\tau \nonumber \\
{\hat{\bf{l}}}_E^{\tau} &=& {1 \over 2} \left[  {{\hat{\bf{q}}} \over m_N} ~ c_3^\tau +i {\hat{\bf{S}}}_\chi~c_{12}^\tau - {{\hat{\bf{q}}} \over  m_N} \times{\hat{\bf{S}}}_\chi  ~c_{13}^\tau-i 
{{\hat{\bf{q}}} \over  m_N} {{\hat{\bf{q}}} \over m_N} \cdot {\hat{\bf{S}}}_\chi  ~c_{15}^\tau \right] \,,
\end{eqnarray}
where ${\hat{\bf{v}}}_{T}^\perp\equiv {\hat{\bf{v}}}^\perp - {\hat{\bf{v}}}_N^\perp$, and ${\hat{\bf{v}}}_N^\perp$ is an operator acting on the $i$th-nucleon space coordinate~\cite{Fitzpatrick:2012ix}. Explicit coordinate space representations for the operators ${\hat{\bf{q}}}$, ${\hat{\bf{v}}}_N^\perp$, and ${\hat{\bf{v}}}_{T}^\perp$ can be found in~\cite{Catena:2015uha}.

Eq.~(\ref{eq:Hx}) accounts for all possible ways of coupling dark matter to the nuclear charges and currents. Dark matter couples to the nuclear vector and axial charges through the operators $\hat{l}_0^\tau$ and $\hat{l}_{0A}^\tau$, and it couples to the nuclear spin, convection and spin-velocity currents through the operators ${\hat{\bf{l}}}_5^{\tau}$, ${\hat{\bf{l}}}_M^{\tau}$ and ${\hat{\bf{l}}}_E^{\tau}$, respectively. Notice however, that the nuclear axial charge does not contribute to scattering cross-sections when nuclear ground states are eigenstates of $P$ and $CP$. 

The operators in Tab.~\ref{tab:operators} contribute to Eq.~(\ref{eq:ls}) through specific combinations of ${\hat{\bf{q}}}$ and ${\hat{\bf{v}}}_{T}^\perp$. Notably, the operators $\hat{\mathcal{O}}_1$ and $\hat{\mathcal{O}}_4$ generate the only momentum and velocity independent terms in $\hat{l}_0^\tau$, $\hat{l}_{0A}^\tau$, ${\hat{\bf{l}}}_5^{\tau}$, ${\hat{\bf{l}}}_M^{\tau}$ and ${\hat{\bf{l}}}_E^{\tau}$.

\subsection{Dark matter scattering from nuclei}
We use the Hamiltonian density~(\ref{eq:Hx}) to calculate the amplitude $\mathcal{M}_{NR}$ for dark matter scattering from nuclei in the Sun. Summing (averaging) $|\mathcal{M}_{NR}|^2$ over final (initial) spin configurations one finds~\cite{Fitzpatrick:2012ix} 
\begin{align}
\langle |\mathcal{M}_{NR}|^2\rangle_{\rm spins} = \frac{4\pi}{2J+1} \sum_{\tau,\tau'} &\bigg[ \sum_{k=M,\Sigma',\Sigma''} R^{\tau\tau'}_k\left(v_T^{\perp 2}, {q^2 \over m_N^2} \right) W_k^{\tau\tau'}(q^2) \nonumber\\
&+{q^{2} \over m_N^2} \sum_{k=\Phi'', \Phi'' M, \tilde{\Phi}', \Delta, \Delta \Sigma'} R^{\tau\tau'}_k\left(v_T^{\perp 2}, {q^2 \over m_N^2}\right) W_k^{\tau\tau'}(q^2) \bigg] \,,
\label{eq:M}
\end{align}  
where the index $k$ extends over (pairs of) nuclear response operators defined below in Eq.~(\ref{eq:multipole}).

The 8 dark matter response functions $R^{\tau\tau'}_k$ depend on matrix elements between eigenstates of $\hat{\bf{S}}_\chi$ of the operators~(\ref{eq:ls}). They are functions of $q^2/m_N^2$ and $v_T^{\perp 2} =w^2-q^2/(4\mu_T^2)$, where $\mu_T$ and $w$ are the reduced dark matter-nucleus mass and relative velocity, respectively. We list the functions $R^{\tau\tau'}_k$ in Appendix~\ref{sec:appDM}. 

The 8 isotope-dependent nuclear response functions $W_k^{\tau\tau'}$ in Eq.~(\ref{eq:M}) depend on the nuclear matrix elements of the charge and current operators in Eq.~(\ref{eq:Hx}). For a given pair of nuclear operators $A_{LM;\tau}$ and $B_{LM;\tau}$, the nuclear response function $W_k^{\tau\tau'}$, $k=AB$, is defined as follows
\begin{equation}
W_{AB}^{\tau \tau^\prime}(q^2)= \sum_{L}  \langle J,T,M_T ||~ A_{L;\tau} (q)~ || J,T,M_T \rangle \langle J,T,M_T ||~ B_{L;\tau^\prime} (q)~ || J,T,M_T \rangle \,,
\label{eq:W}
\end{equation}
where the ket $|J,T,M_T \rangle$ represents a nuclear state of spin $J$, isospin $T$, and isospin magnetic quantum number $M_T$, and $M_J$ is the nuclear spin magnetic quantum number. Here we use the notation $W_{AB}^{\tau \tau^\prime}\equiv W_{A}^{\tau \tau^\prime}$, for $A=B$. 
The reduction operation in Eq.~(\ref{eq:W}) is done via the Wigner-Eckart theorem:
\begin{equation}
\langle J,M_J |\,{A}_{LM;\tau}\,|J,M_J\rangle =(-1)^{J-M_J}\left(
\begin{array}{ccc} J&L&J\\
-M_J&M&M_J 
\end{array} 
\right)
\langle  J  ||\,{A}_{L;\tau}\,|| J  \rangle \,.
\label{eq:red}
\end{equation}
In Eq.~(\ref{eq:M}), the operators $A_{LM;\tau}$ and $B_{LM;\tau}$ correspond to the following nuclear response operators
\begin{eqnarray}
M_{LM;\tau}(q) &=& \sum_{i=1}^{A} M_{LM}(q {\bf{r}}_i) t^{\tau}_{(i)}\nonumber\\
\Sigma'_{LM;\tau}(q) &=& -i \sum_{i=1}^{A} \left[ \frac{1}{q} \overrightarrow{\nabla}_{{\bf{r}}_i} \times {\bf{M}}_{LL}^{M}(q {\bf{r}}_i)  \right] \cdot \vec{\sigma}_i \, t^{\tau}_{(i)}\nonumber\\
\Sigma''_{LM;\tau}(q) &=&\sum_{i=1}^{A} \left[ \frac{1}{q} \overrightarrow{\nabla}_{{\bf{r}}_i} M_{LM}(q {\bf{r}}_i)  \right] \cdot \vec{\sigma}_i \, t^{\tau}_{(i)}\nonumber\\
\Delta_{LM;\tau}(q) &=&\sum_{i=1}^{A}  {\bf{M}}_{LL}^{M}(q {\bf{r}}_i) \cdot \frac{1}{q}\overrightarrow{\nabla}_{{\bf{r}}_i} t^{\tau}_{(i)} \nonumber\\
\tilde{\Phi}^{\prime}_{LM;\tau}(q) &=& \sum_{i=1}^A \left[ \left( {1 \over q} \overrightarrow{\nabla}_{{\bf{r}}_i} \times {\bf{M}}_{LL}^M(q {\bf{r}}_i) \right) \cdot \left(\vec{\sigma}_i \, \times {1 \over q} \overrightarrow{\nabla}_{{\bf{r}}_i} \right) + {1 \over 2} {\bf{M}}_{LL}^M(q {\bf{r}}_i) \cdot \vec{\sigma}_i \, \right]~t^\tau_{(i)} \nonumber \\
\Phi^{\prime \prime}_{LM;\tau}(q ) &=& i  \sum_{i=1}^A\left( {1 \over q} \overrightarrow{\nabla}_{{\bf{r}}_i}  M_{LM}(q {\bf{r}}_i) \right) \cdot \left(\vec{\sigma}_i \, \times \
{1 \over q} \overrightarrow{\nabla}_{{\bf{r}}_i}  \right)~t^\tau_{(i)} \,,
\label{eq:multipole}
\end{eqnarray}
where ${\bf{M}}_{LL}^{M}(q {\bf{r}}_i)=j_{L}(q r_i){\bf Y}^M_{LL1}(\Omega_{{\bf{r}}_i})$ , and $M_{LM}(q {\bf{r}}_i)=j_{L}(q r_i)Y_{LM}(\Omega_{{\bf{r}}_i})$. The vector spherical harmonics, ${\bf Y}^M_{LL1}(\Omega_{{\bf{r}}_i})$, are defined in terms of Clebsch-Gordan coefficients and scalar spherical harmonics:
\begin{equation}
{\bf Y}^M_{LL'1}(\Omega_{{\bf{r}}_i}) = \sum_{m\lambda} \langle L'm1\lambda|L'1LM \rangle
Y_{L'm}(\Omega_{{\bf{r}}_i}) \, {\bf e}_\lambda \,,
\end{equation}
where ${\bf e}_\lambda$ is a spherical unit vector basis. The 6 nuclear response operators in Eqs.~(\ref{eq:multipole}) arise from the multipole expansion of the nuclear charges and currents in Eq.~(\ref{eq:Hx}). The multipole index $L$ in Eq.~(\ref{eq:multipole}) is bounded from above: $L \le 2J$. 

The isotope-dependent nuclear response functions $W_k^{\tau\tau'}$ must be calculated through detailed nuclear structure calculations for the most abundant elements in the Sun. 
In our analysis of dark matter annihilation signals from the Sun, we adopt the nuclear response functions derived in~\cite{Catena:2015uha} through numerical shell model calculations carried out for the 16 most abundant elements in the Sun.   
 
Combining Eqs.~(\ref{eq:M}), (\ref{eq:W}) and (\ref{eq:R}), we can finally write the differential cross-section for dark matter scattering from nuclei of type $i$ and mass $m_i$
\begin{equation}
\frac{{\rm d} \sigma_i(w^2,E_R)}{{\rm d} E_R} = \frac{m_i}{2\pi w^2} \,\langle |\mathcal{M}_{NR}|^2\rangle_{\rm spins}\,,
\label{eq:sigma}
\end{equation} 
which is in general a function of the nuclear recoil energy $E_R=q^2/(2 m_i)$, and of the dark matter-nucleus relative velocity. 

\section{Dark matter annihilation signals from the Sun}
\label{sec:signals}
Given the scattering cross-section~(\ref{eq:sigma}), we now review how to calculate the rate of dark matter capture by the Sun, and the neutrino-induced muon flux at neutrino telescopes from dark matter annihilation in the Sun. 
\subsection{Capture}
Weakly interacting dark matter particles of the Milky Way dark matter halo have a finite probability of elastically scattering from nuclei in the Sun.
A dark matter particle becomes gravitationally bound to the Sun, when traveling through the solar medium it scatters to a velocity smaller than the local escape velocity. 
The rate of scattering from a velocity $w$ to a velocity less than the escape velocity $v(R)$ at a distance $R$ from the Sun's centre is given by~\cite{Gould:1987ir}
\begin{equation}
\Omega_{v}^{-}(w)= \sum_i n_i w\,\Theta\left( \frac{\mu_i}{\mu^2_{+,i}} - \frac{u^2}{w^2} \right)\int_{E u^2/w^2}^{E \mu_i/\mu_{+,i}^2} {\rm d}E_R\,\frac{{\rm d}\sigma_{i}\left(w^2,E_R\right)}{{\rm d}E_R}\,,
\label{eq:omega}
\end{equation}
where $E=m_\chi w^2/2$, is the dark matter particle initial kinetic energy, $d\sigma_i/dE_R$ is the differential cross-section for dark matter scattering from nuclei of mass $m_i$ and density in the Sun $n_i(R)$, and $u$ is the velocity of the dark matter particle at $R\rightarrow \infty$, where the Sun's gravitational potential is negligible. $\Omega_{v}^{-}(w)$ depends on the radial coordinate $R$, in that $w=\sqrt{u^2+v(R)^2}$. The sum in the scattering rate (\ref{eq:omega}) extends over the most abundant elements in the Sun, and the dimensionless parameters $\mu_i$ and $\mu_{\pm,i}$  are defined as follows 
\begin{equation}
\mu_i\equiv \frac{m_\chi}{m_i}\, \qquad\qquad \mu_{\pm,i}\equiv \frac{\mu_i\pm1}{2}\,.
\end{equation}
For a population of galactic dark matter particles with speed distribution at infinity given by $f(u)$, the differential rate of capture per unit volume is given by~\cite{Gould:1987ir}
\begin{equation}
\frac{{\rm d} C}{{\rm d}V} = \int_{0}^{\infty} {\rm d}u\, \frac{f(u)}{u}\, w\Omega_{v}^{-}(w) \,.
\label{eq:drate}
\end{equation}
Integrating (\ref{eq:drate}) over a sphere of radius $R_{\rm \odot}$, where $R_{\rm \odot}$ is the solar radius, one obtains the total rate of dark matter capture by the Sun
\begin{equation}
C = 4\pi \int_{0}^{R_{\odot}} {\rm d} R\, R^2\,\frac{{\rm d} C (R)}{{\rm d}V}\,.
\label{eq:rate}
\end{equation}
In this work, we consider the 16 most abundant elements in the Sun, i.e. H, $^{3}$He, $^{4}$He, $^{12}$C, $^{14}$N, $^{16}$O, $^{20}$Ne, $^{23}$Na, $^{24}$Mg, $^{27}$Al, $^{28}$Si, $^{32}$S, $^{40}$Ar, $^{40}$Ca, $^{56}$Fe, and $^{58}$Ni,  using the densities $n_i(R)$, and the velocity $v(R)$ implemented in the {\sffamily darksusy} code~\cite{Gondolo:2004sc}. At the same time, we assume the so-called standard halo model~\cite{Freese:2012xd}, with a Maxwell-Boltzmann speed distribution for $f(u)$, and a local standard of rest velocity of 220 km~s$^{-1}$~\cite{Bozorgnia:2013pua,Catena:2011kv,Catena:2009mf}.

\subsection{Annihilation}
Once captured, dark matter particles undergo subsequent scatters 
and sink into the centre of the Sun. In a variety of models~\cite{Jungman:1995df,Bergstrom:2000pn,Bertone:2004pz,Catena:2013pka}, in the Sun dark matter annihilates into Standard Model final states, producing a flux of energetic neutrinos escaping the solar medium, and potentially observable at neutrino telescopes.

The differential neutrino flux from dark matter annihilation in the Sun is given by~\cite{Jungman:1995df}
\begin{equation}
\frac{{\rm d \Phi_\nu}}{{\rm d} E_\nu} = \frac{\Gamma_A}{4\pi D^2} \sum_{f} B_{\chi}^{f} \, \frac{{\rm d} N_\nu^f}{{\rm d} E_\nu} \,,
\label{eq:nuflux}
\end{equation}
where $E_{\nu}$ is the neutrino energy, $B_{\chi}^{f}$ the branching ratio for dark matter pair annihilation into the final state $f$, $D$ the distance of the observer to the center of the Sun, and ${\rm d} N_\nu^f/{\rm d} E_\nu$ is the neutrino energy spectrum at the detector from dark matter annihilation into the final state $f$.

The annihilation rate, $\Gamma_A$, in Eq.~(\ref{eq:nuflux}) is defined as follows 
\begin{equation}
\Gamma_A=\frac{1}{2}\int d^3{\bf x} \,n^2({\bf x}) \,\langle \sigma_{\rm ann} v_{\rm rel} \rangle \,,
\label{eq:Gamma}
\end{equation}
where $n({\bf x})$ is the dark matter space density in the Sun, and $\langle \sigma_{\rm ann} v_{\rm rel} \rangle$ the thermally averaged dark matter annihilation cross-section times relative velocity. Eq.~(\ref{eq:Gamma}) implies the relation $\Gamma_A=C_A N^2/2$, where $N$ is the number of dark matter particles in the Sun.  The constant $C_A$ is given by 
\begin{equation}
C_A = \langle \sigma_{\rm ann} v_{\rm rel} \rangle \frac{V_2}{V_1^2}\,,
\label{eq:CA}
\end{equation}
with 
\begin{equation}
V_1 = \int d^3{\bf x} \,\frac{n({\bf x})}{n_0} \,; \qquad \qquad V_2 = \int d^3{\bf x} \,\frac{n^2({\bf x})}{n_0^2}\,,
\end{equation}
and $n_0$ equal to the dark matter density at the core of the Sun.

The number of dark matter particles $N$ obeys the following differential equation in the time variable $t$
\begin{equation}
\dot{N} = C - C_A N^2 \,,
\label{eq:N}
\end{equation}
which admits the general solution 
\begin{equation}
\Gamma_A= \frac{C_A N^2}{2} = \frac{C}{2}\tanh^2\left(\sqrt{C C_A}t\right) \,.
\end{equation}
In our analysis we assume $\sqrt{C C_A}t_{\odot}\gg 1$, where $t_\odot$ is the age of the Sun. This assumption implies $\Gamma_A=C/2$, which corresponds to equilibrium between capture and annihilation of dark matter in the Sun.

Neutrinos from dark matter annihilation in the Sun can be detected at neutrino telescopes observing an upward muon flux induced by charged-current neutrino interactions with nuclei in the material surrounding the detector. The differential neutrino-induced muon flux at the detector is given by
\begin{equation}
\frac{{\rm d}\Phi_{\mu}}{{\rm d} E_\mu} = N_T \int_{E_\mu^{\rm th}}^{\infty} {\rm d} E_\nu
\int_0^{\infty} {\rm d} \lambda \int_{E_\mu}^{E_\nu} {\rm d} E_\mu^{\prime} \,\mathcal{P}(E_\mu,E_\mu^{\prime};\lambda)\, \frac{{\rm d}\sigma_{{\rm CC} }(E_\nu,E_\mu^{\prime})}{{\rm d} E_{\mu}^{\prime}}  \,\frac{{\rm d \Phi_\nu}}{{\rm d} E_\nu} \,,
\label{eq:Phi}
\end{equation}
where $E_\mu^{\rm th}$ is the experimental threshold energy, $N_T$ is the number of nucleons per cubic centimeter, $\lambda$ is the muon range, $\mathcal{P}(E_\mu,E_\mu^{\prime};\lambda)$ is the probability for a muon of initial energy $E_{\mu}^{\prime}$ to have a final energy $E_\mu$ after traveling a distance $\lambda$ inside the detector, and ${\rm d}\sigma_{{\rm CC}}/{\rm d} E_\mu^{\prime}$ is the weak differential cross-section for production of a muon of energy $E_\mu^{\prime}$.

\begin{table}
  \centering
  \begin{tabular}[!h]{|c|c|c|c|}
\hline
    Element& Average mass fraction &  Element& Average mass fraction\\
\hline
\hline
H& 0.684&${}^{24}$Mg&7.30$\times10^{-4}$\\   
\hline
${}^{4}$He& 0.298 &${}^{27}$Al&6.38$\times10^{-5}$\\  
\hline
${}^3$He& 3.75$\times10^{-4}$ &${}^{28}$Si&7.95$\times10^{-4}$\\
\hline
${}^{12}$C& 2.53$\times10^{-3}$&${}^{32}$S&5.48$\times10^{-4}$ \\
\hline
${}^{14}$N& 1.56$\times10^{-3}$&${}^{40}$Ar&8.04$\times10^{-5}$\\ 
\hline
${}^{16}$O&8.50$\times10^{-3}$&${}^{40}$Ca&7.33$\times10^{-5}$\\
\hline
${}^{20}$Ne&1.92$\times10^{-3}$&${}^{56}$Fe&1.42$\times10^{-3}$\\
\hline
${}^{23}$Na&3.94$\times10^{-5}$&${}^{58}$Ni&8.40$\times10^{-5}$\\
\hline
  \end{tabular}
  \caption{List of average mass fractions for the 16 most abundant elements in the Sun as implemented in the {\sffamily darksusy} program~\cite{Gondolo:2004sc} (see also~\cite{Bahcall:2004pz}).}
\label{tab:massfrac}
\end{table}

We evaluate Eq.~(\ref{eq:Phi}) using muon yields generated by {\sffamily WimpSim}~\cite{Blennow:2007tw}, and pre-tabulated in {\sffamily darksusy}~\cite{Gondolo:2004sc}. Notice that {\sffamily WimpSim} also accounts for the angular dependence of the upward muon flux, not included in Eq.~(\ref{eq:Phi}) for simplicity. At the same time, {\sffamily WimpSim} assumes a Gaussian space distribution for dark matter in the Sun, which is not generically true when the dark matter-nucleon scattering cross-section is momentum or velocity dependent~\cite{Vincent:2013lua}. Departures from this assumption are however negligible for a dark matter particle mass larger than about 30 GeV (see Fig.~3 in~\cite{Vincent:2013lua}).  

In computing the neutrino-induced muon flux, we calculate the rate of dark matter capture by the Sun, and the relevant dark matter-nucleus scattering cross-sections using our own routines and nuclear response functions~\cite{Catena:2015uha}. We then compare our predictions with neutrino telescope observations as explained in the next section.

\begin{figure}[t]
\begin{center}
\begin{minipage}[t]{0.49\linewidth}
\centering
\includegraphics[width=\textwidth]{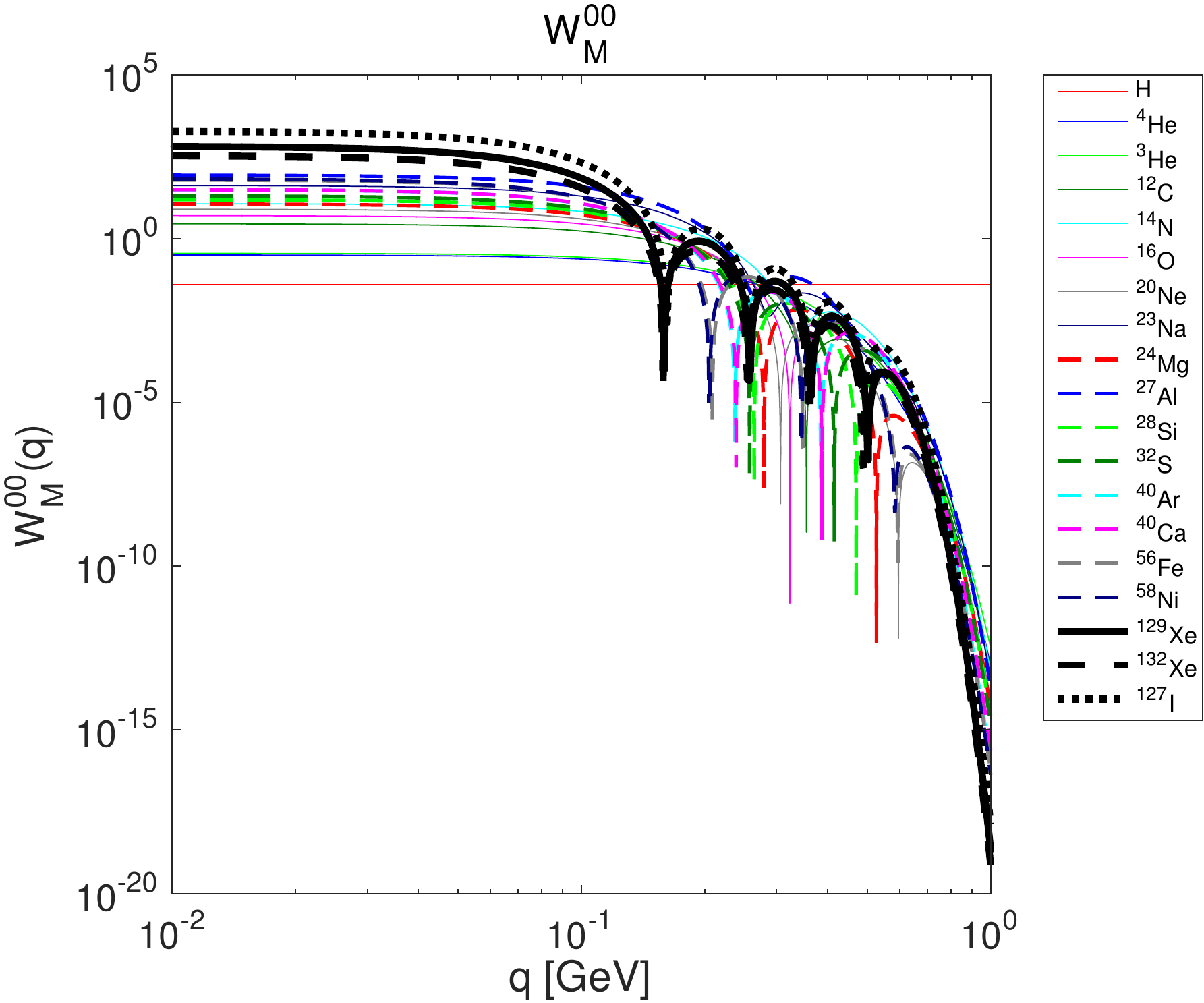}
\end{minipage}
\begin{minipage}[t]{0.49\linewidth}
\centering
\includegraphics[width=\textwidth]{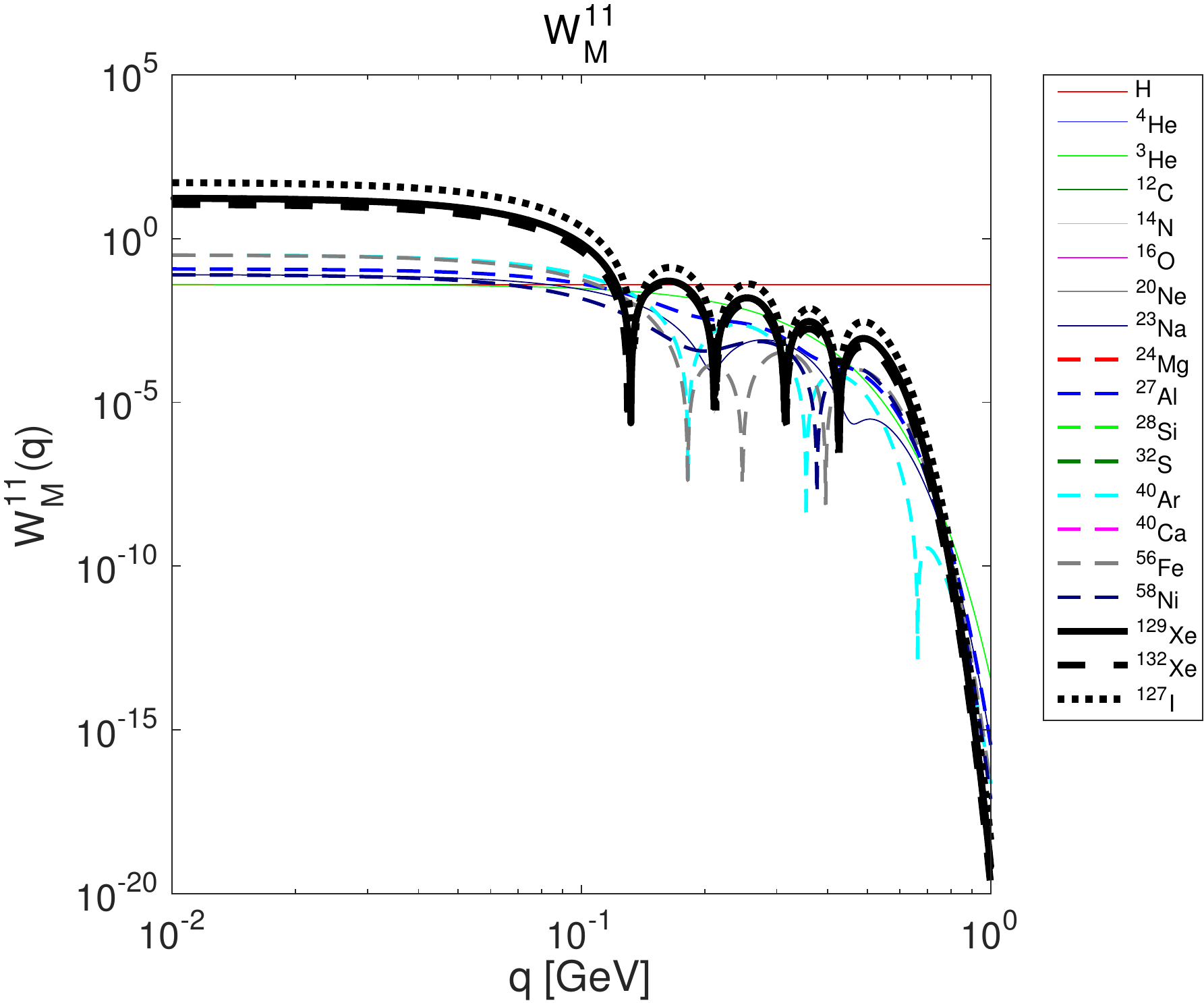}
\end{minipage}
\begin{minipage}[t]{0.49\linewidth}
\centering
\includegraphics[width=\textwidth]{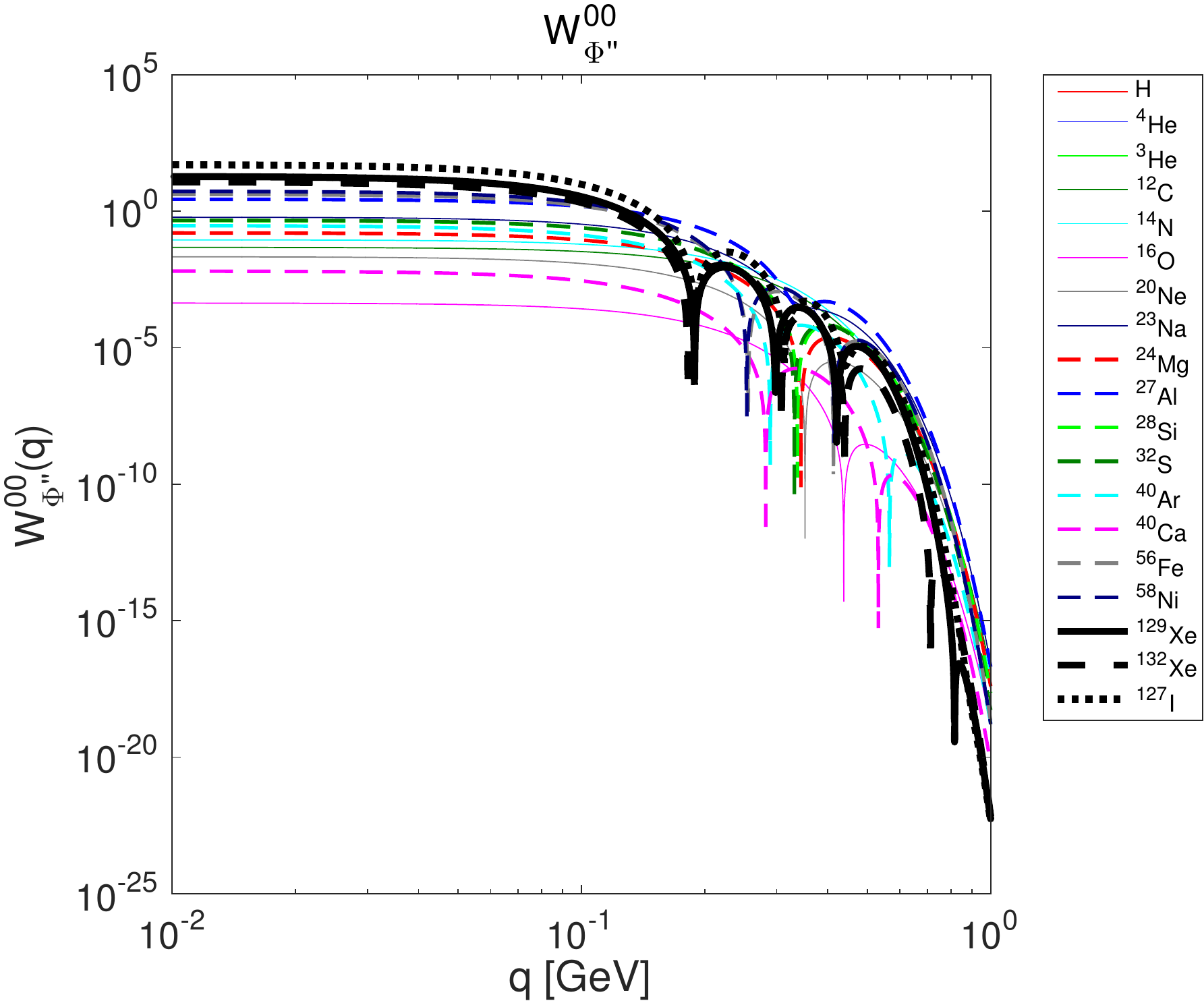}
\end{minipage}
\begin{minipage}[t]{0.49\linewidth}
\centering
\includegraphics[width=\textwidth]{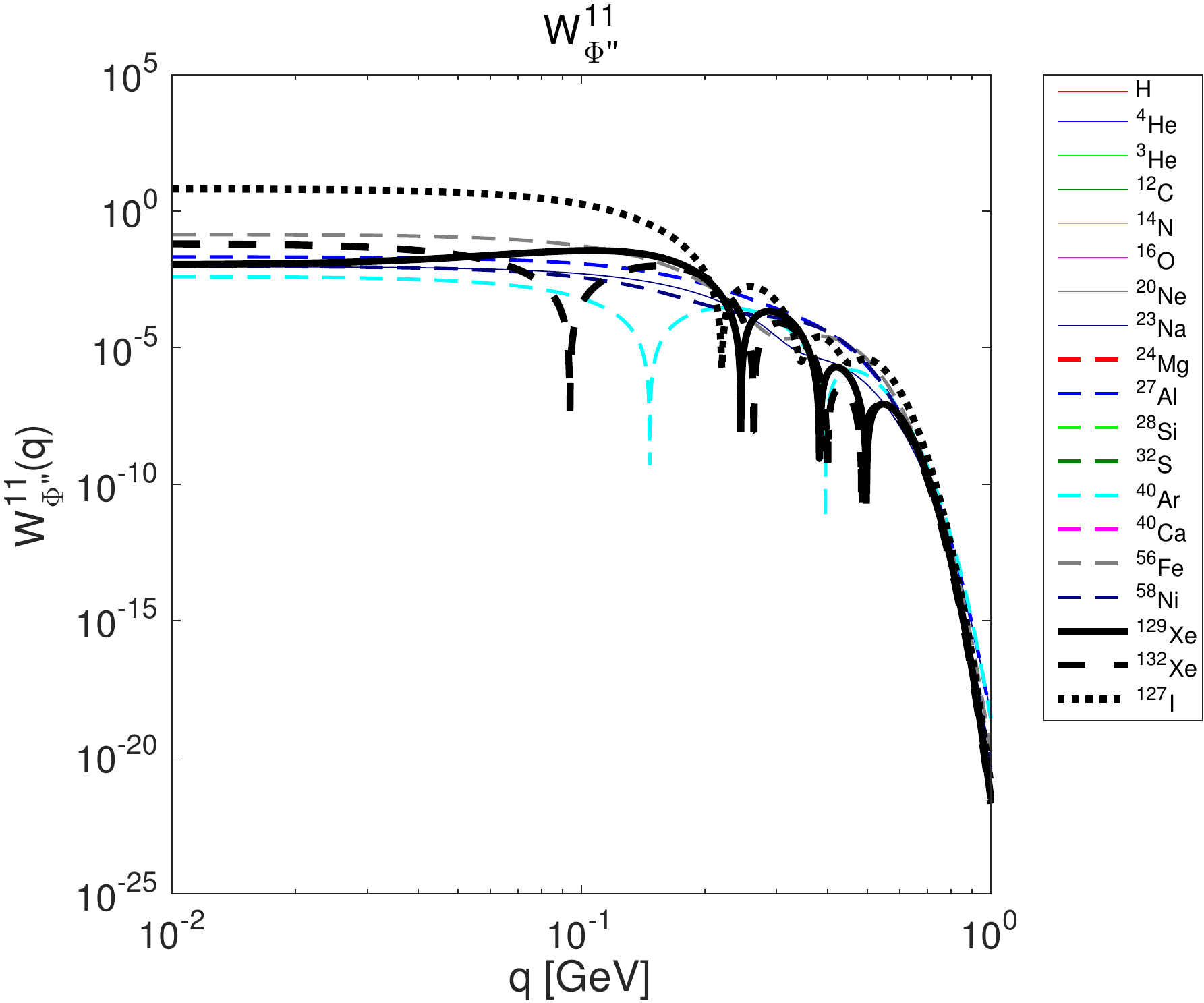}
\end{minipage}
\end{center}
\caption{Nuclear response functions $W_{M}^{\tau\tau'}$ and $W_{\Phi''}^{\tau\tau'}$ as a function of the momentum transfer $q$, for $\tau\neq\tau'$, and for the 16 most abundant elements in the Sun, as well as for Xe and I. Conventions for colors and lines are those in the legends.}
\label{fig:W1}
\end{figure}
\begin{figure}[t]
\begin{center}
\begin{minipage}[t]{0.49\linewidth}
\centering
\includegraphics[width=\textwidth]{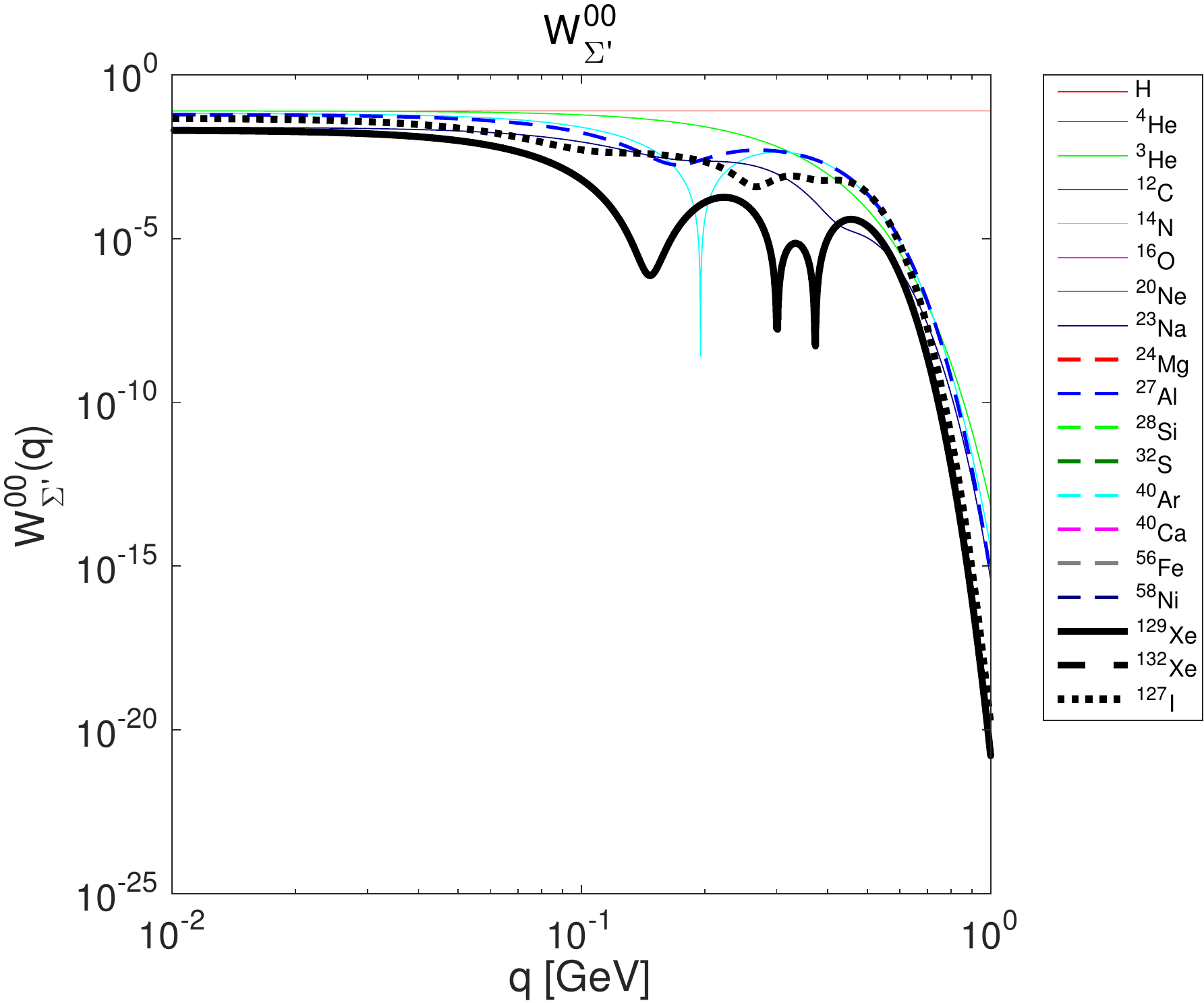}
\end{minipage}
\begin{minipage}[t]{0.49\linewidth}
\centering
\includegraphics[width=\textwidth]{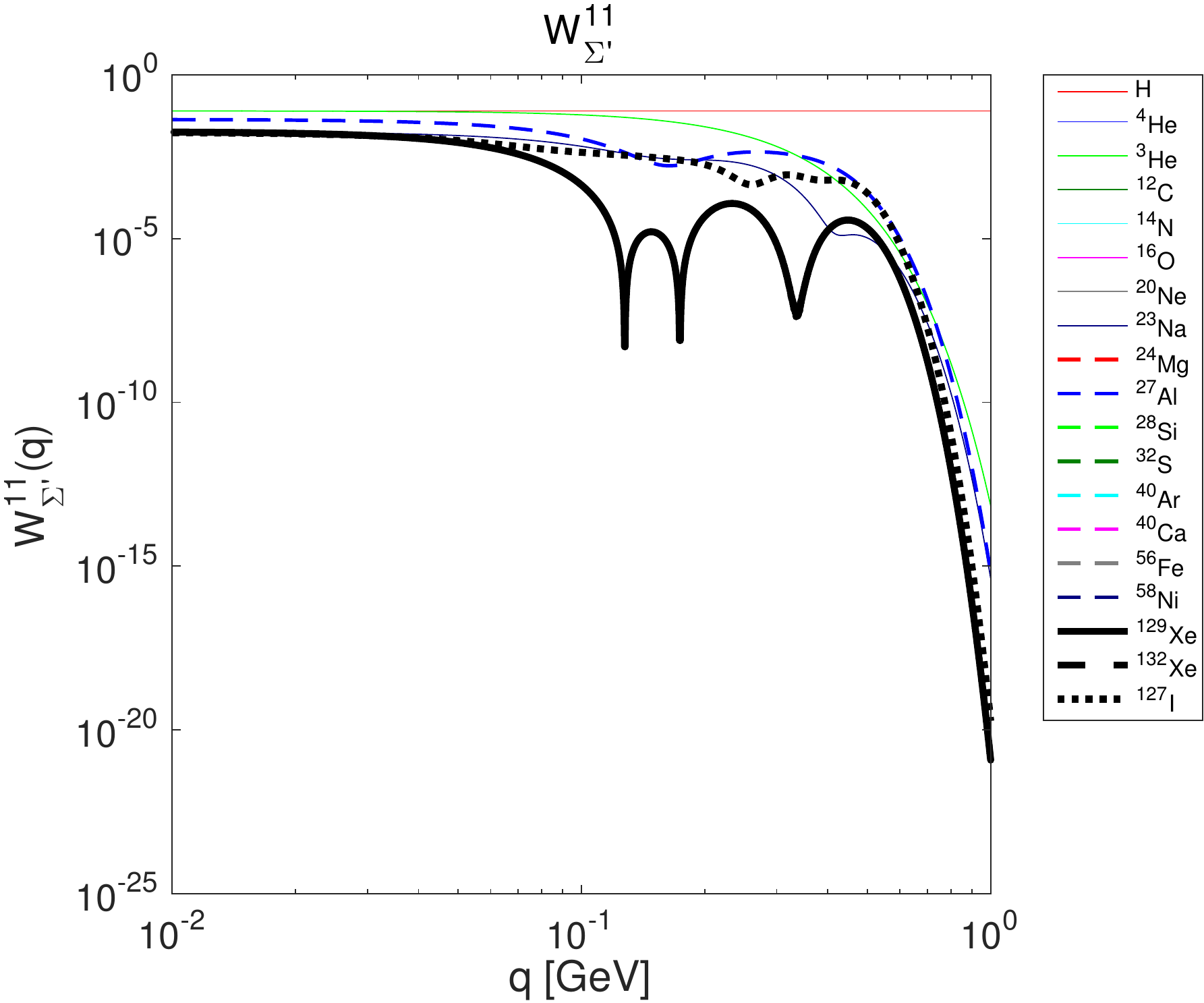}
\end{minipage}
\begin{minipage}[t]{0.49\linewidth}
\centering
\includegraphics[width=\textwidth]{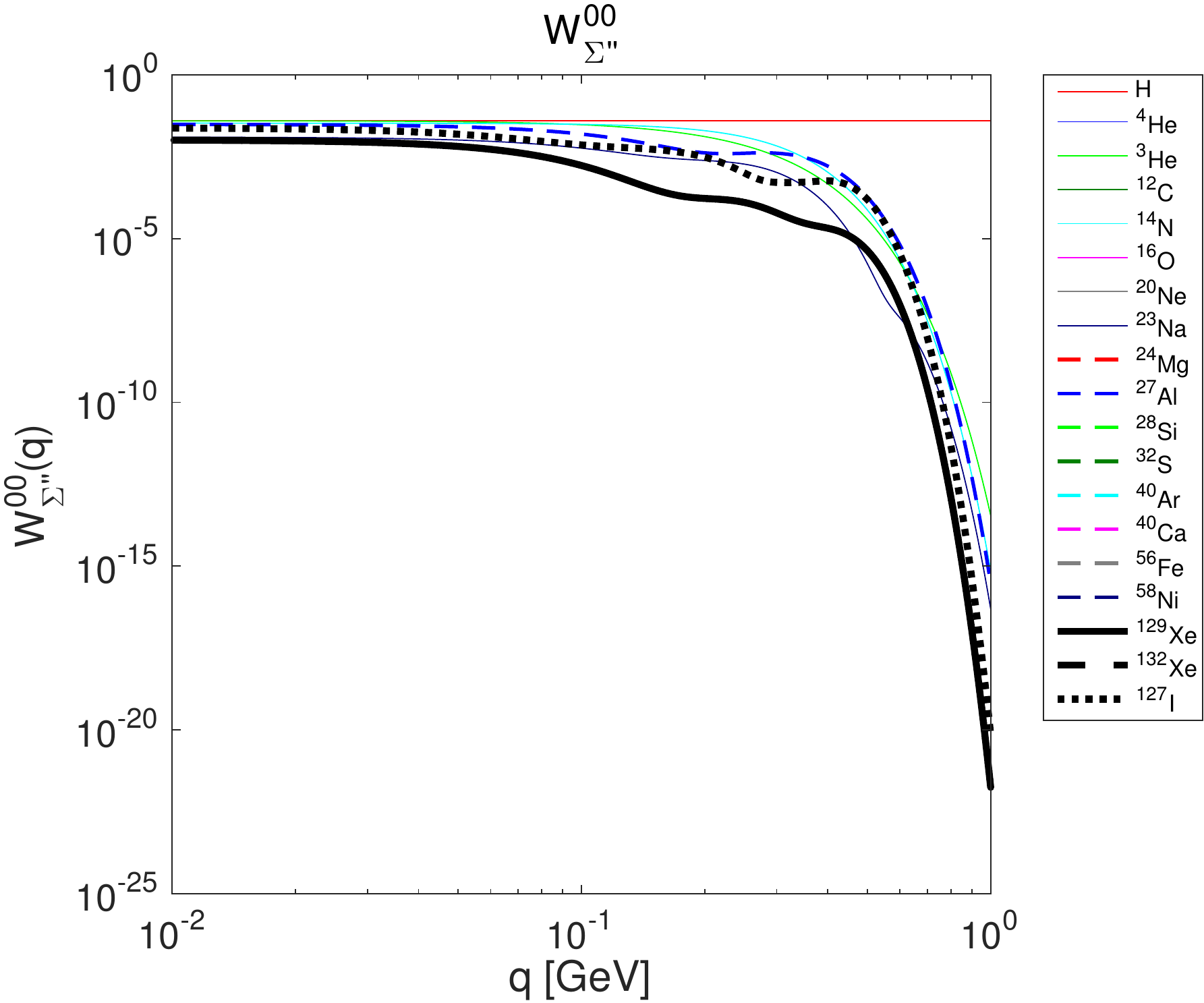}
\end{minipage}
\begin{minipage}[t]{0.49\linewidth}
\centering
\includegraphics[width=\textwidth]{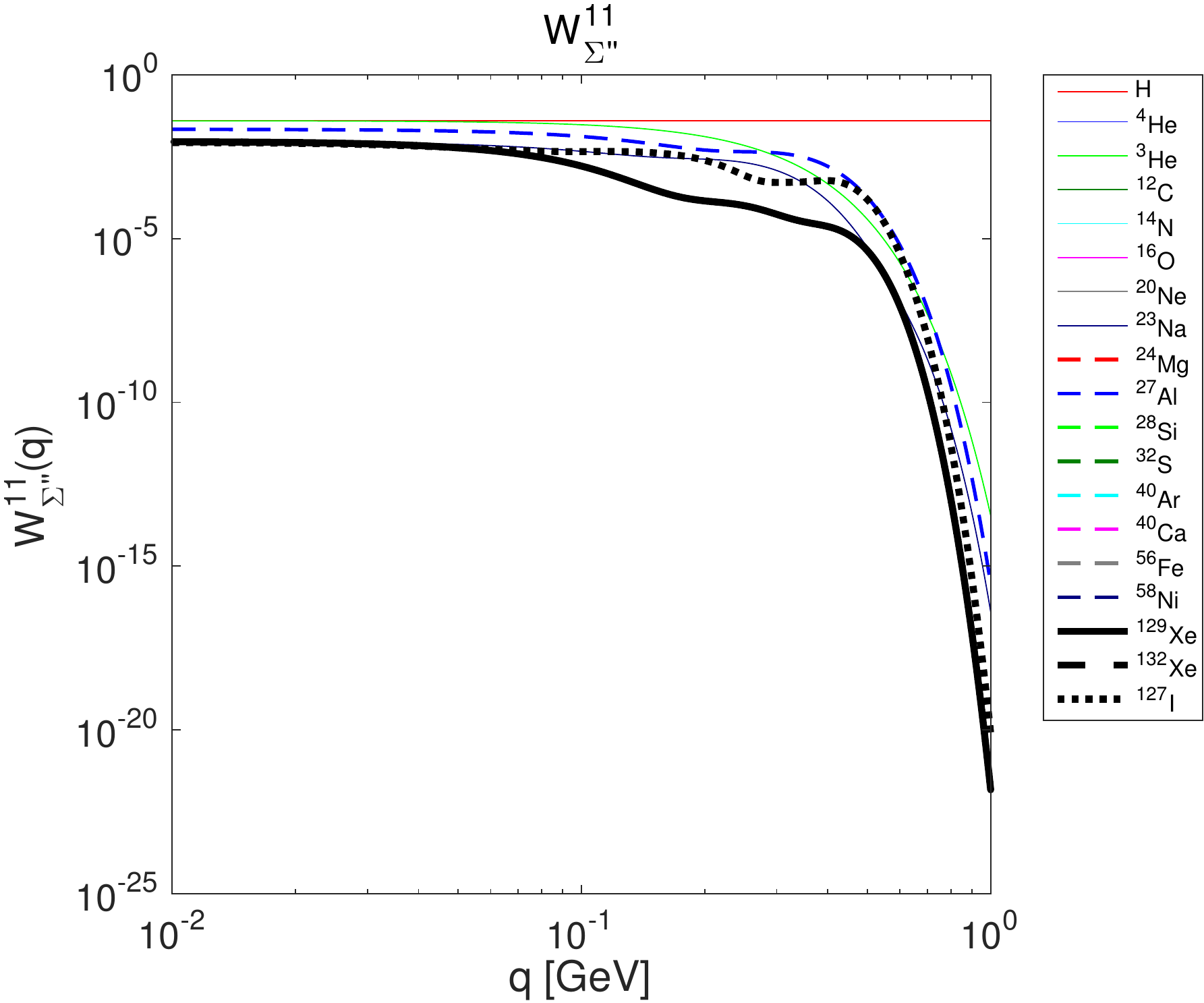}
\end{minipage}
\end{center}
\caption{Same as for Fig.~\ref{fig:W1}, but now for the nuclear response functions $W_{\Sigma'}^{\tau\tau'}$ and $W_{\Sigma''}^{\tau\tau'}$.}
\label{fig:W2}
\end{figure}
\begin{figure}[t]
\begin{center}
\begin{minipage}[t]{0.49\linewidth}
\centering
\includegraphics[width=\textwidth]{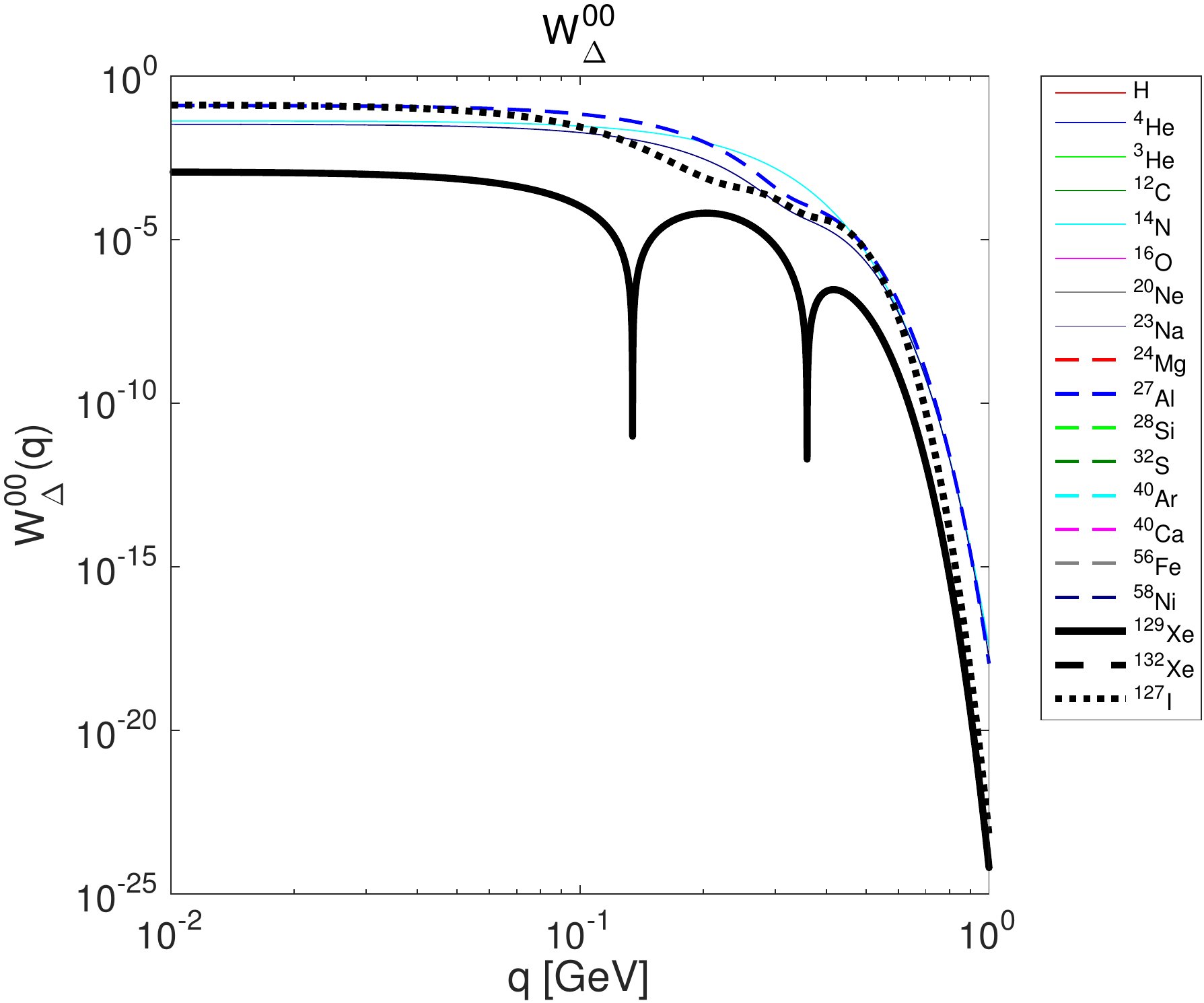}
\end{minipage}
\begin{minipage}[t]{0.49\linewidth}
\centering
\includegraphics[width=\textwidth]{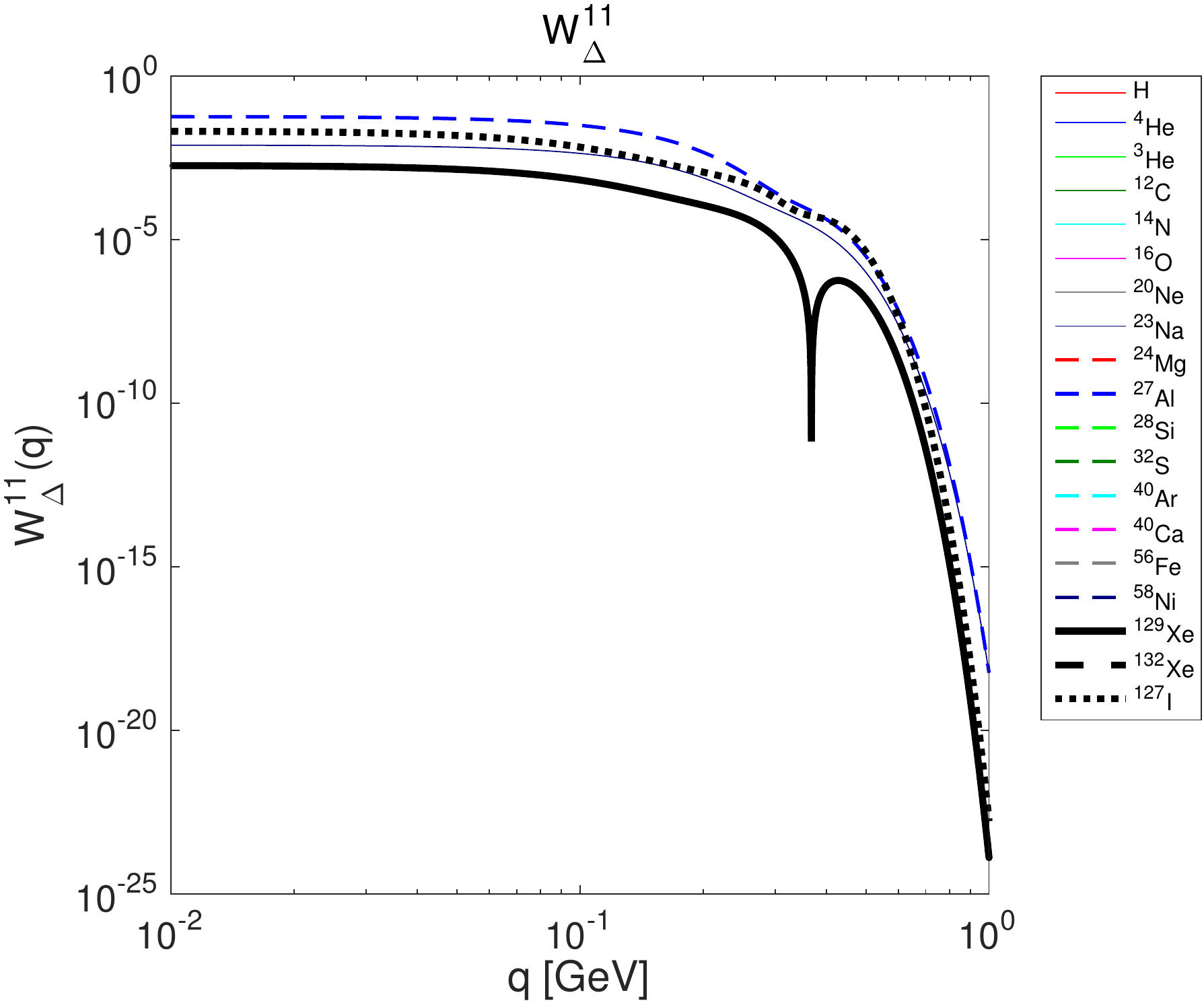}
\end{minipage}
\begin{minipage}[t]{0.49\linewidth}
\centering
\includegraphics[width=\textwidth]{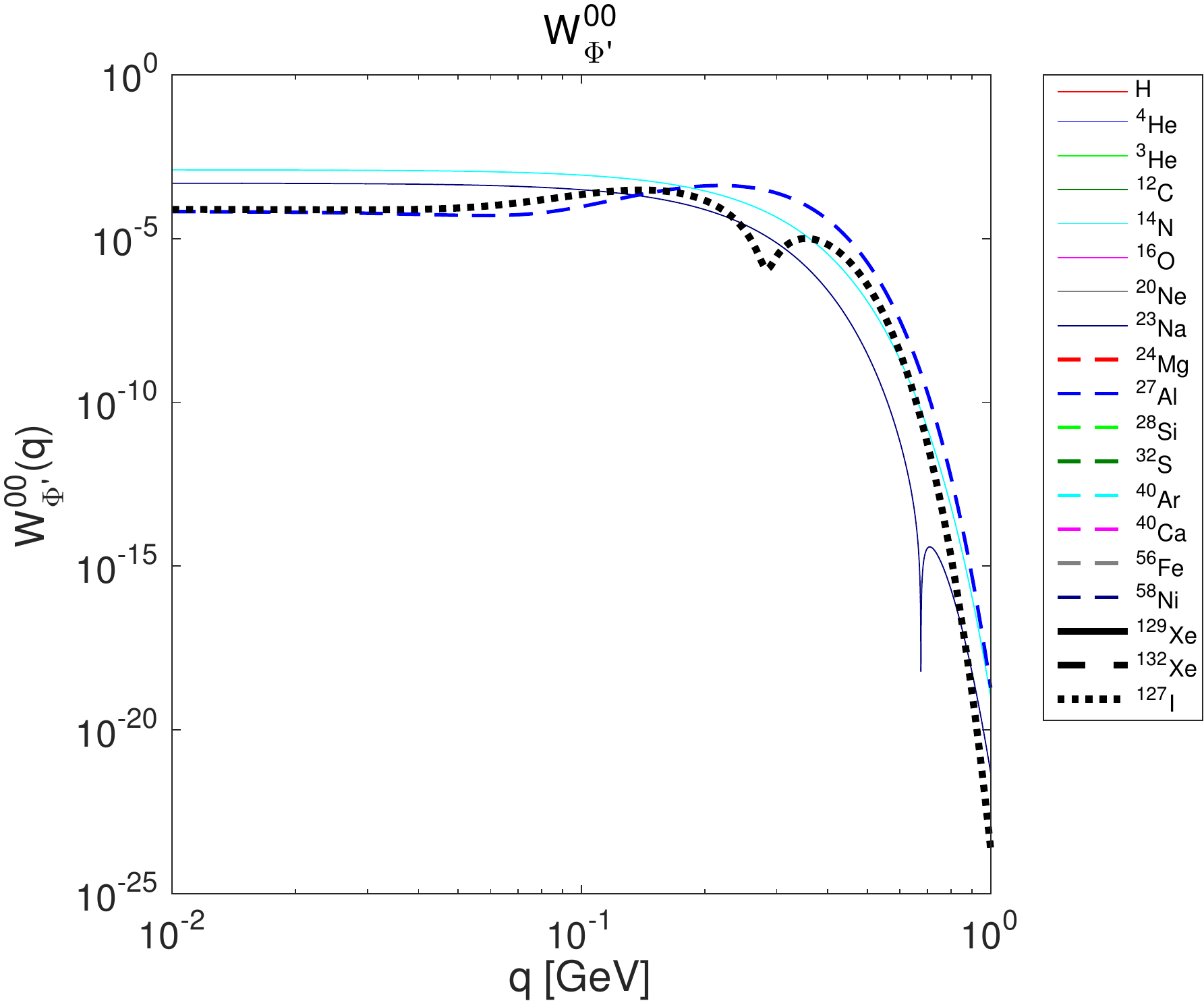}
\end{minipage}
\begin{minipage}[t]{0.49\linewidth}
\centering
\includegraphics[width=\textwidth]{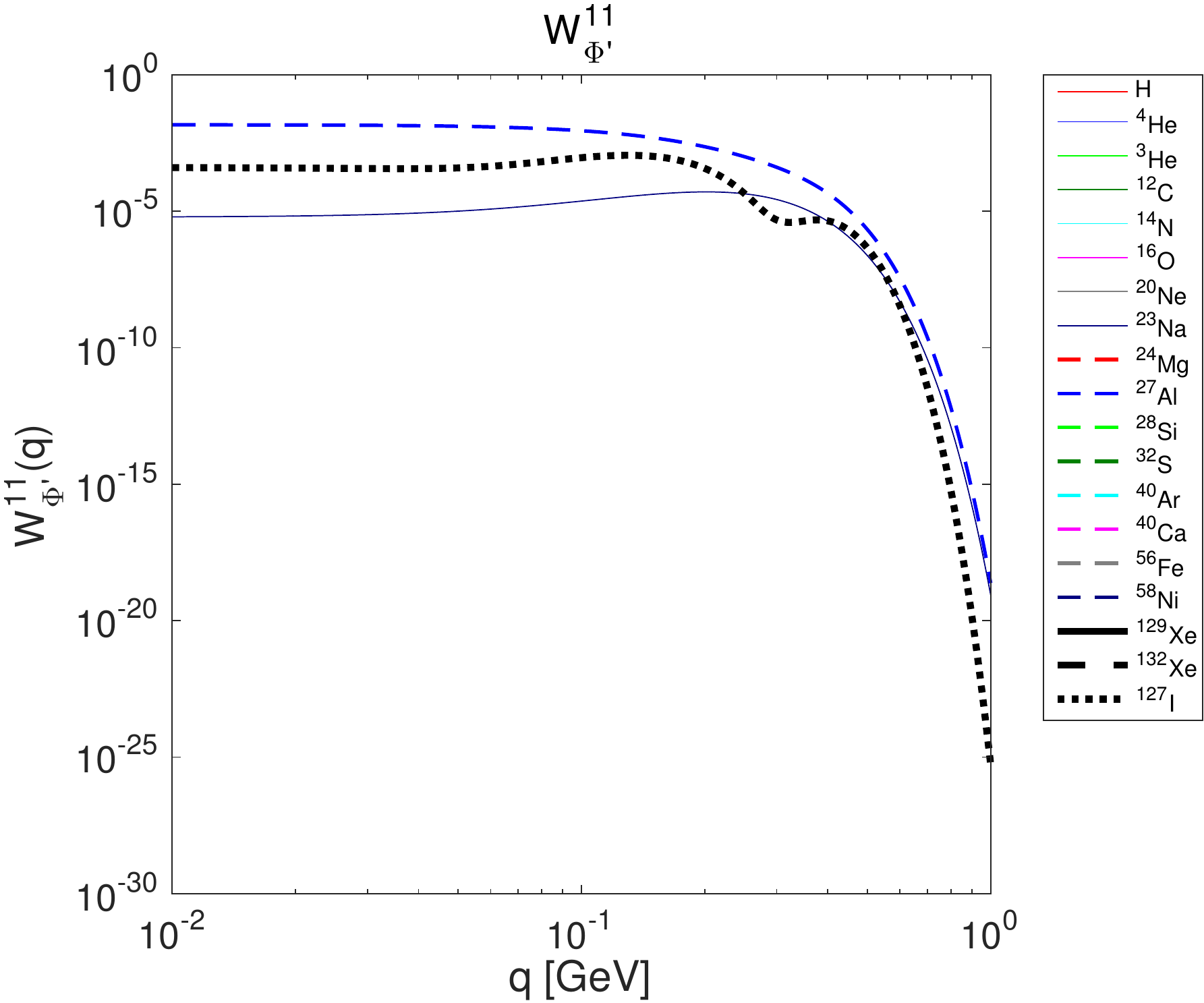}
\end{minipage}
\end{center}
\caption{Same as for Fig.~\ref{fig:W1}, but now for the nuclear response functions $W_{\Delta}^{\tau\tau'}$ and $W_{\tilde{\Phi}'}^{\tau\tau'}$.}
\label{fig:W3}
\end{figure}
\label{sec:obdme}
\begin{figure}[t]
\begin{center}
\begin{minipage}[t]{0.49\linewidth}
\centering
\includegraphics[width=\textwidth]{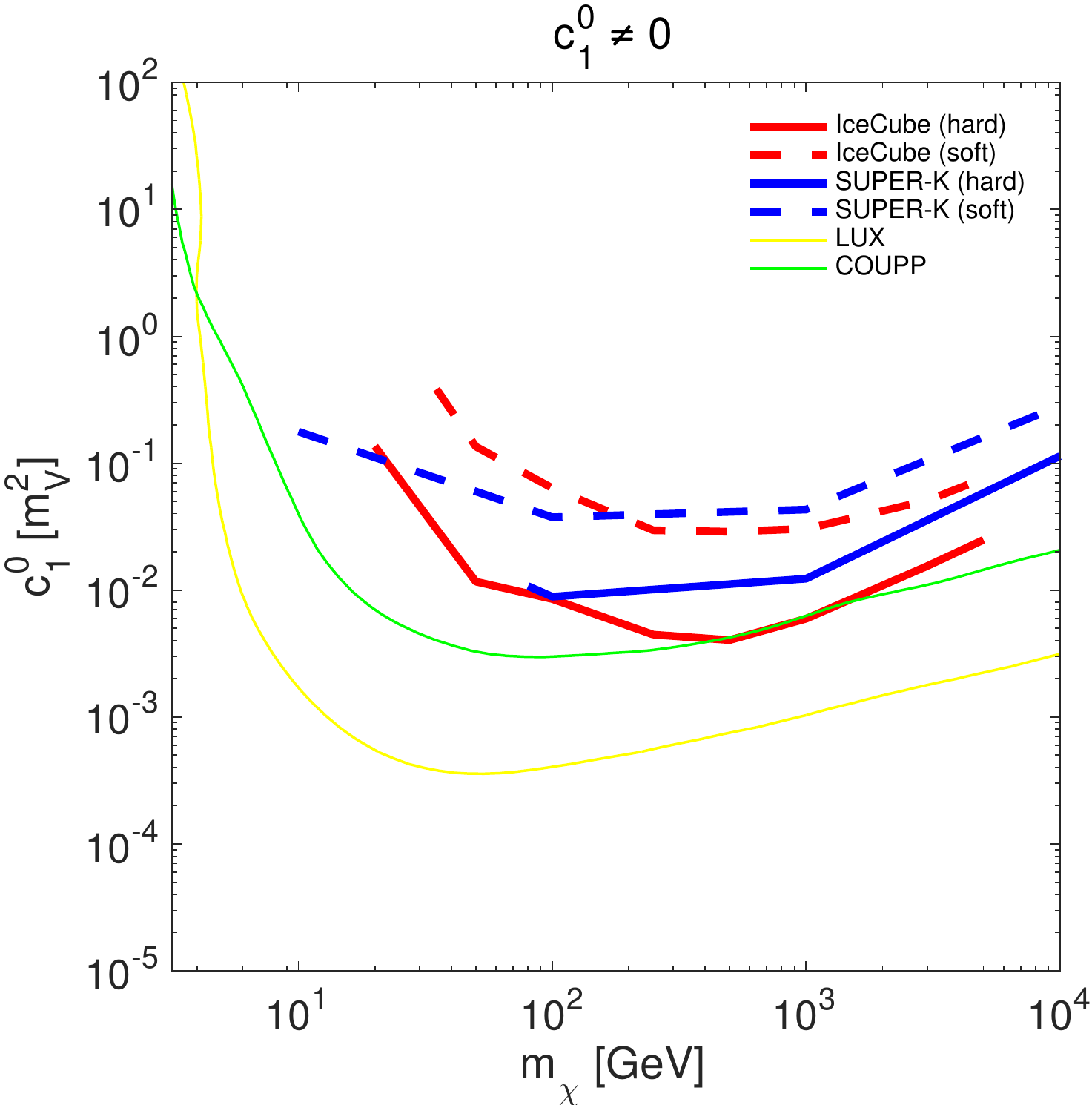}
\end{minipage}
\begin{minipage}[t]{0.49\linewidth}
\centering
\includegraphics[width=\textwidth]{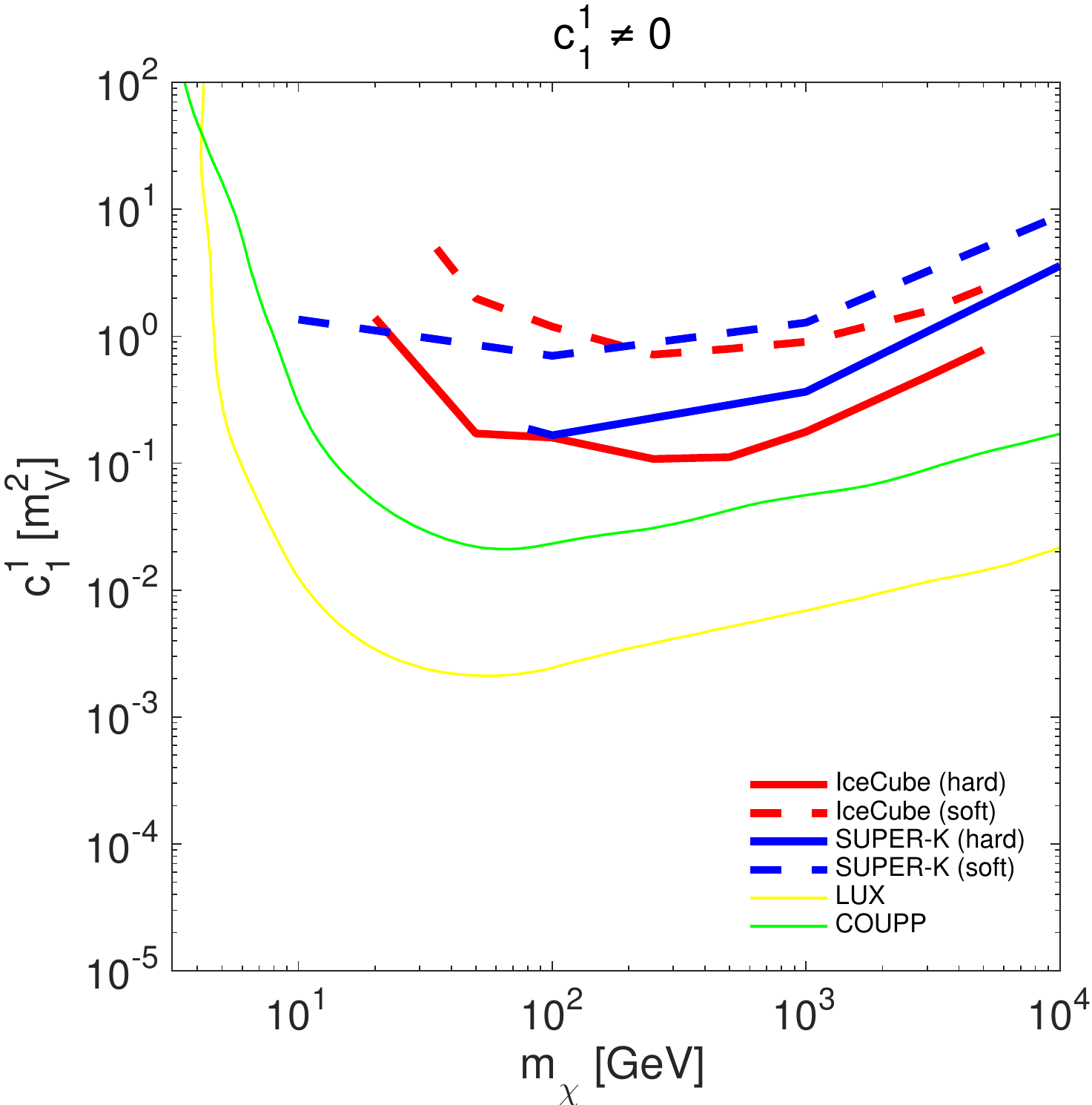}
\end{minipage}
\begin{minipage}[t]{0.49\linewidth}
\centering
\includegraphics[width=\textwidth]{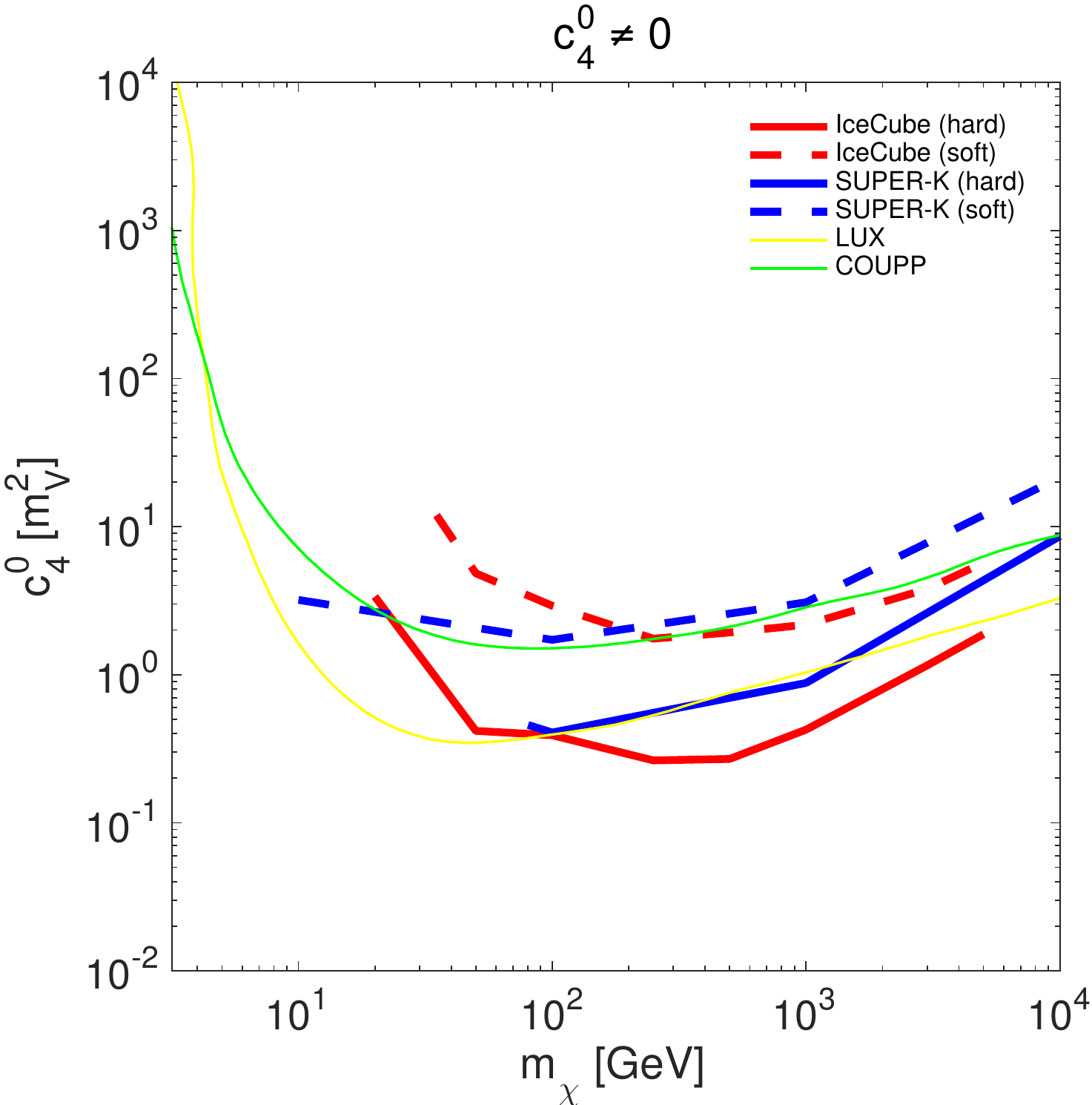}
\end{minipage}
\begin{minipage}[t]{0.49\linewidth}
\centering
\includegraphics[width=\textwidth]{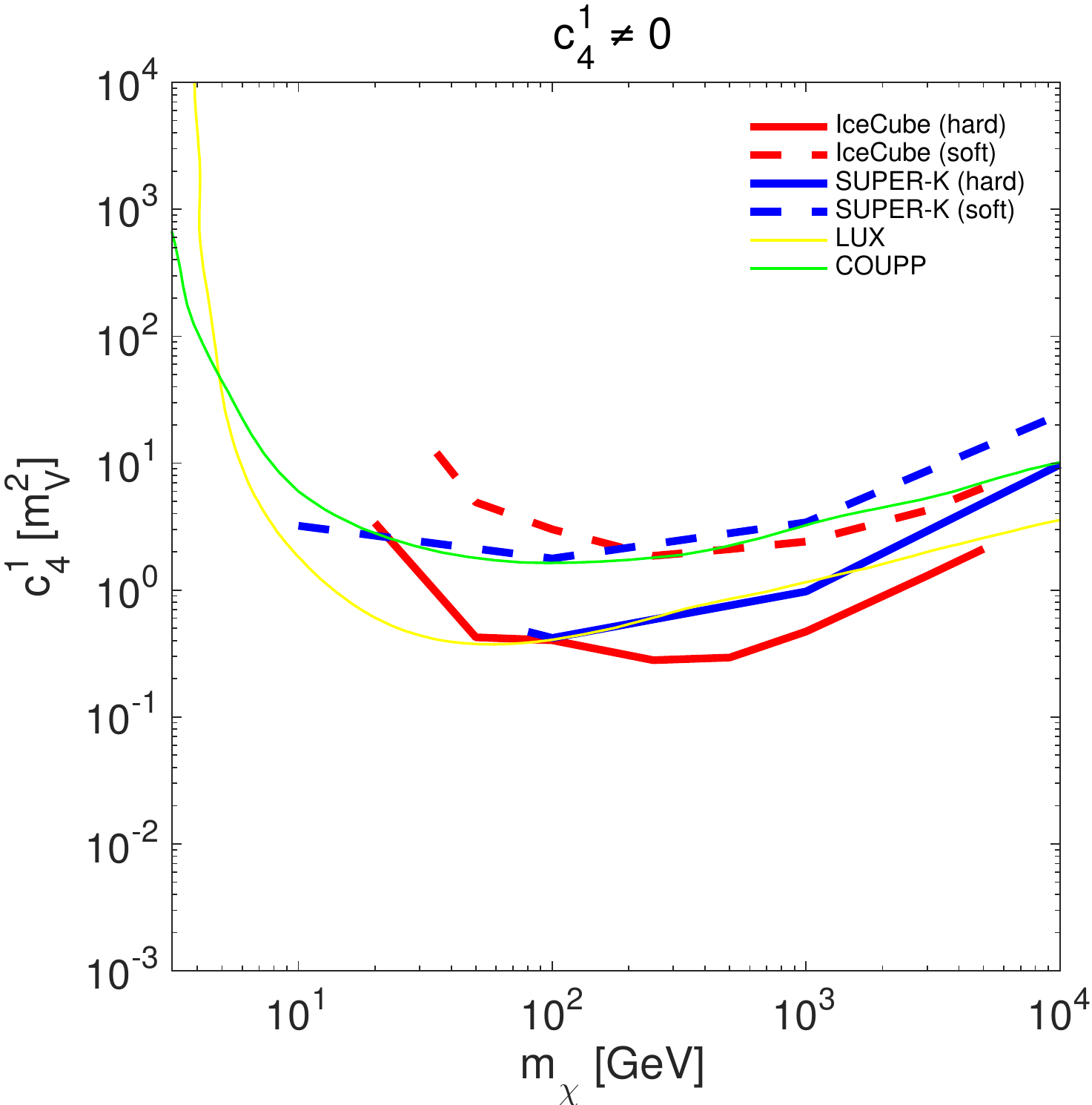}
\end{minipage}
\end{center}
\caption{Exclusion limits on the isoscalar and isovector coupling constants corresponding to the operators $\hat{\mathcal{O}}_1$ and $\hat{\mathcal{O}}_4$. Limits are presented at the 90\% confidence level. Solid red (blue) contours correspond to an analysis of the IceCube (SUPER-K) data which assumes dark matter pair annihilation into $W^+W^-$ (or into $\tau^+\tau^-$, see text at the beginning of Sec.~\ref{sec:analysis} for more details). Dashed red (blue) contours refer to an analysis of the IceCube (SUPER-K) data which assumes dark matter pair annihilation into $b\bar{b}$. For comparison, we also report the 2D 90\% credible regions that we obtain from LUX (yellow) and COUPP (green). Coupling constants are expressed in units of $m_V^2=246.2$~GeV.}
\label{fig:c1c4}
\end{figure}
\begin{figure}[t]
\begin{center}
\begin{minipage}[t]{0.49\linewidth}
\centering
\includegraphics[width=\textwidth]{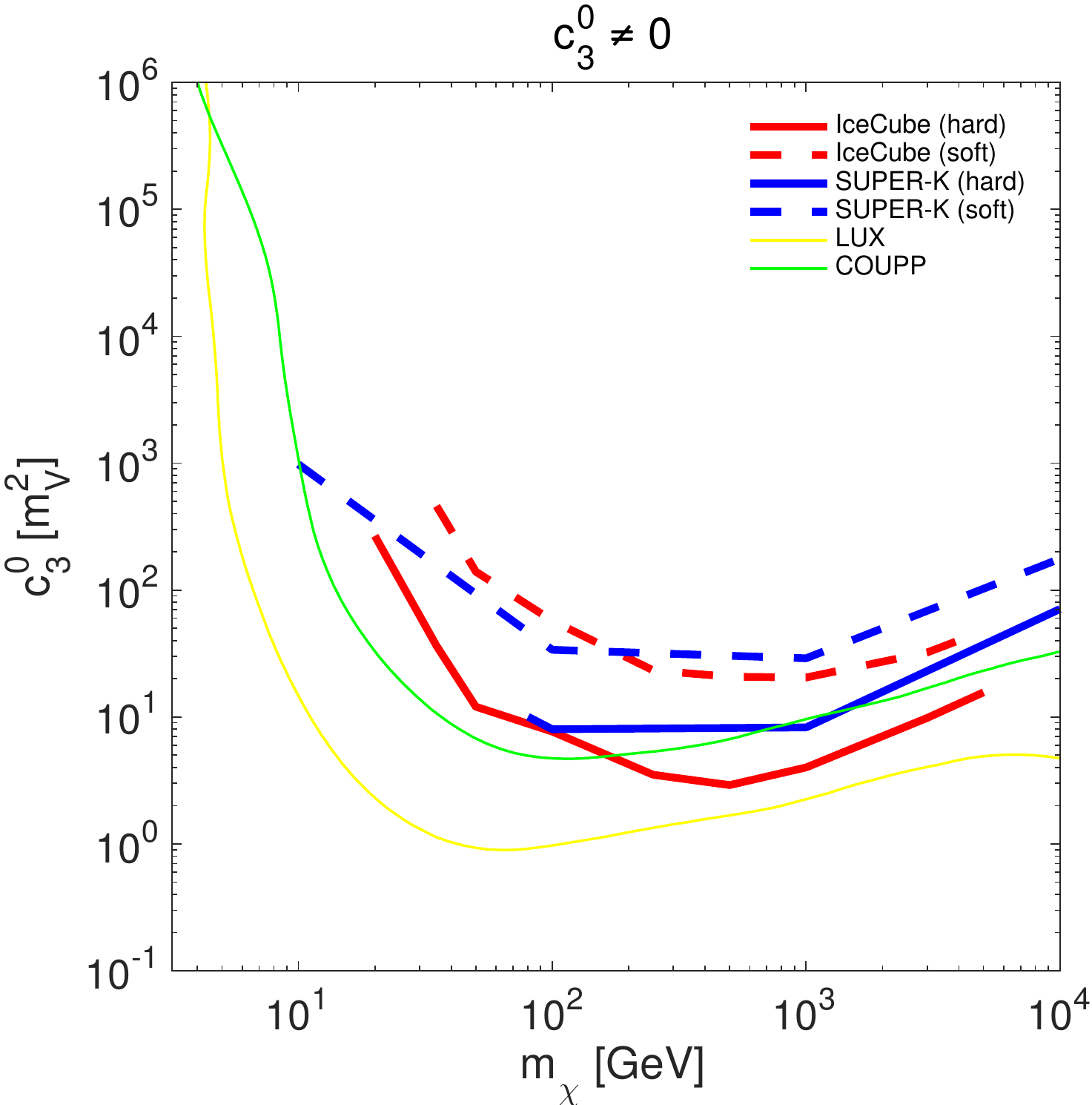}
\end{minipage}
\begin{minipage}[t]{0.49\linewidth}
\centering
\includegraphics[width=\textwidth]{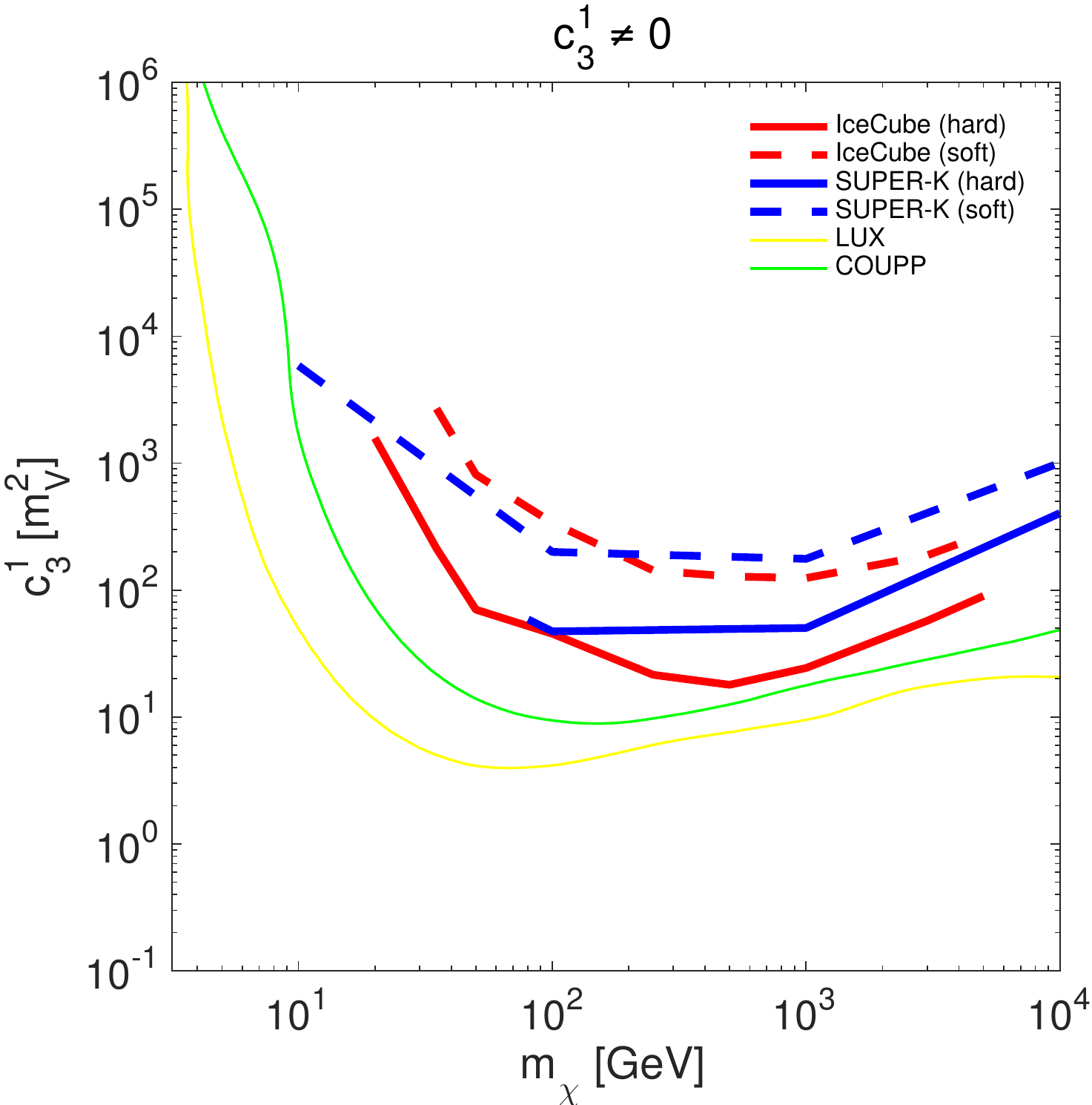}
\end{minipage}
\begin{minipage}[t]{0.49\linewidth}
\centering
\includegraphics[width=\textwidth]{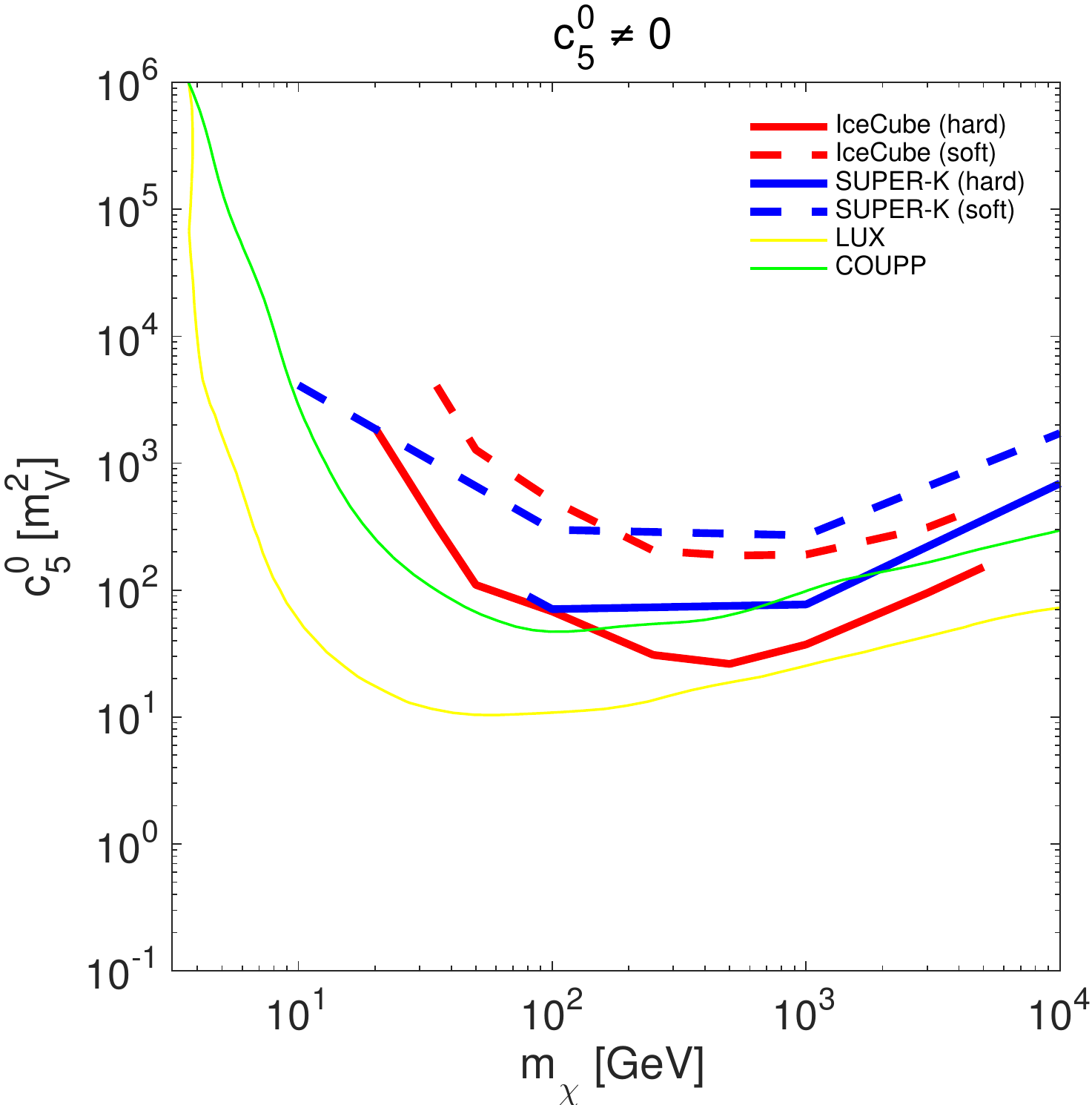}
\end{minipage}
\begin{minipage}[t]{0.49\linewidth}
\centering
\includegraphics[width=\textwidth]{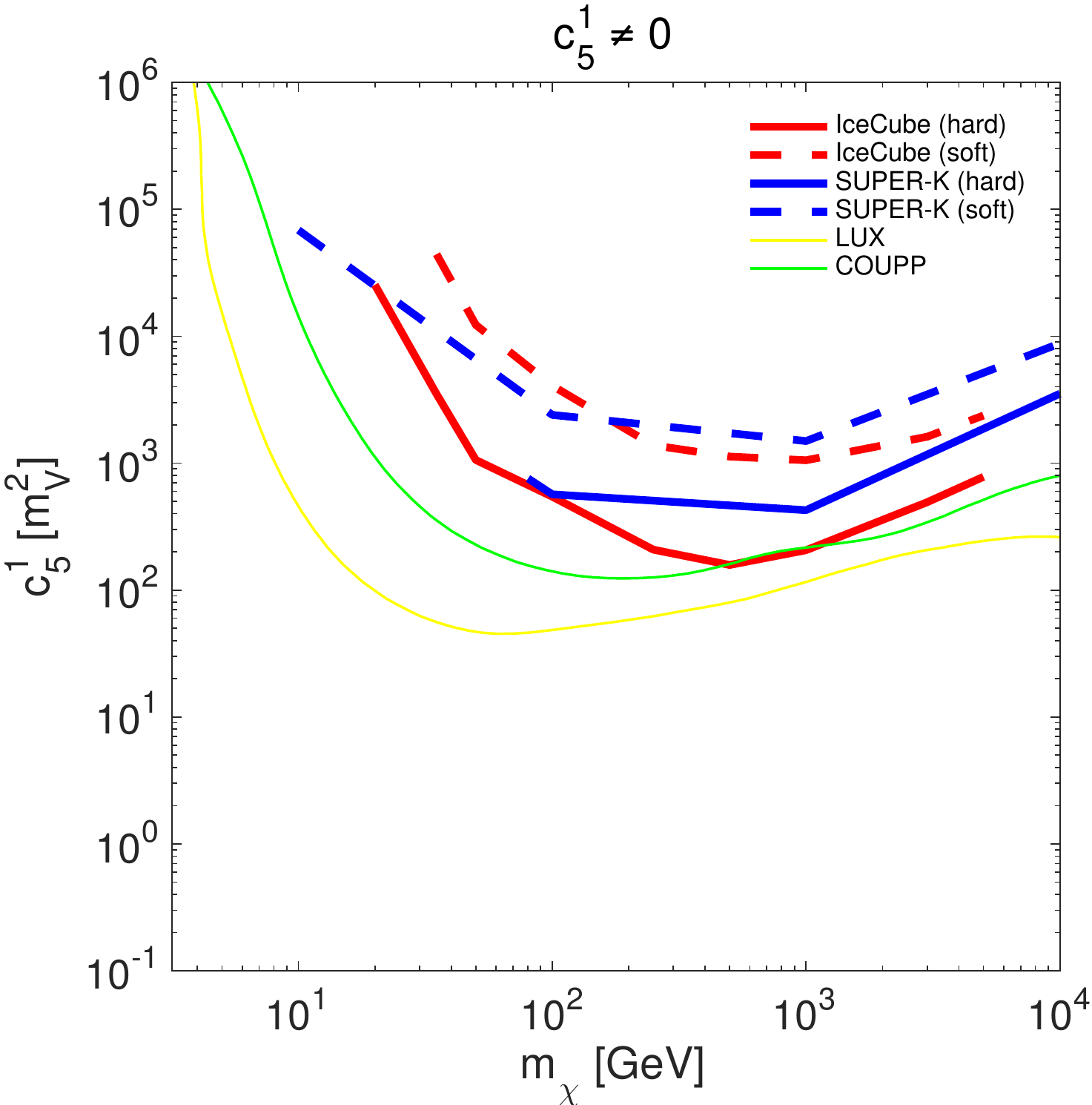}
\end{minipage}
\end{center}
\caption{Same as for Fig.~\ref{fig:c1c4}, but now for the operators $\hat{\mathcal{O}}_{3}$ and $\hat{\mathcal{O}}_{5}$.}
\label{fig:c3c5}
\end{figure}
\begin{figure}[t]
\begin{center}
\begin{minipage}[t]{0.49\linewidth}
\centering
\includegraphics[width=\textwidth]{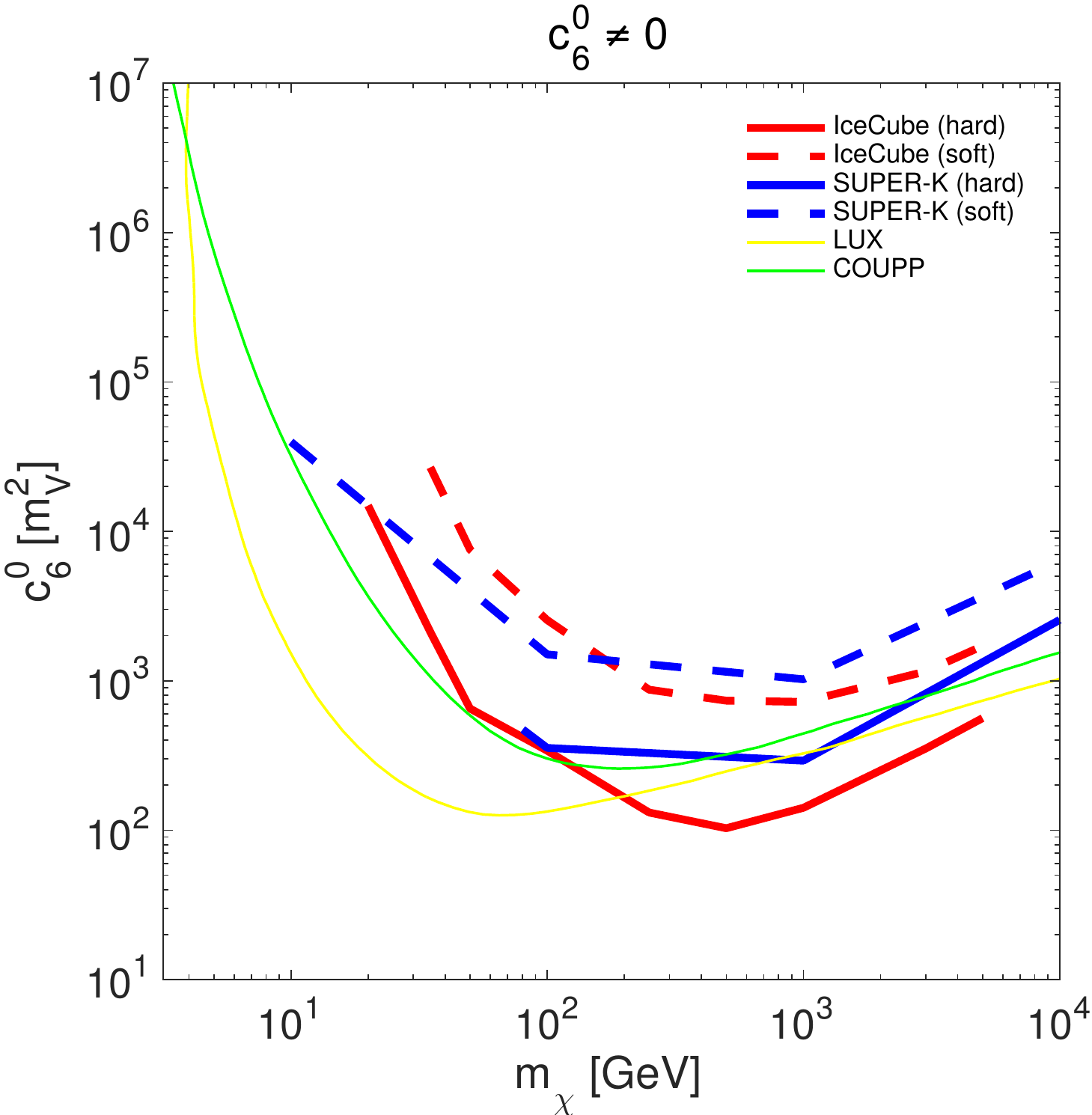}
\end{minipage}
\begin{minipage}[t]{0.49\linewidth}
\centering
\includegraphics[width=\textwidth]{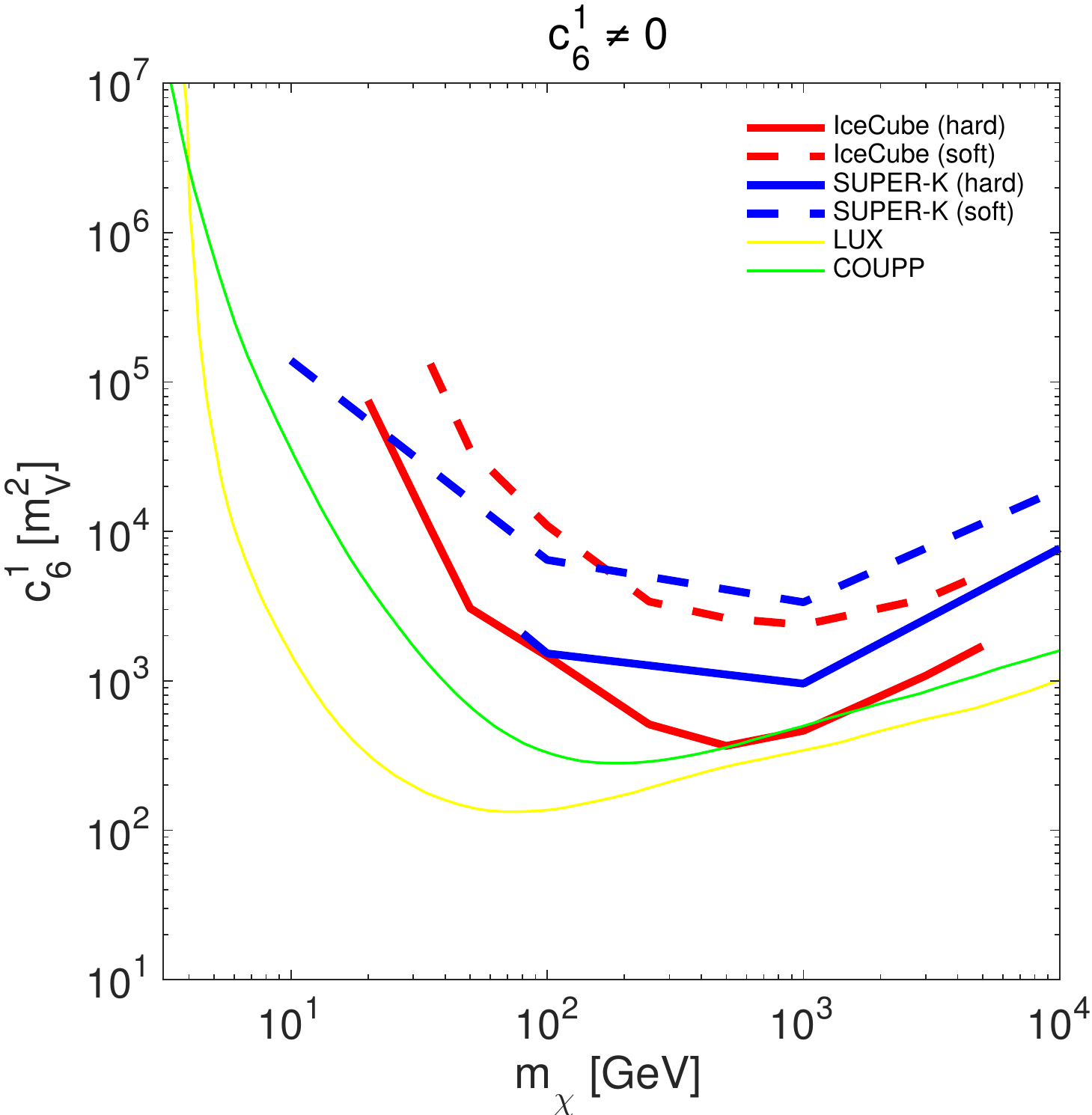}
\end{minipage}
\begin{minipage}[t]{0.49\linewidth}
\centering
\includegraphics[width=\textwidth]{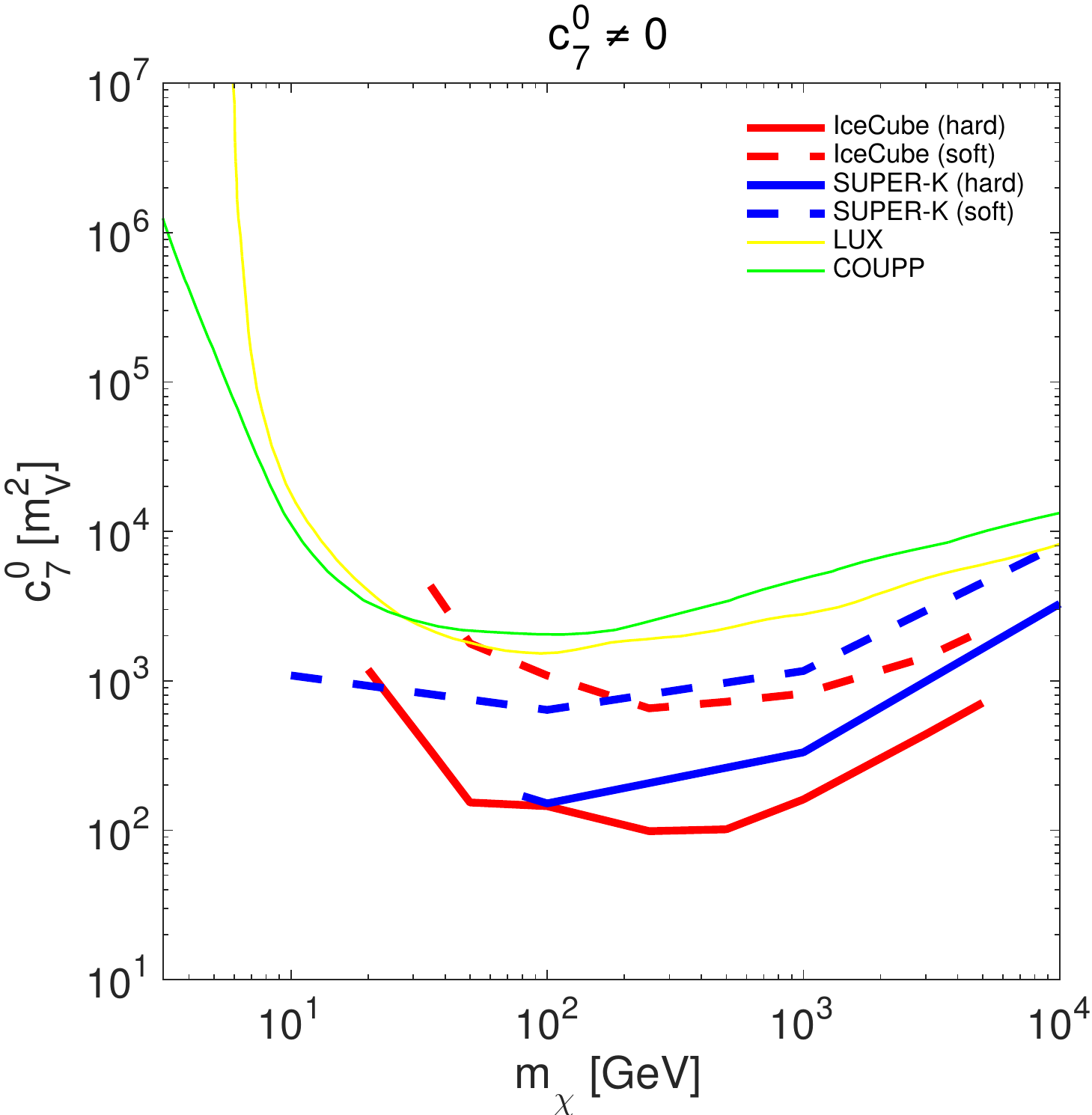}
\end{minipage}
\begin{minipage}[t]{0.49\linewidth}
\centering
\includegraphics[width=\textwidth]{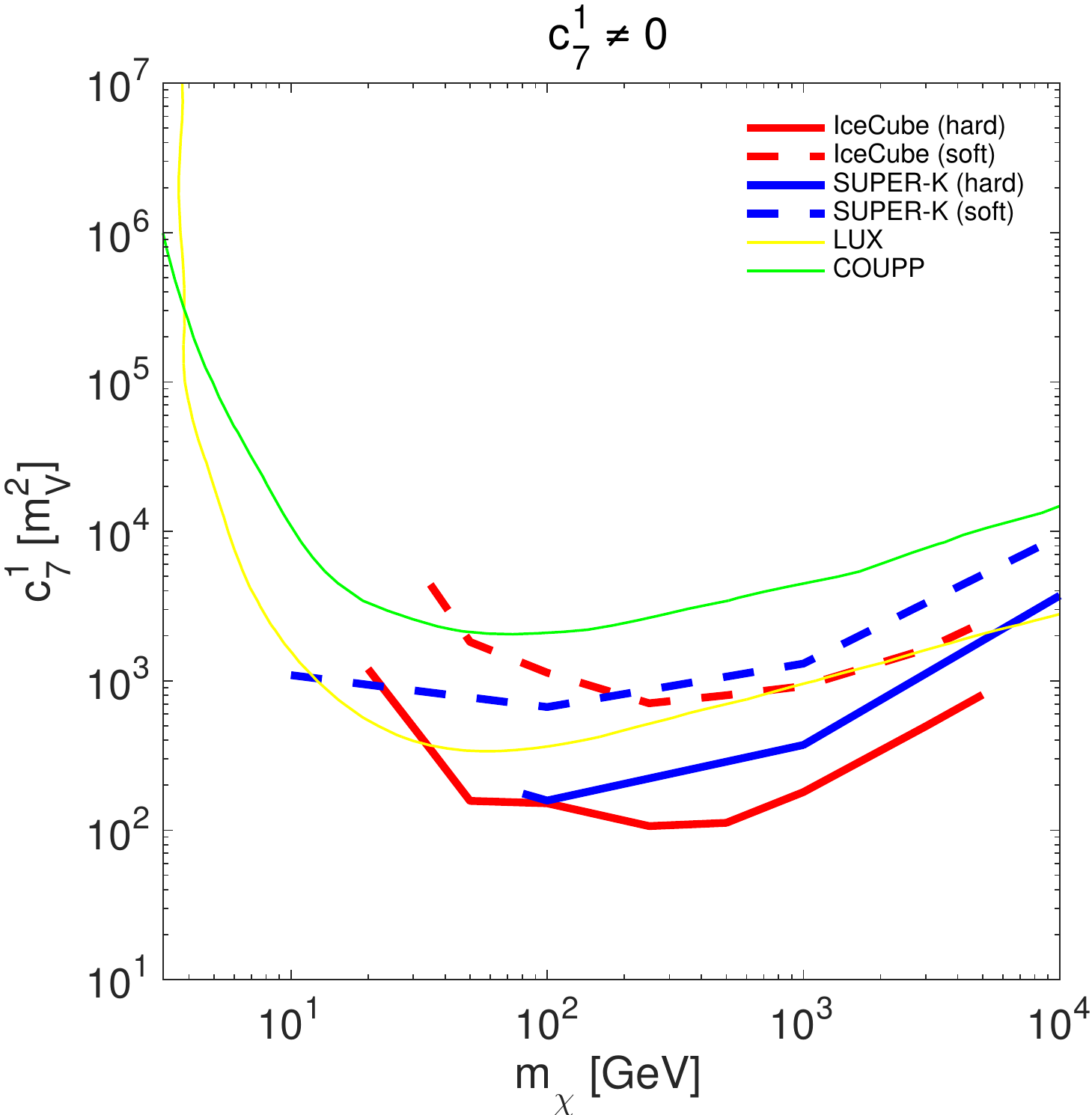}
\end{minipage}
\end{center}
\caption{Same as for Fig.~\ref{fig:c1c4}, but now for the operators $\hat{\mathcal{O}}_{6}$ and $\hat{\mathcal{O}}_{7}$.}
\label{fig:c6c7}
\end{figure}
\begin{figure}[t]
\begin{center}
\begin{minipage}[t]{0.49\linewidth}
\centering
\includegraphics[width=\textwidth]{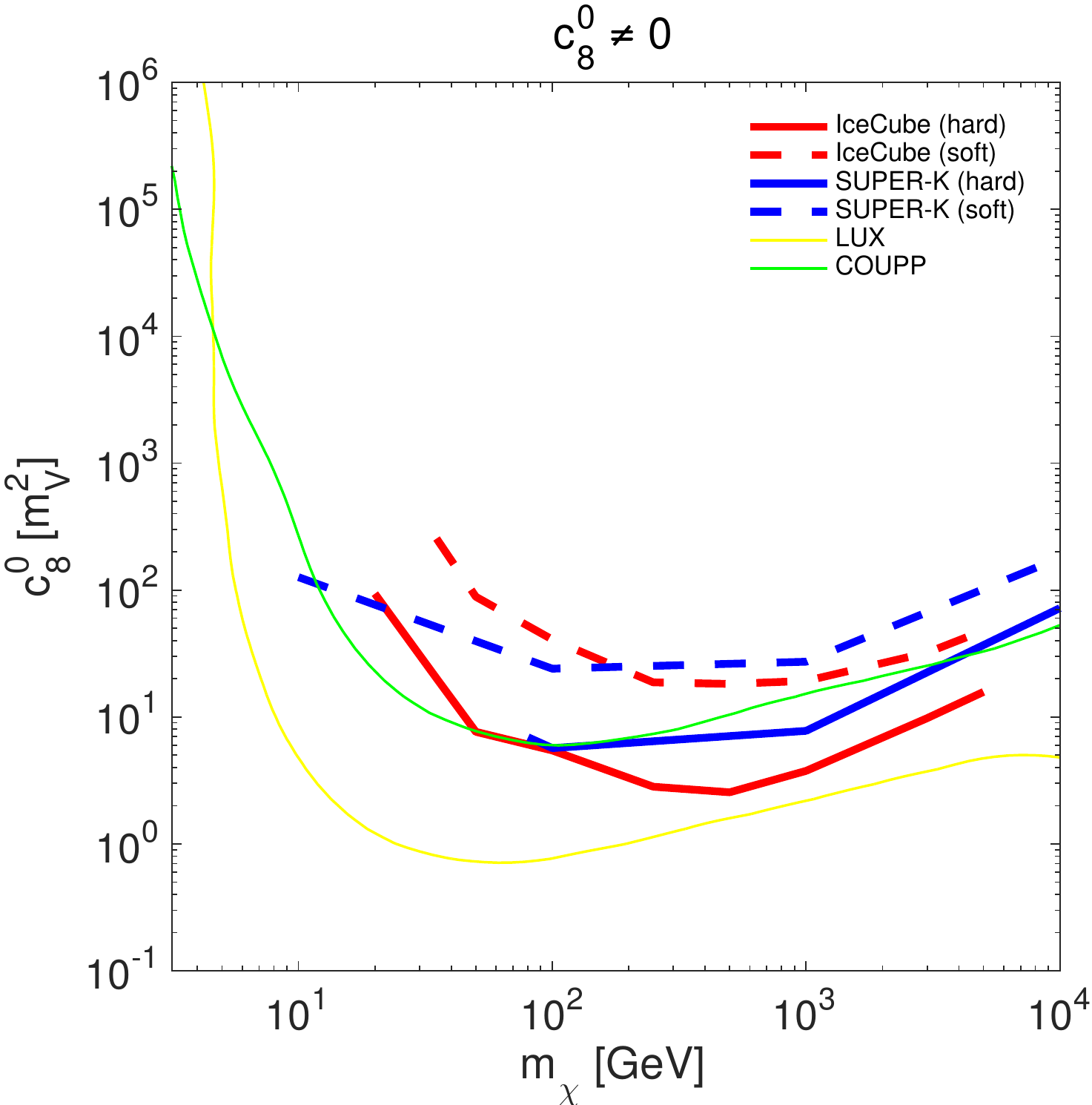}
\end{minipage}
\begin{minipage}[t]{0.49\linewidth}
\centering
\includegraphics[width=\textwidth]{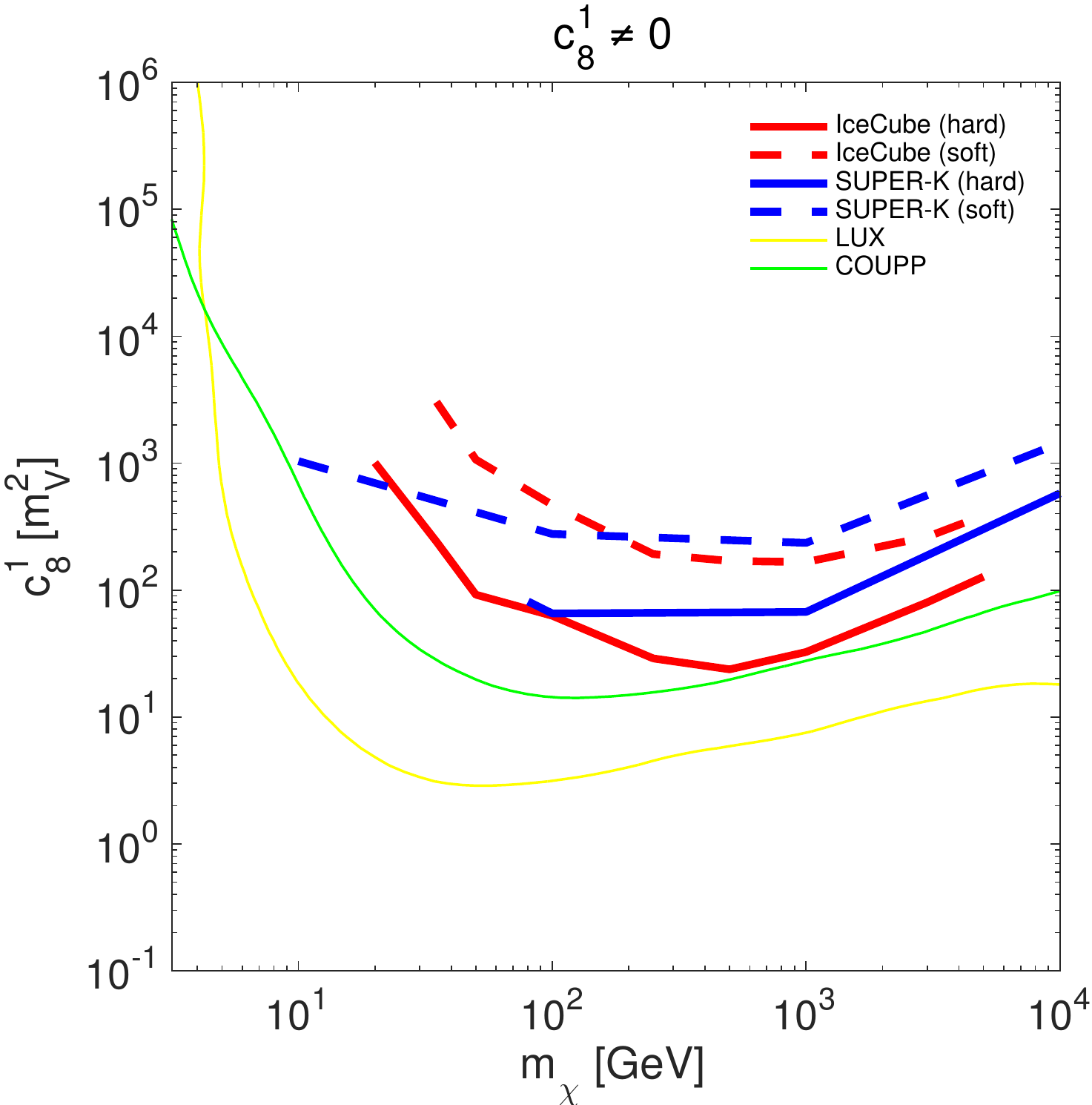}
\end{minipage}
\begin{minipage}[t]{0.49\linewidth}
\centering
\includegraphics[width=\textwidth]{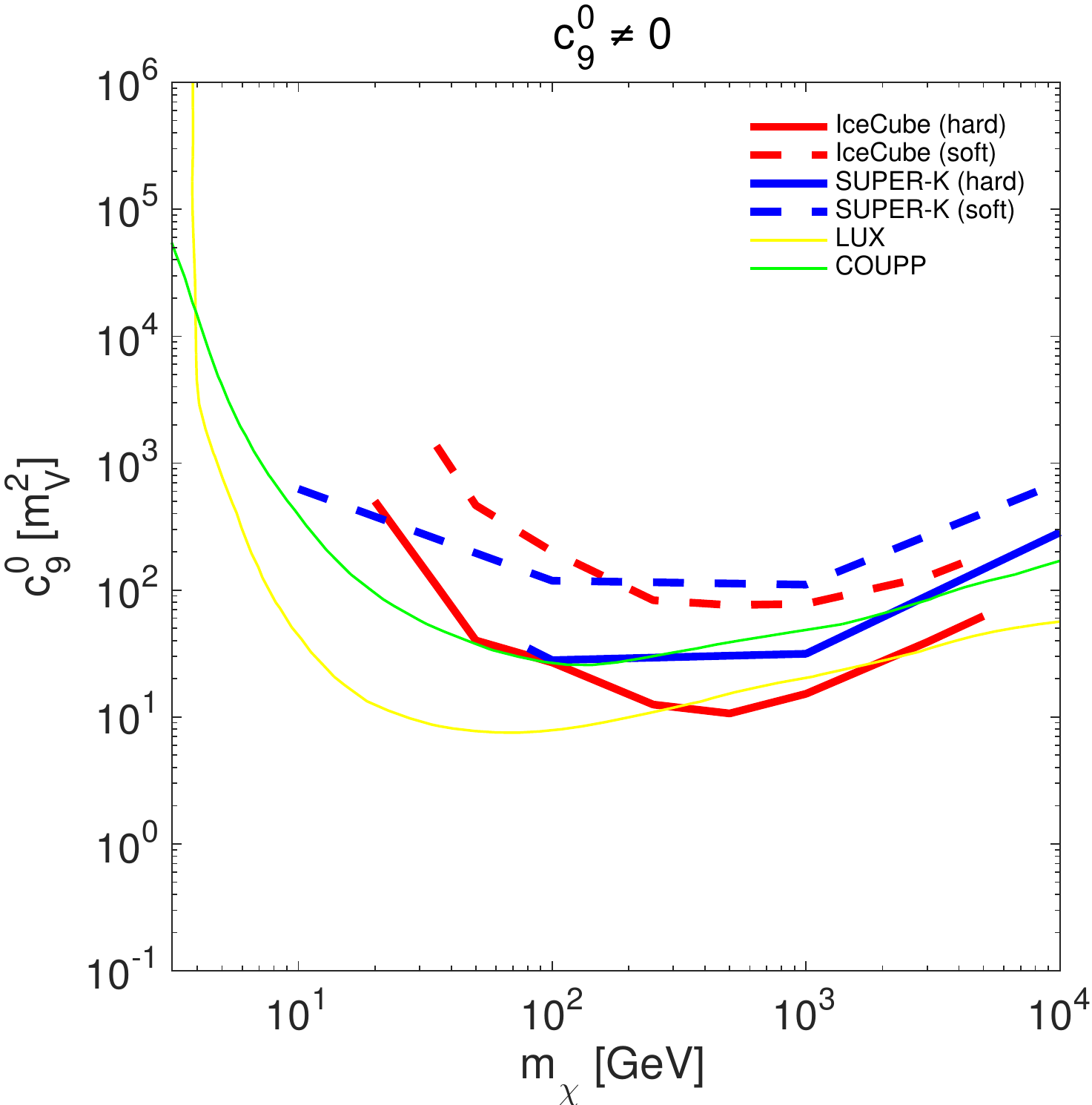}
\end{minipage}
\begin{minipage}[t]{0.49\linewidth}
\centering
\includegraphics[width=\textwidth]{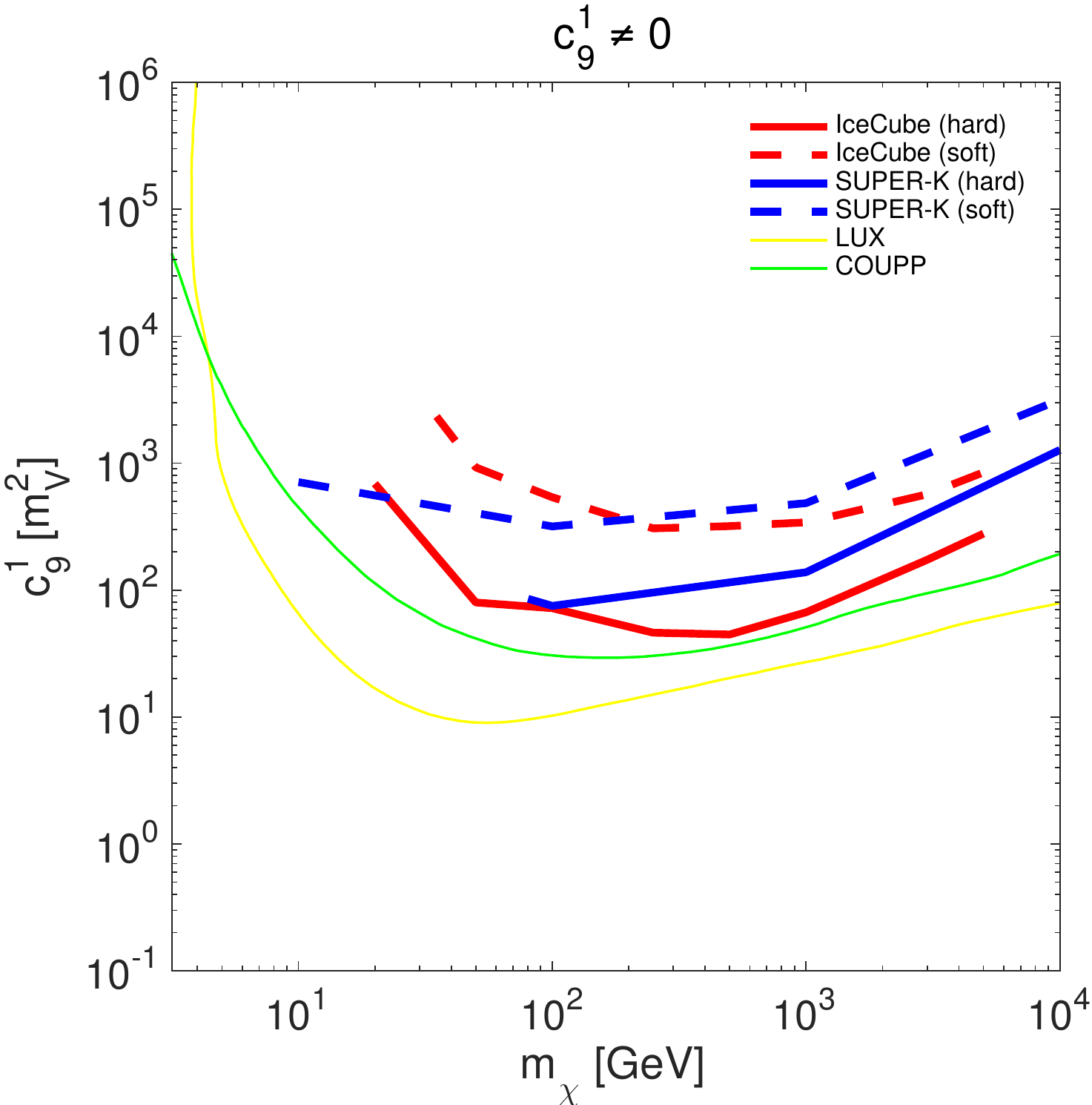}
\end{minipage}
\end{center}
\caption{Same as for Fig.~\ref{fig:c1c4}, but now for the operators $\hat{\mathcal{O}}_{8}$ and $\hat{\mathcal{O}}_{9}$.}
\label{fig:c8c9}
\end{figure}
\begin{figure}[t]
\begin{center}
\begin{minipage}[t]{0.49\linewidth}
\centering
\includegraphics[width=\textwidth]{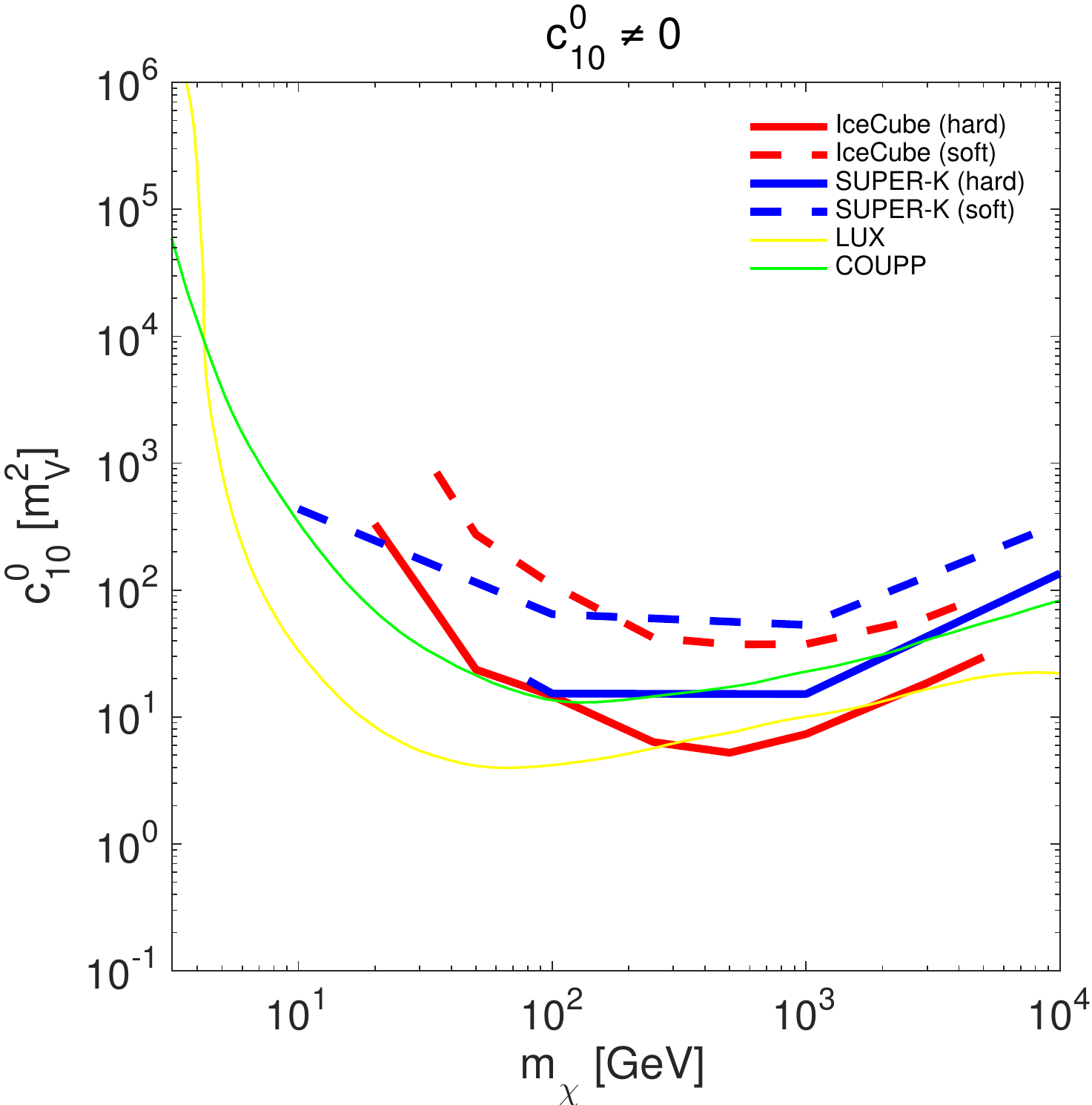}
\end{minipage}
\begin{minipage}[t]{0.49\linewidth}
\centering
\includegraphics[width=\textwidth]{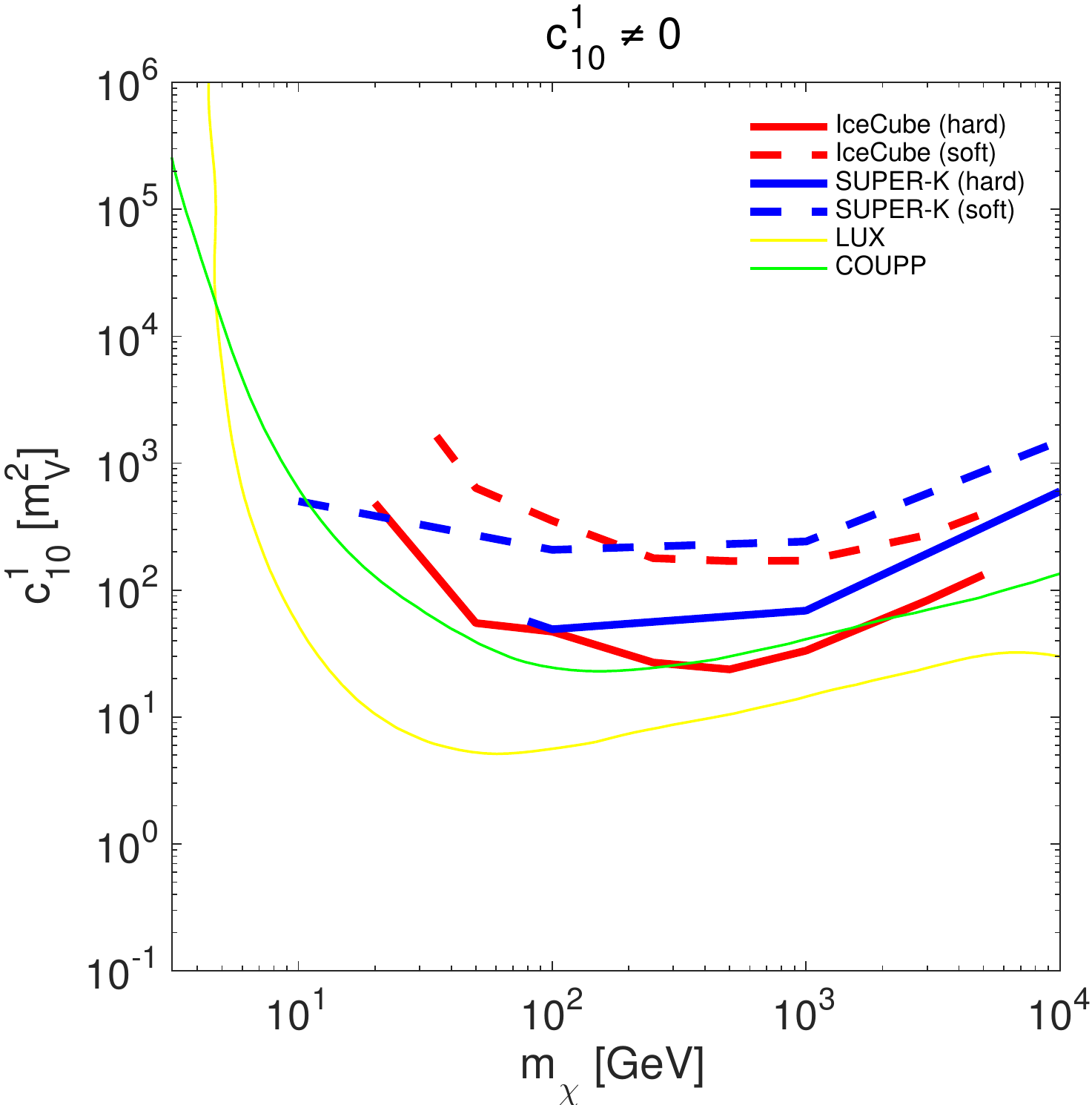}
\end{minipage}
\begin{minipage}[t]{0.49\linewidth}
\centering
\includegraphics[width=\textwidth]{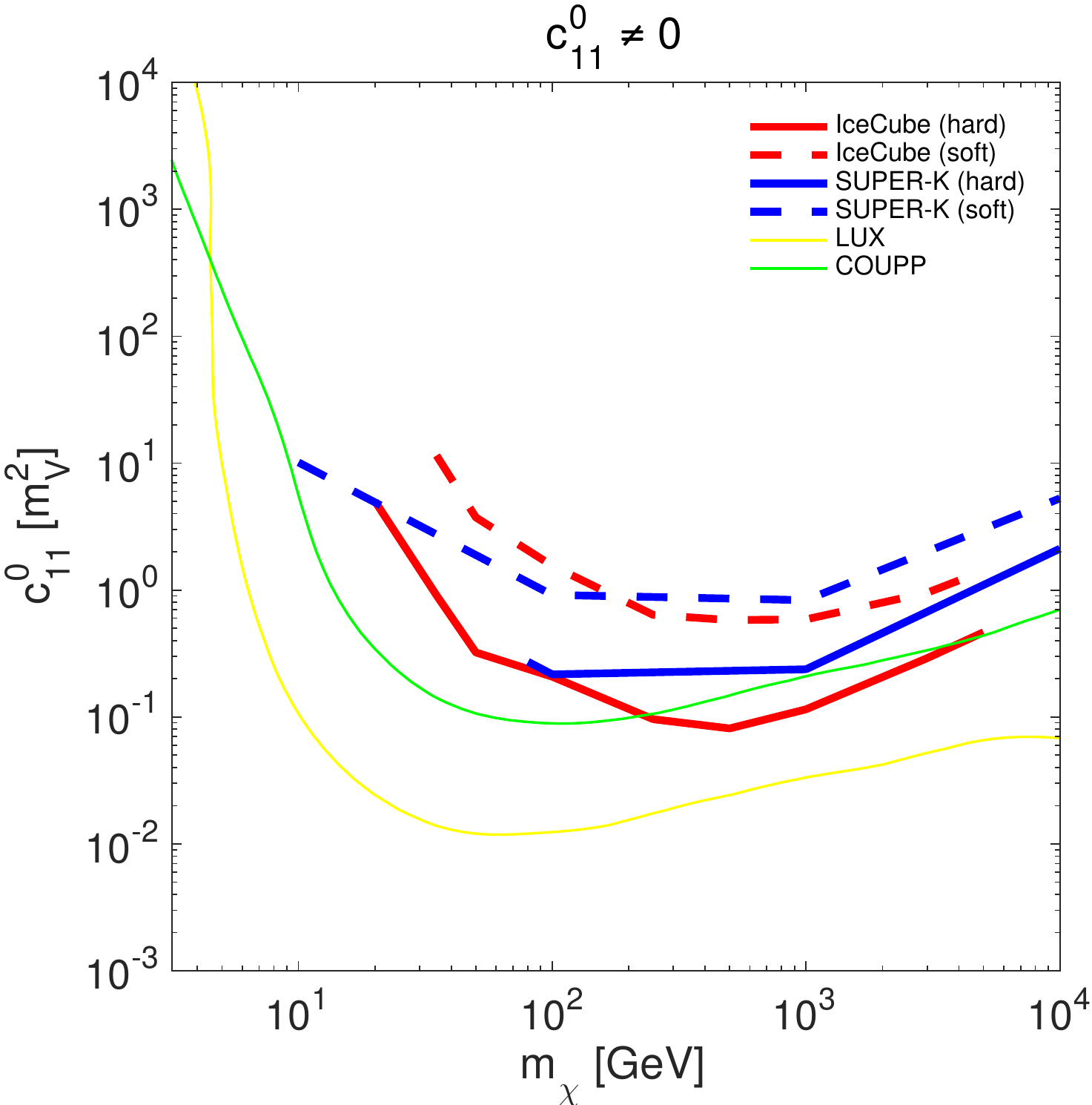}
\end{minipage}
\begin{minipage}[t]{0.49\linewidth}
\centering
\includegraphics[width=\textwidth]{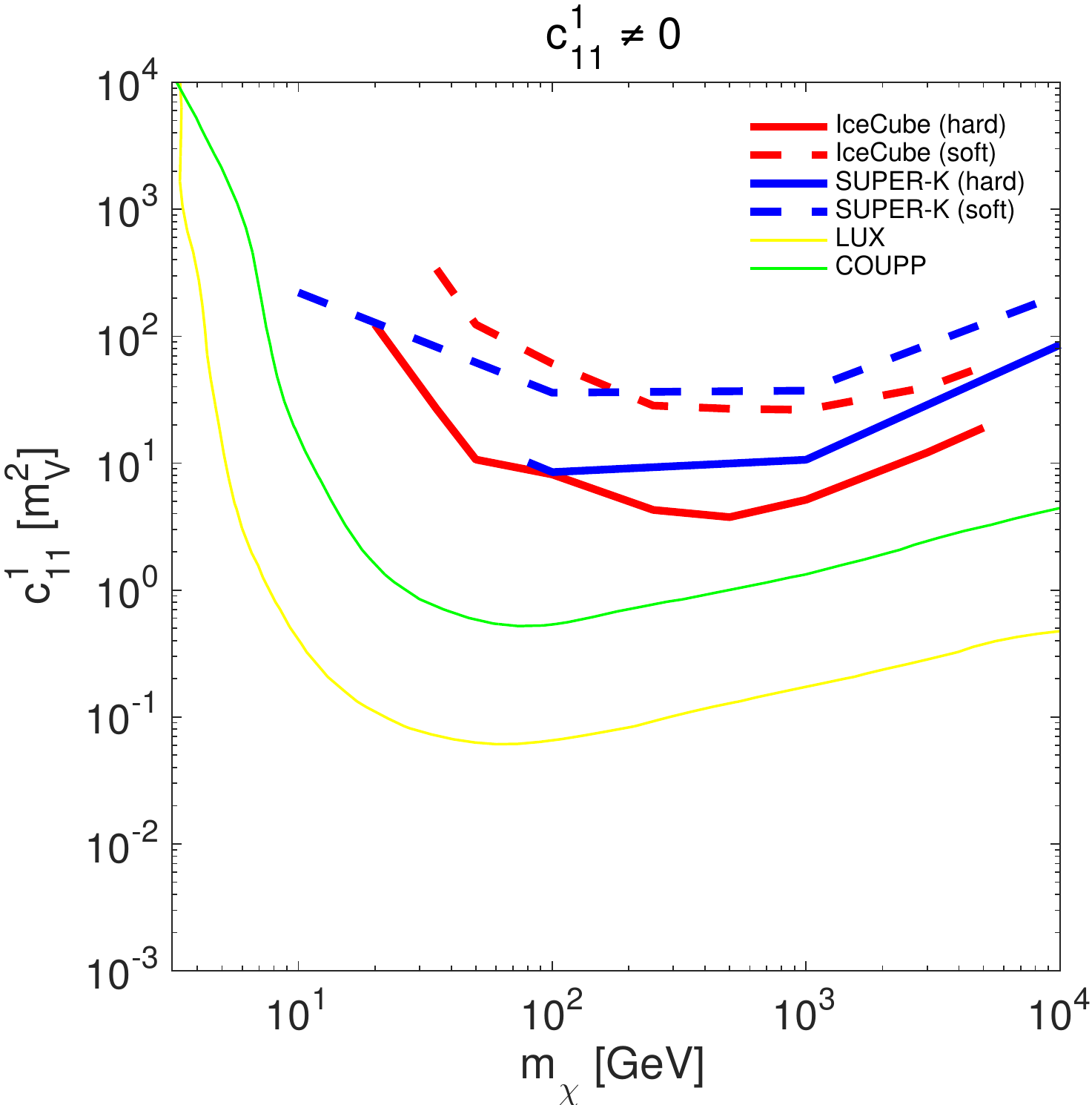}
\end{minipage}
\end{center}
\caption{Same as for Fig.~\ref{fig:c1c4}, but now for the operators $\hat{\mathcal{O}}_{10}$ and $\hat{\mathcal{O}}_{11}$.}
\label{fig:c10c11}
\end{figure}
\begin{figure}[t]
\begin{center}
\begin{minipage}[t]{0.49\linewidth}
\centering
\includegraphics[width=\textwidth]{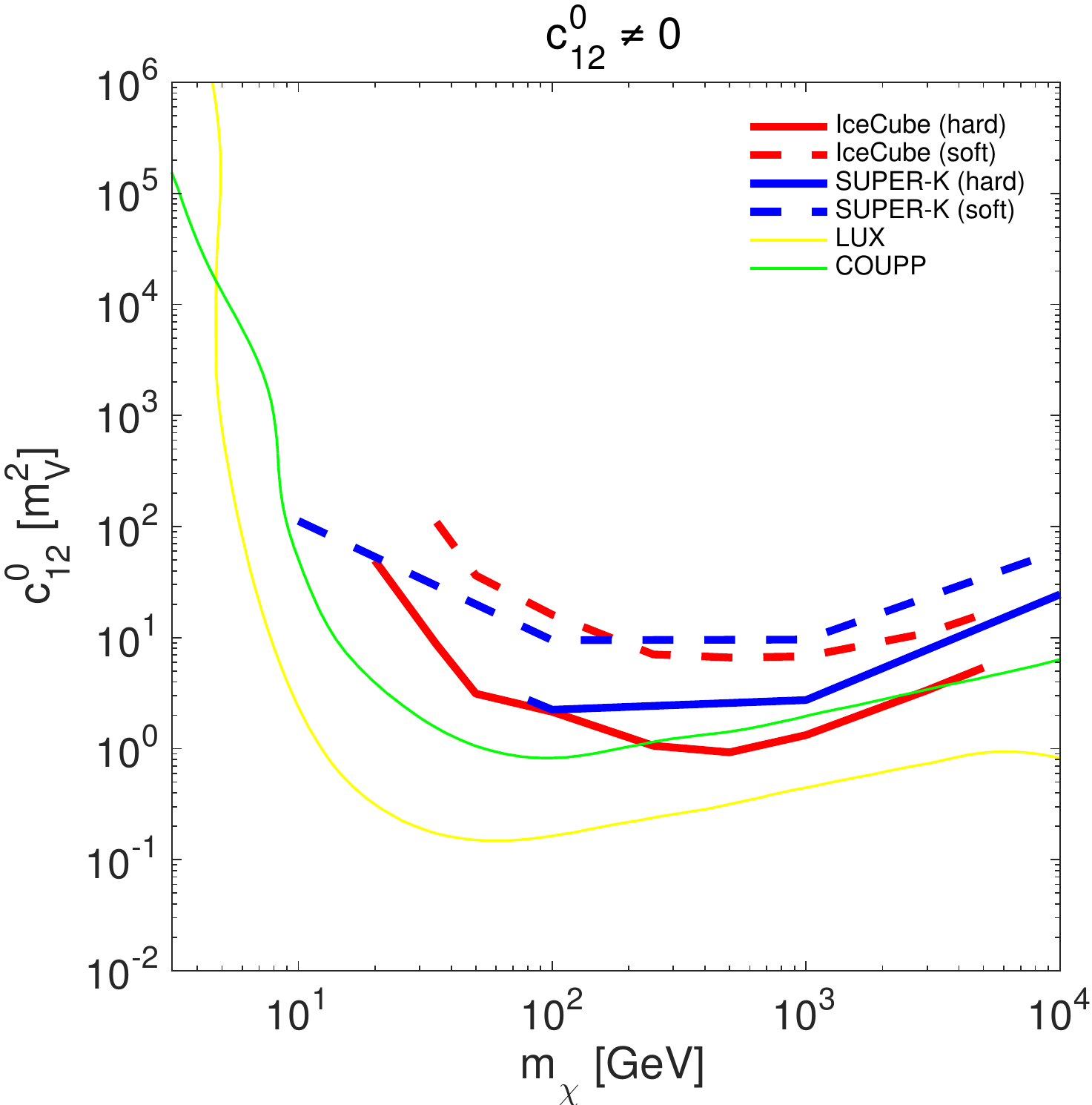}
\end{minipage}
\begin{minipage}[t]{0.49\linewidth}
\centering
\includegraphics[width=\textwidth]{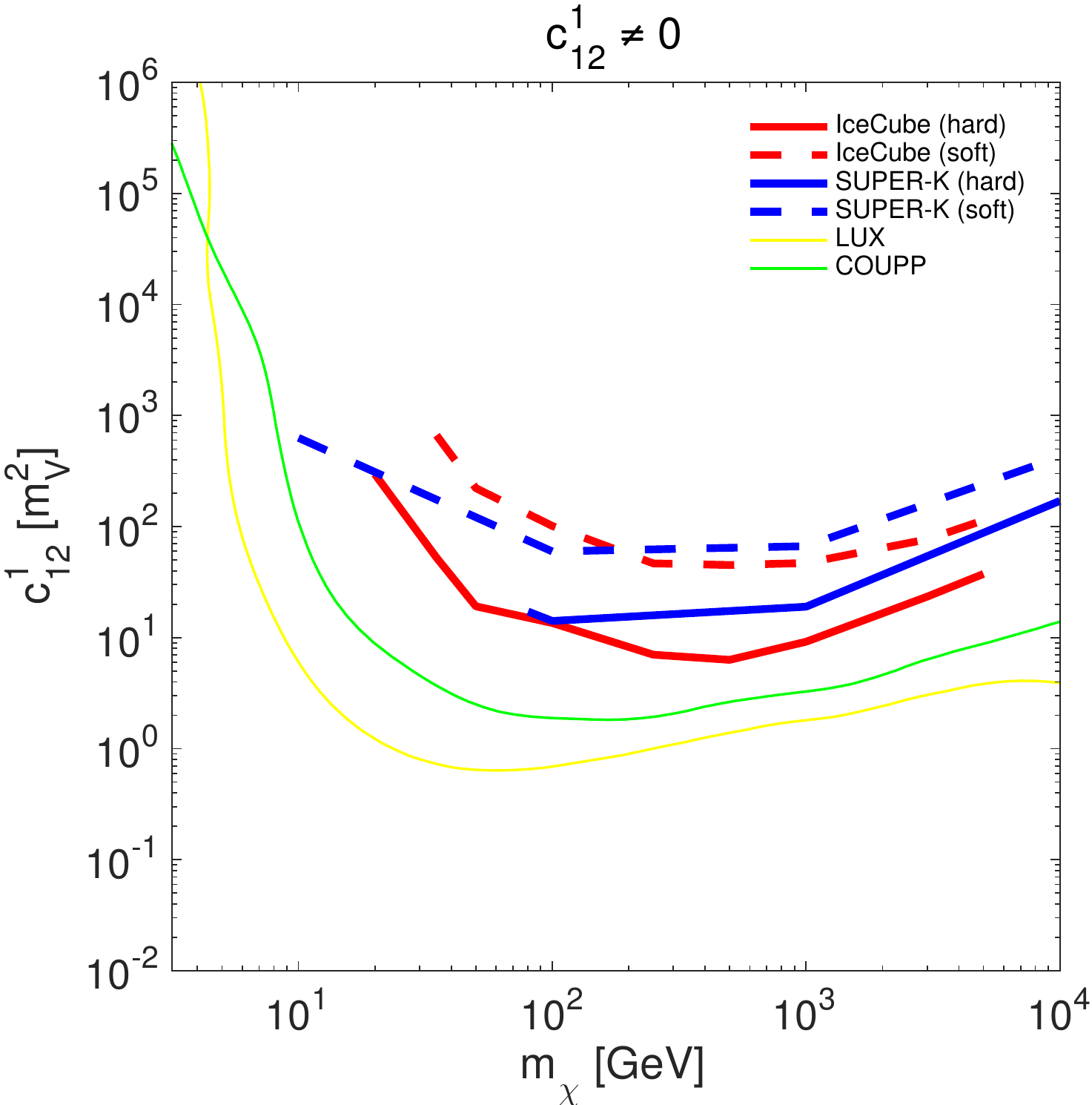}
\end{minipage}
\begin{minipage}[t]{0.49\linewidth}
\centering
\includegraphics[width=\textwidth]{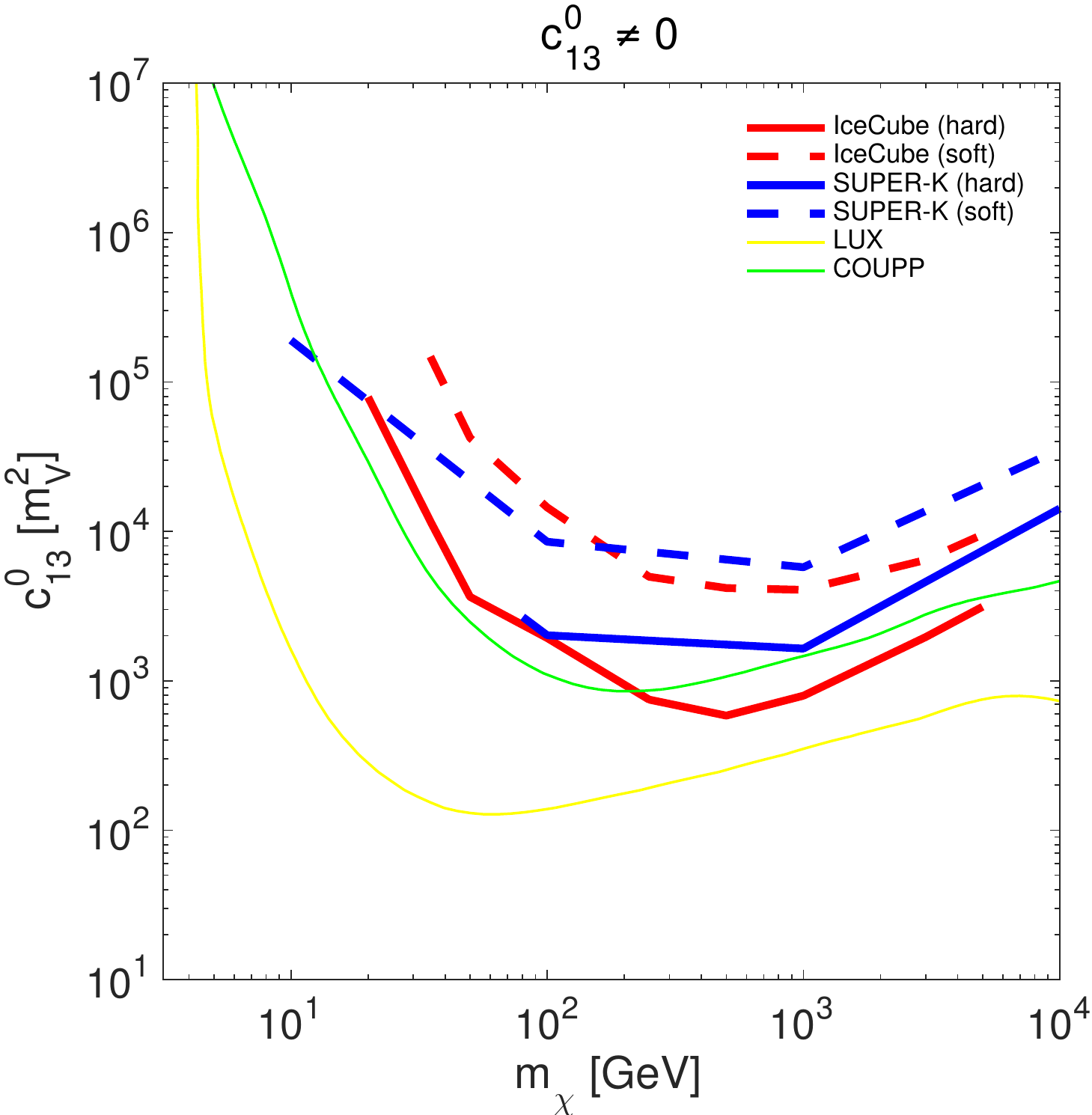}
\end{minipage}
\begin{minipage}[t]{0.49\linewidth}
\centering
\includegraphics[width=\textwidth]{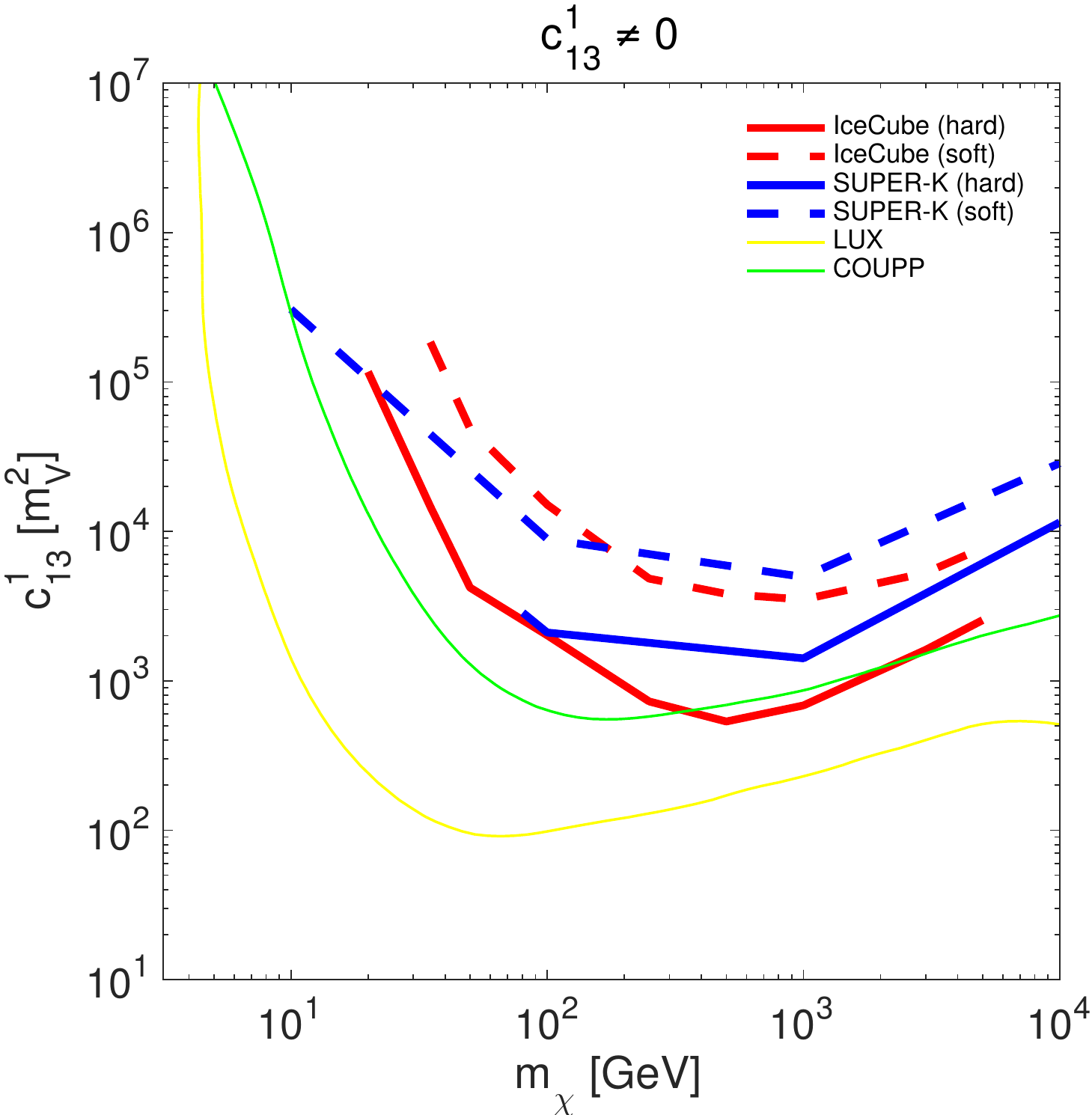}
\end{minipage}
\end{center}
\caption{Same as for Fig.~\ref{fig:c1c4}, but now for the operators $\hat{\mathcal{O}}_{12}$ and $\hat{\mathcal{O}}_{13}$.}
\label{fig:c12c13}
\end{figure}
\begin{figure}[t]
\begin{center}
\begin{minipage}[t]{0.49\linewidth}
\centering
\includegraphics[width=\textwidth]{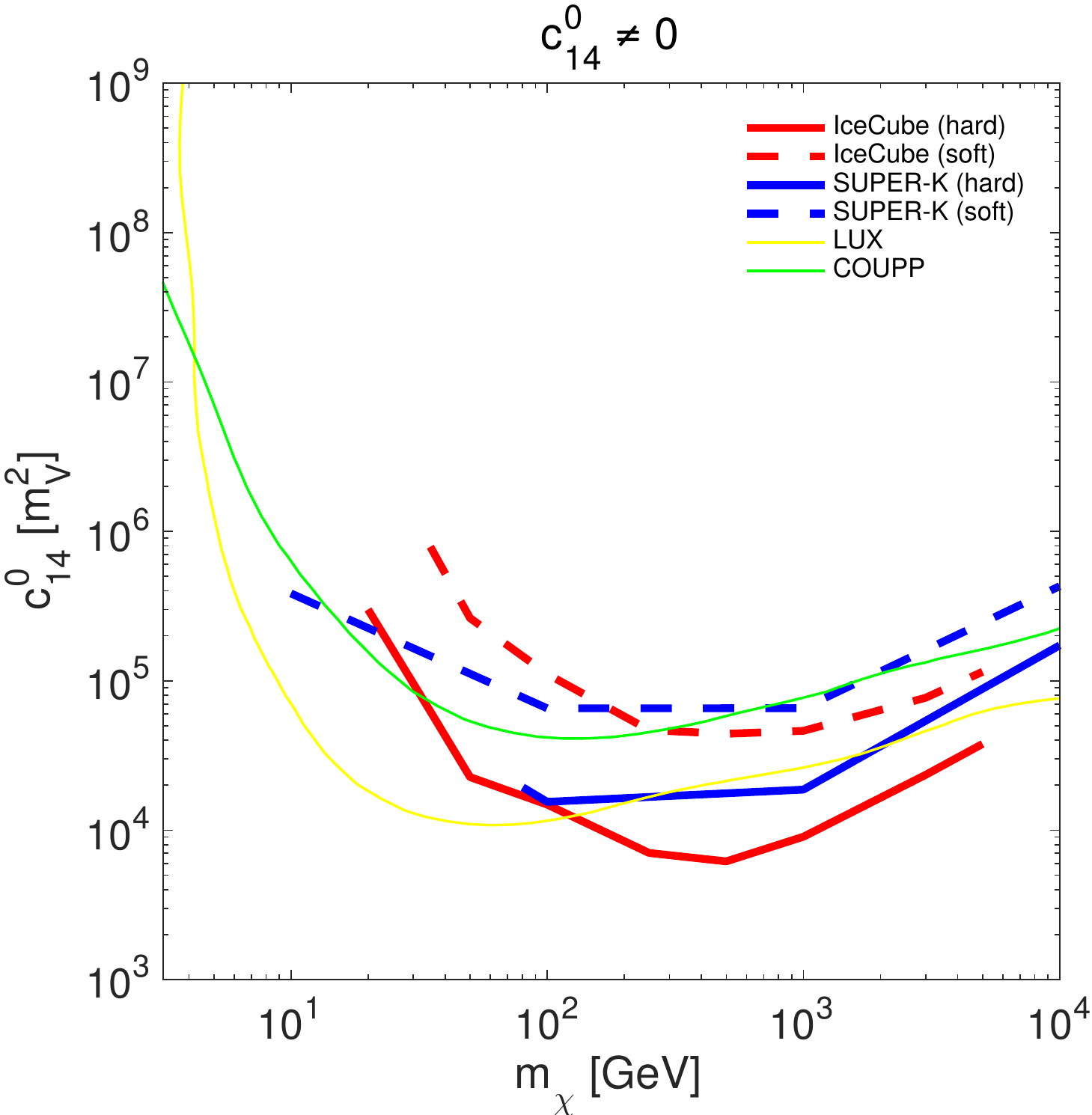}
\end{minipage}
\begin{minipage}[t]{0.49\linewidth}
\centering
\includegraphics[width=\textwidth]{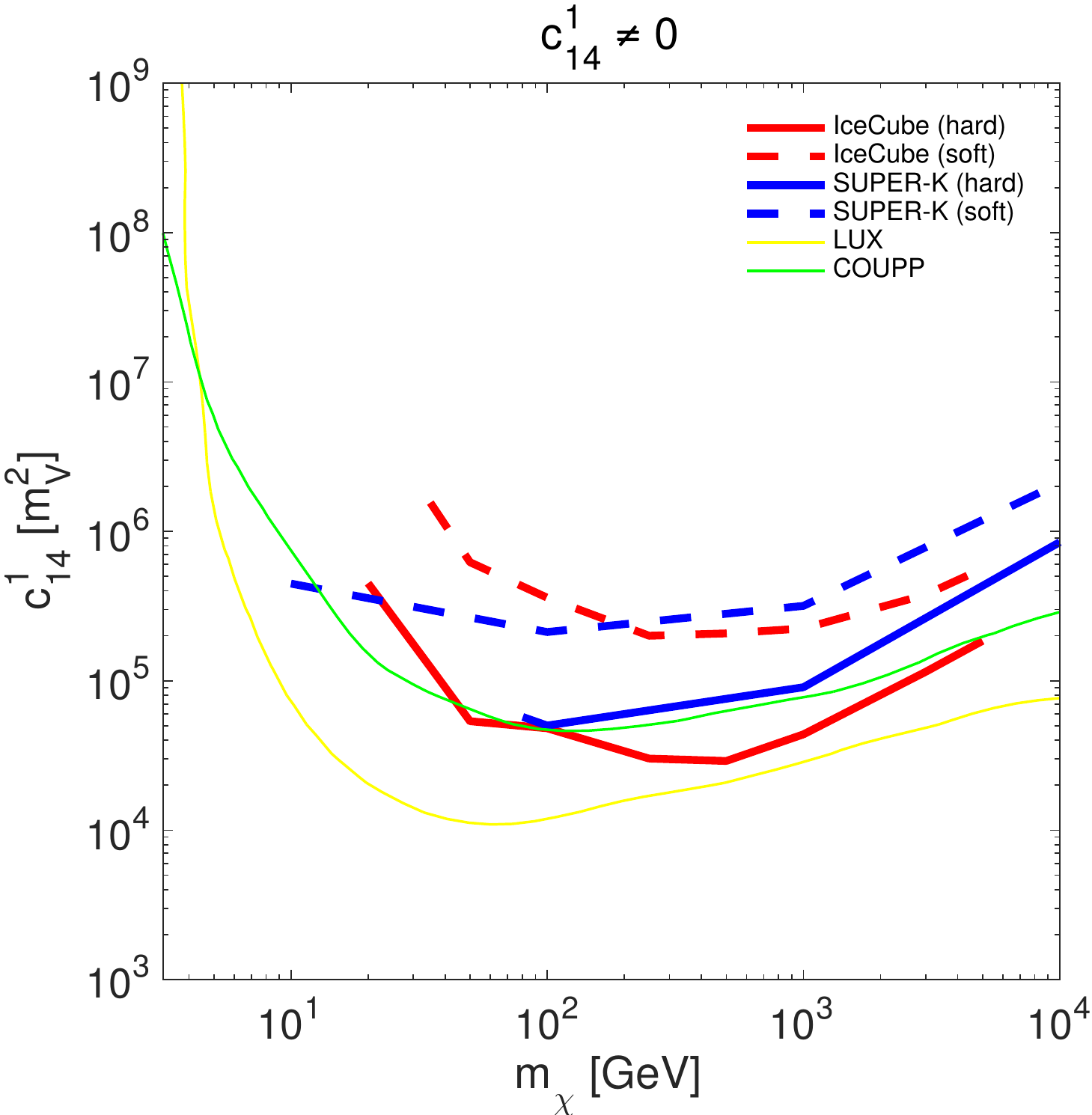}
\end{minipage}
\begin{minipage}[t]{0.49\linewidth}
\centering
\includegraphics[width=\textwidth]{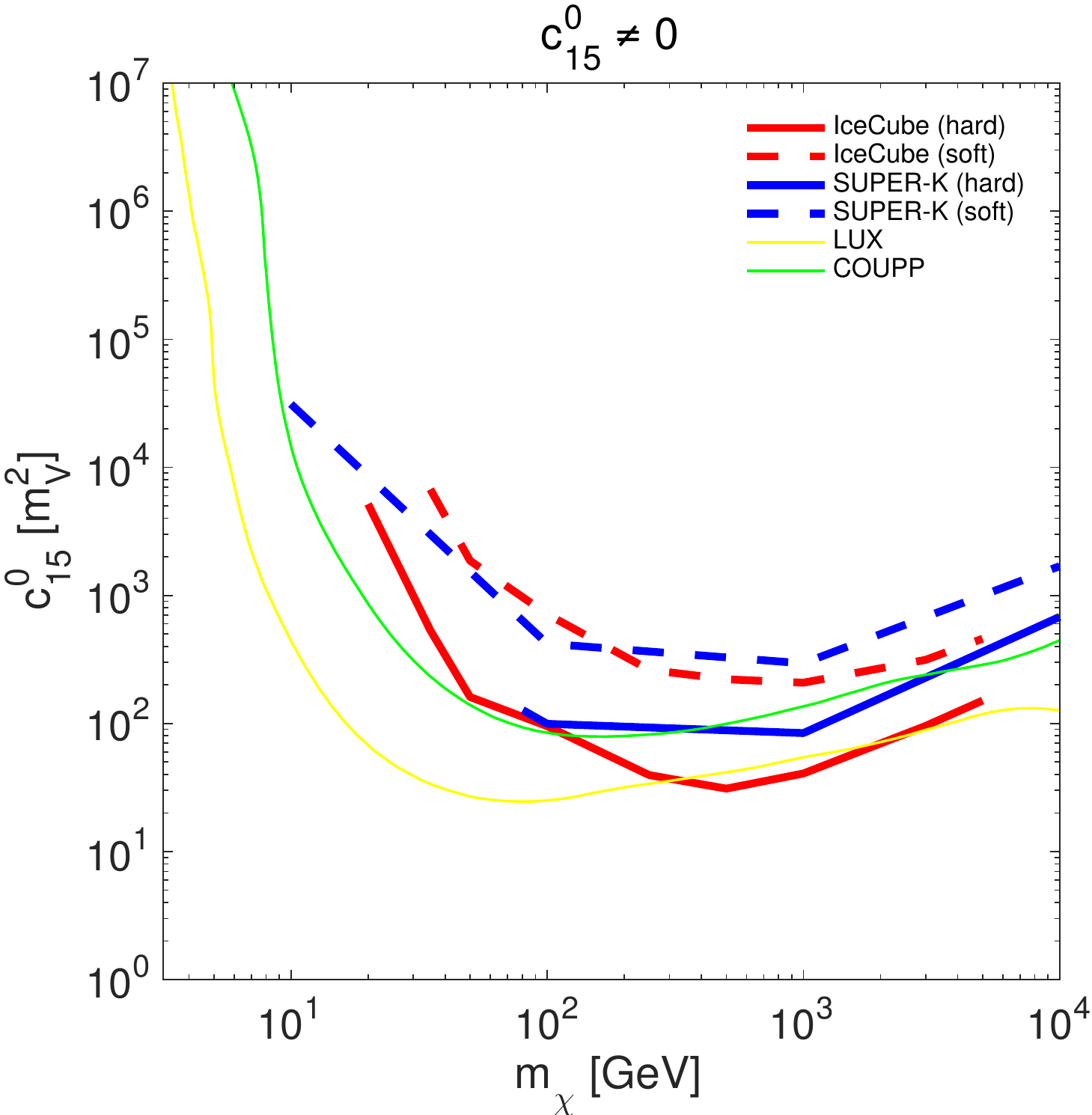}
\end{minipage}
\begin{minipage}[t]{0.49\linewidth}
\centering
\includegraphics[width=\textwidth]{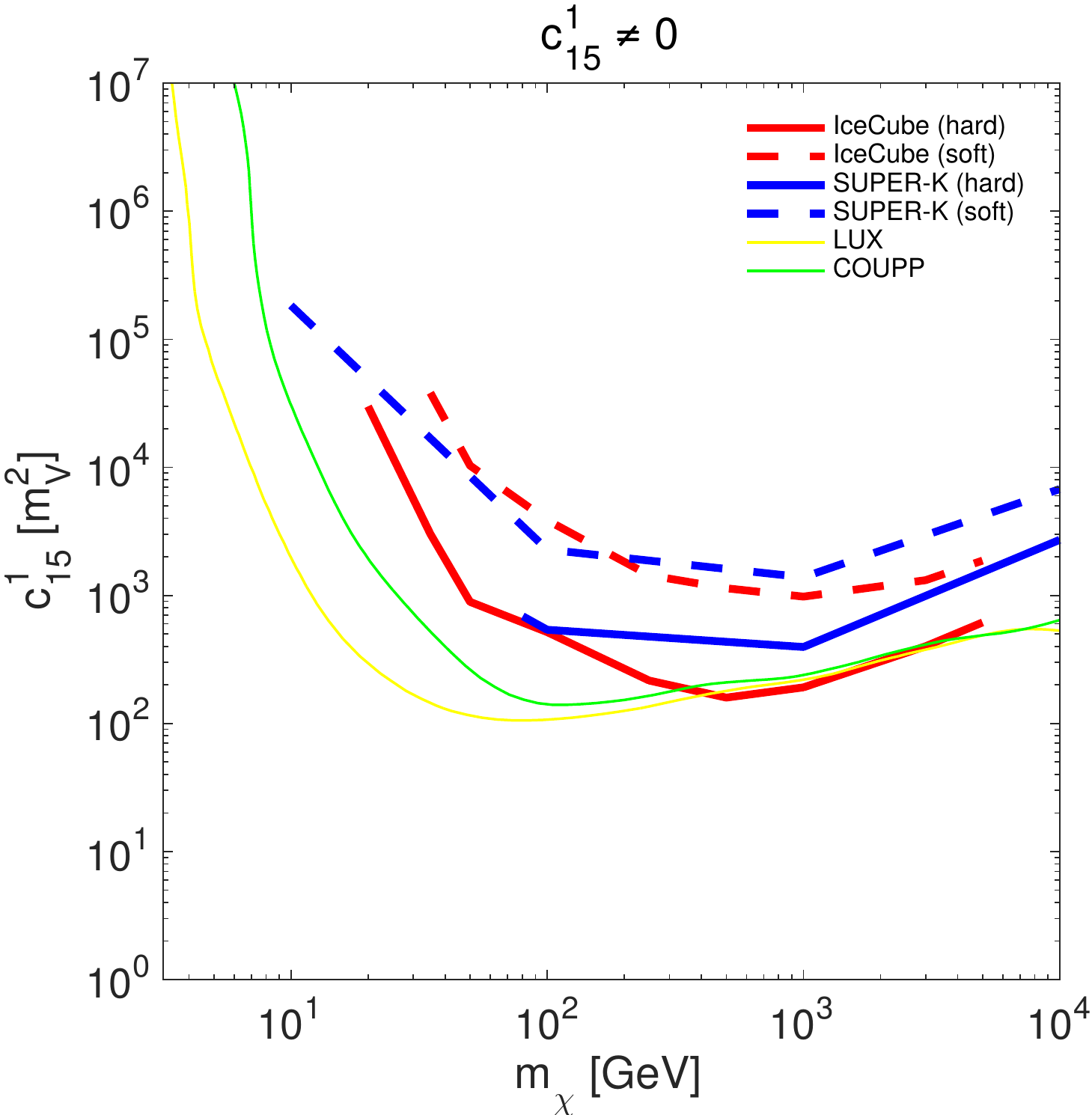}
\end{minipage}
\end{center}
\caption{Same as for Fig.~\ref{fig:c1c4}, but now for the operators $\hat{\mathcal{O}}_{14}$ and $\hat{\mathcal{O}}_{15}$.}
\label{fig:c14c15}
\end{figure}
\begin{figure}[t]
\begin{center}
\begin{minipage}[t]{0.49\linewidth}
\centering
\includegraphics[width=\textwidth]{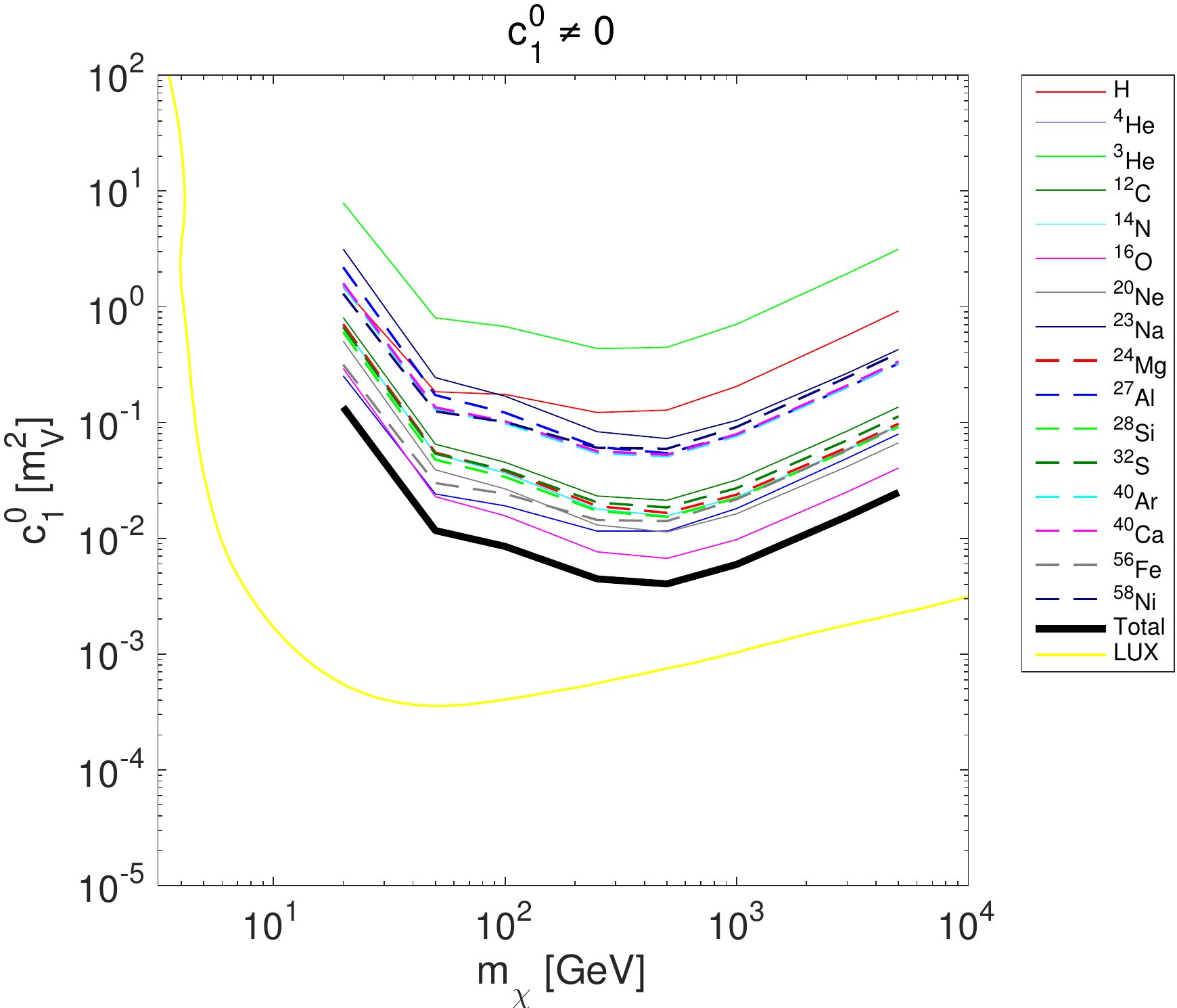}
\end{minipage}
\begin{minipage}[t]{0.49\linewidth}
\centering
\includegraphics[width=\textwidth]{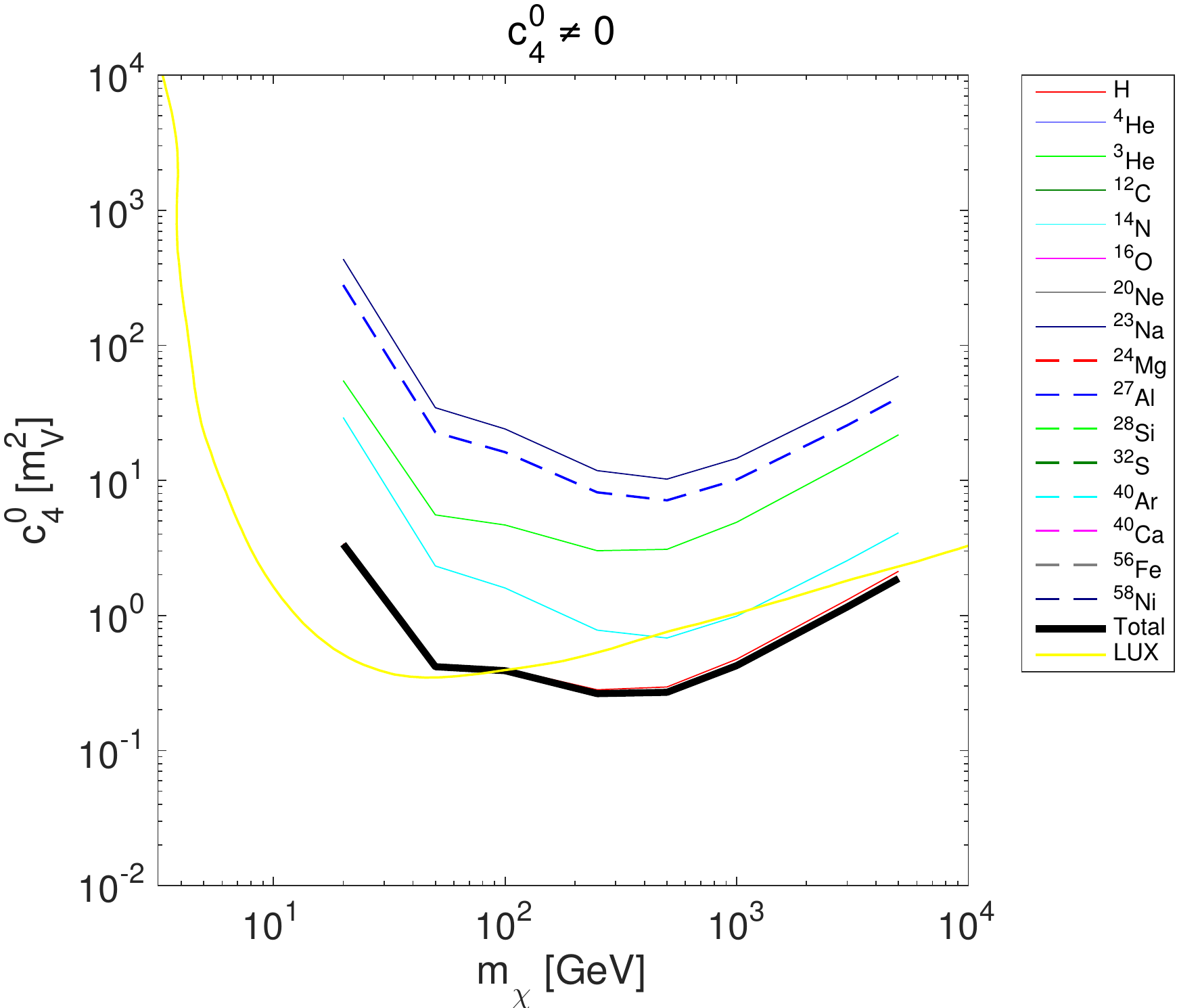}
\end{minipage}
\begin{minipage}[t]{0.49\linewidth}
\centering
\includegraphics[width=\textwidth]{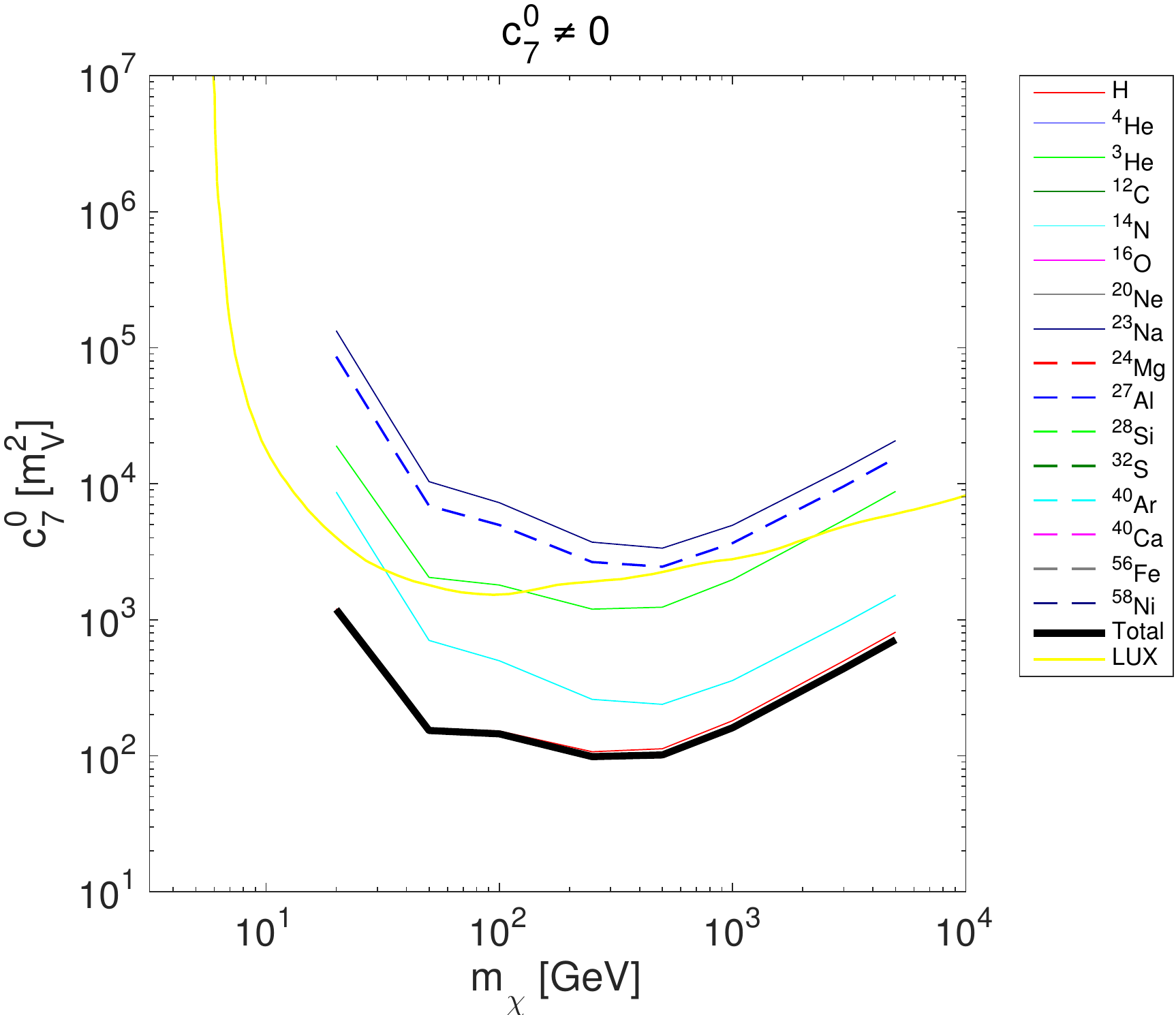}
\end{minipage}
\begin{minipage}[t]{0.49\linewidth}
\centering
\includegraphics[width=\textwidth]{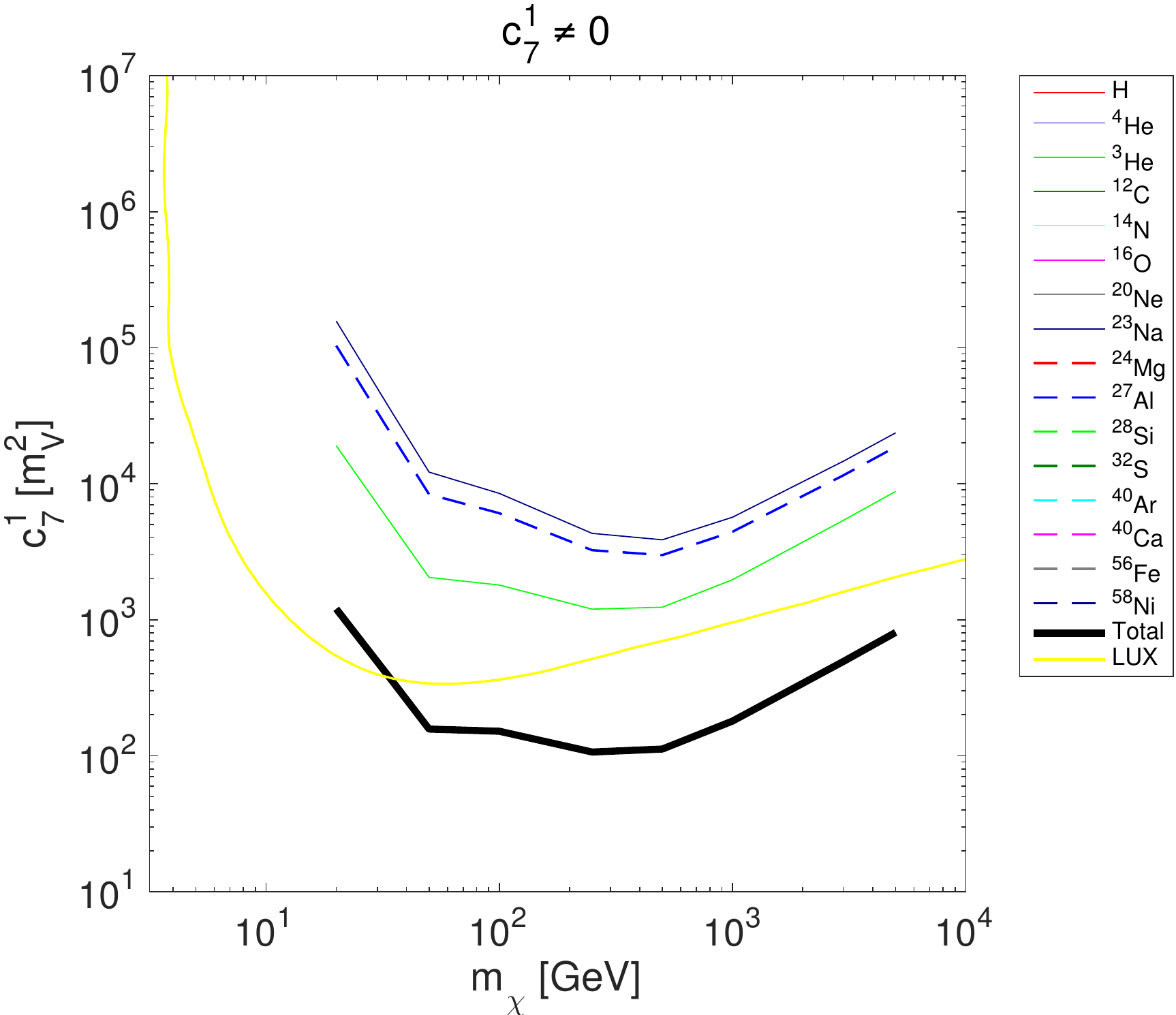}
\end{minipage}
\end{center}
\caption{Isotope-dependent exclusion limits on the coupling constants $c_1^0$ (top left panel), $c_4^0$ (top right panel), $c_7^0$ (bottom left panel), and $c_7^1$ (bottom right panel) from a hard spectrum analysis of the IceCube/Deepcore data. Colored lines assume dark matter scattering from the specific isotope in the legend. Black lines correspond to total exclusion limits, i.e. including all isotopes. For comparison, in each panel we also report the 2D credible region that we obtain from the LUX experiment (yellow thick line). Limits are presented at the 90\% confidence level. Though spin-independent, we also include the isoscalar component of $\hat{\mathcal{O}}_1$ for reference. We also include the isovector component of $\hat{\mathcal{O}}_7$, since limits from neutrino telescopes on this operator are particularly competitive.}
\label{fig:i1}
\end{figure}
\begin{figure}[t]
\begin{center}
\begin{minipage}[t]{0.49\linewidth}
\centering
\includegraphics[width=\textwidth]{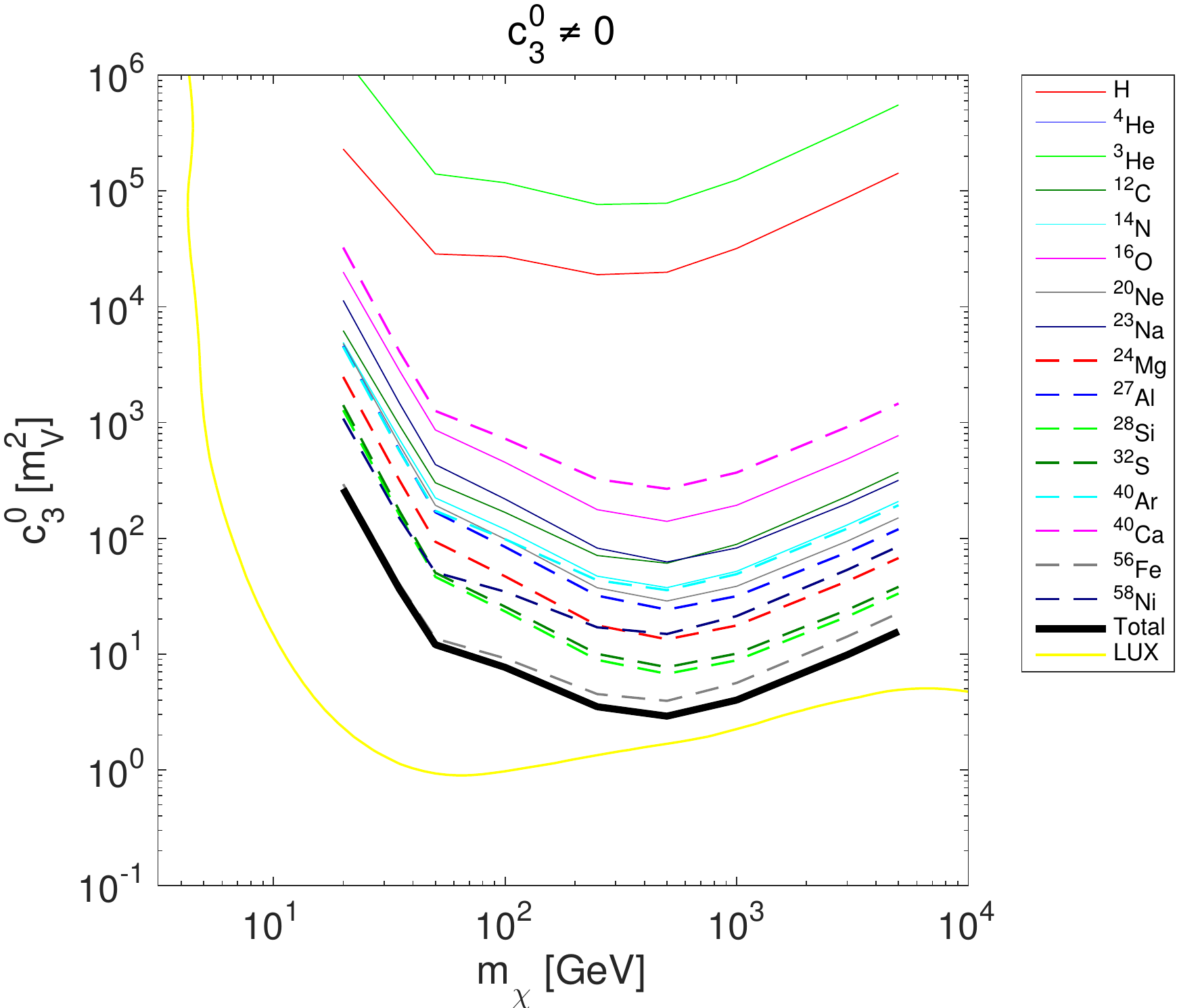}
\end{minipage}
\begin{minipage}[t]{0.49\linewidth}
\centering
\includegraphics[width=\textwidth]{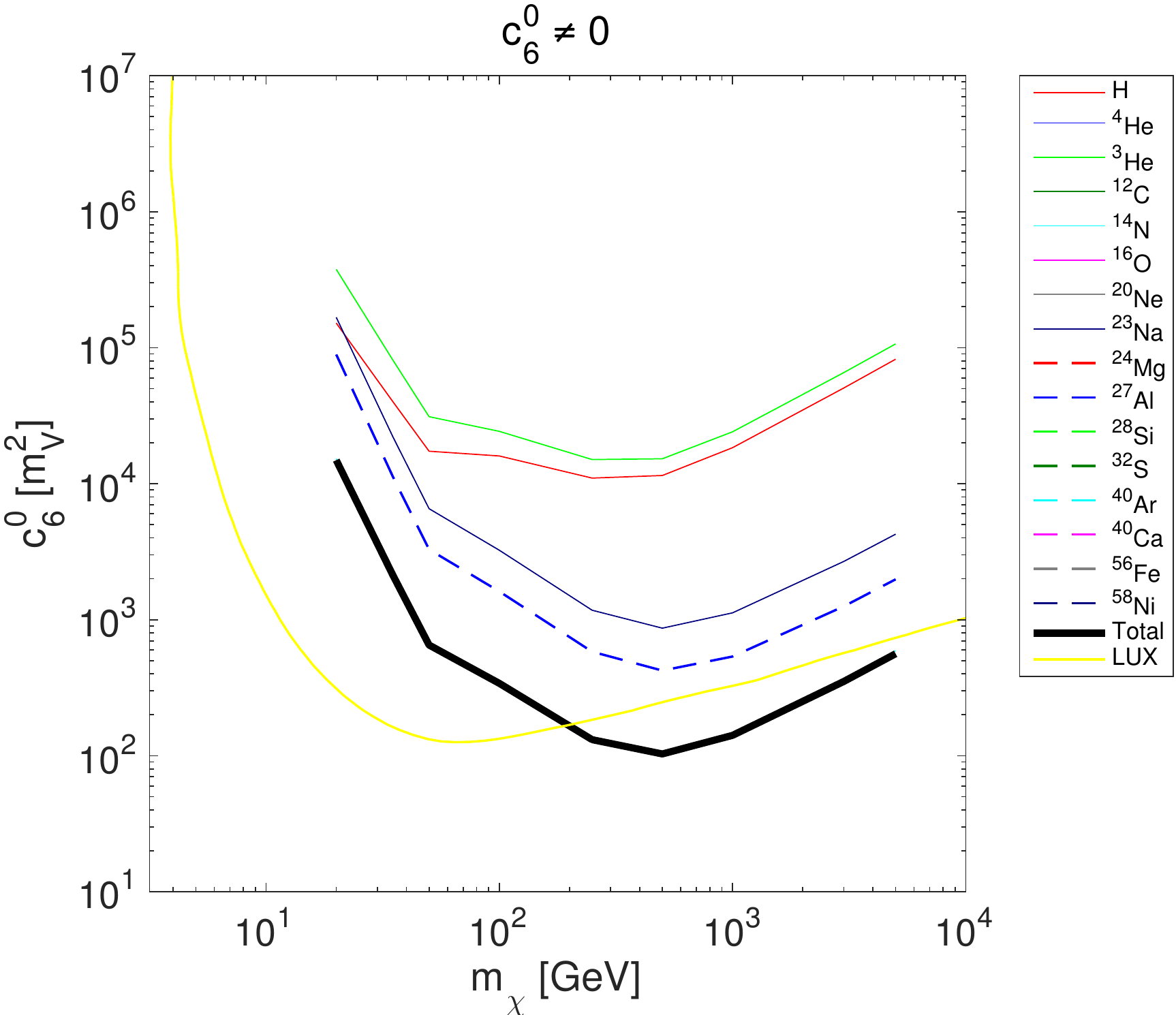}
\end{minipage}
\begin{minipage}[t]{0.49\linewidth}
\centering
\includegraphics[width=\textwidth]{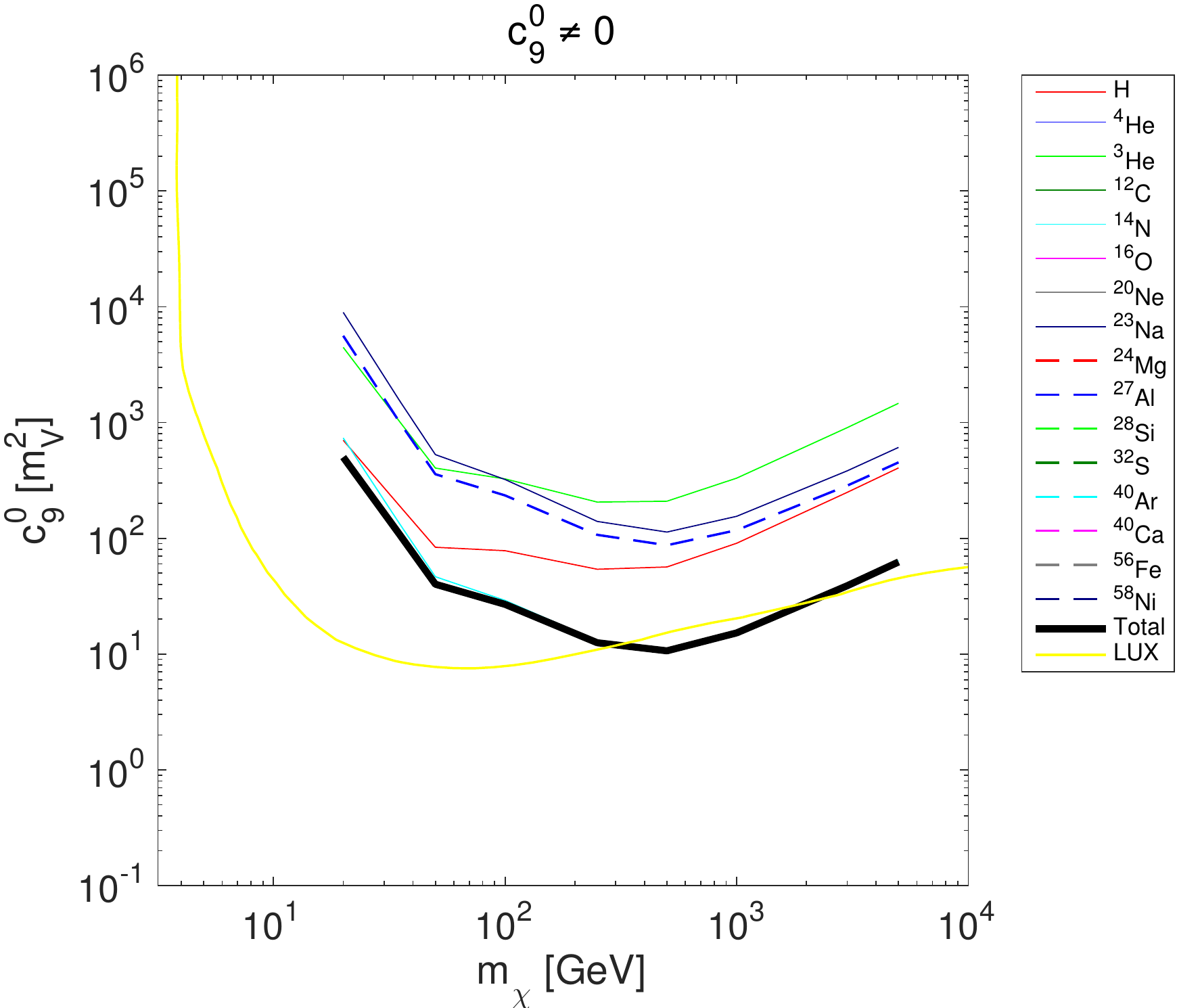}
\end{minipage}
\begin{minipage}[t]{0.49\linewidth}
\centering
\includegraphics[width=\textwidth]{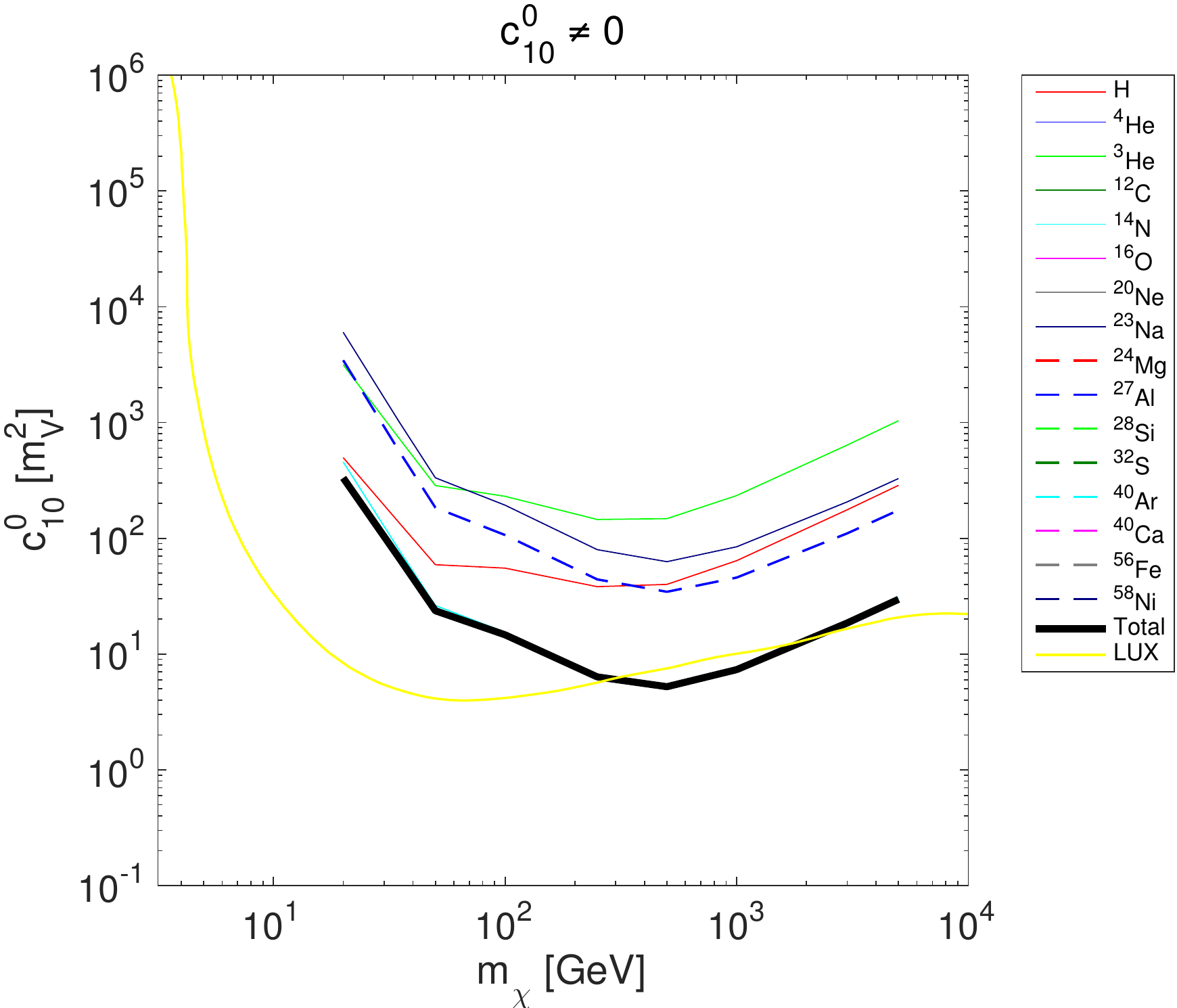}
\end{minipage}
\end{center}
\caption{Same as for Fig.~\ref{fig:i1}, but now for the coupling constant $c^0_{3}$,  $c^0_{6}$, $c^0_{9}$, and $c^0_{10}$.}
\label{fig:i2}
\end{figure}
\begin{figure}[t]
\begin{center}
\begin{minipage}[t]{0.49\linewidth}
\centering
\includegraphics[width=\textwidth]{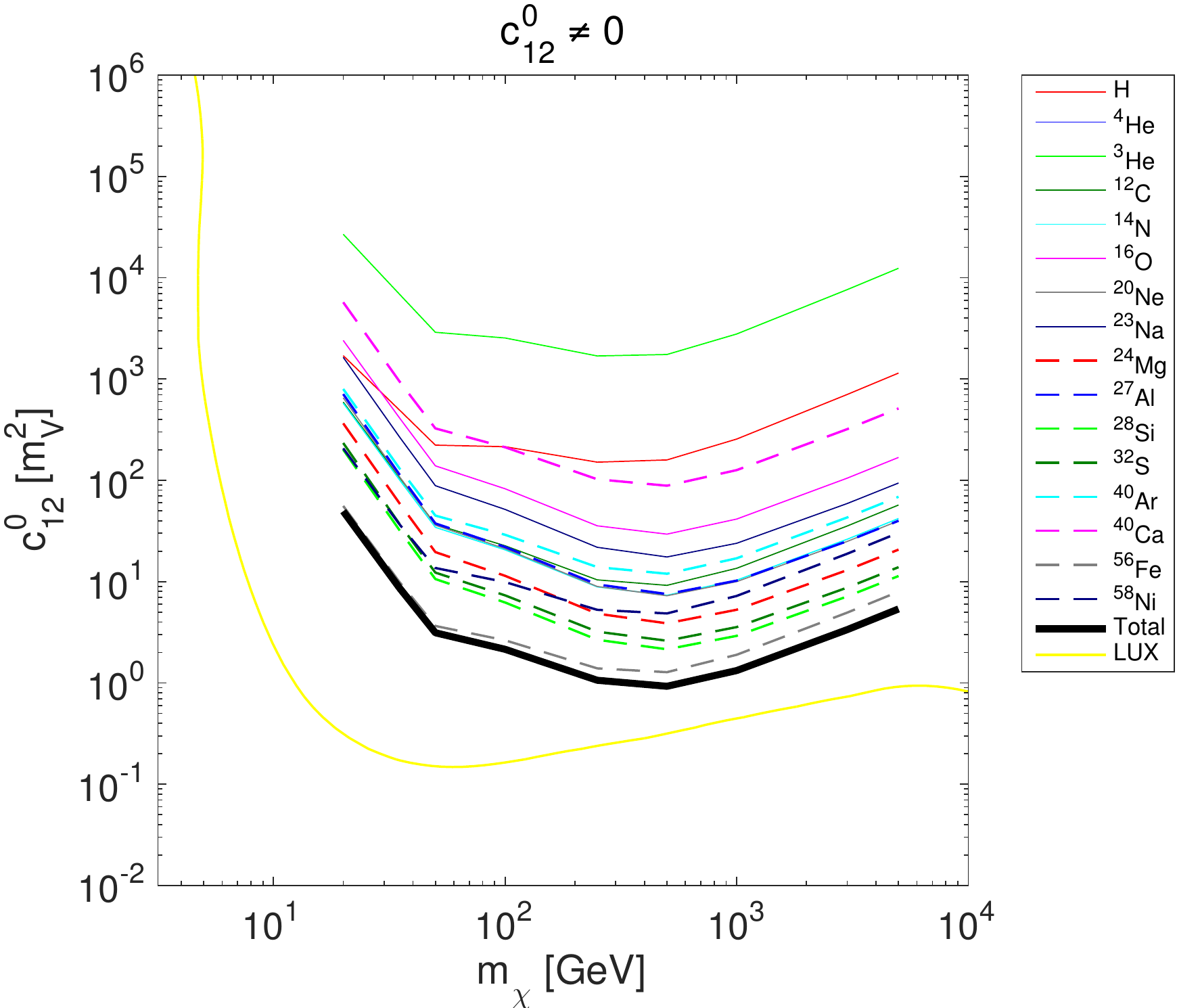}
\end{minipage}
\begin{minipage}[t]{0.49\linewidth}
\centering
\includegraphics[width=\textwidth]{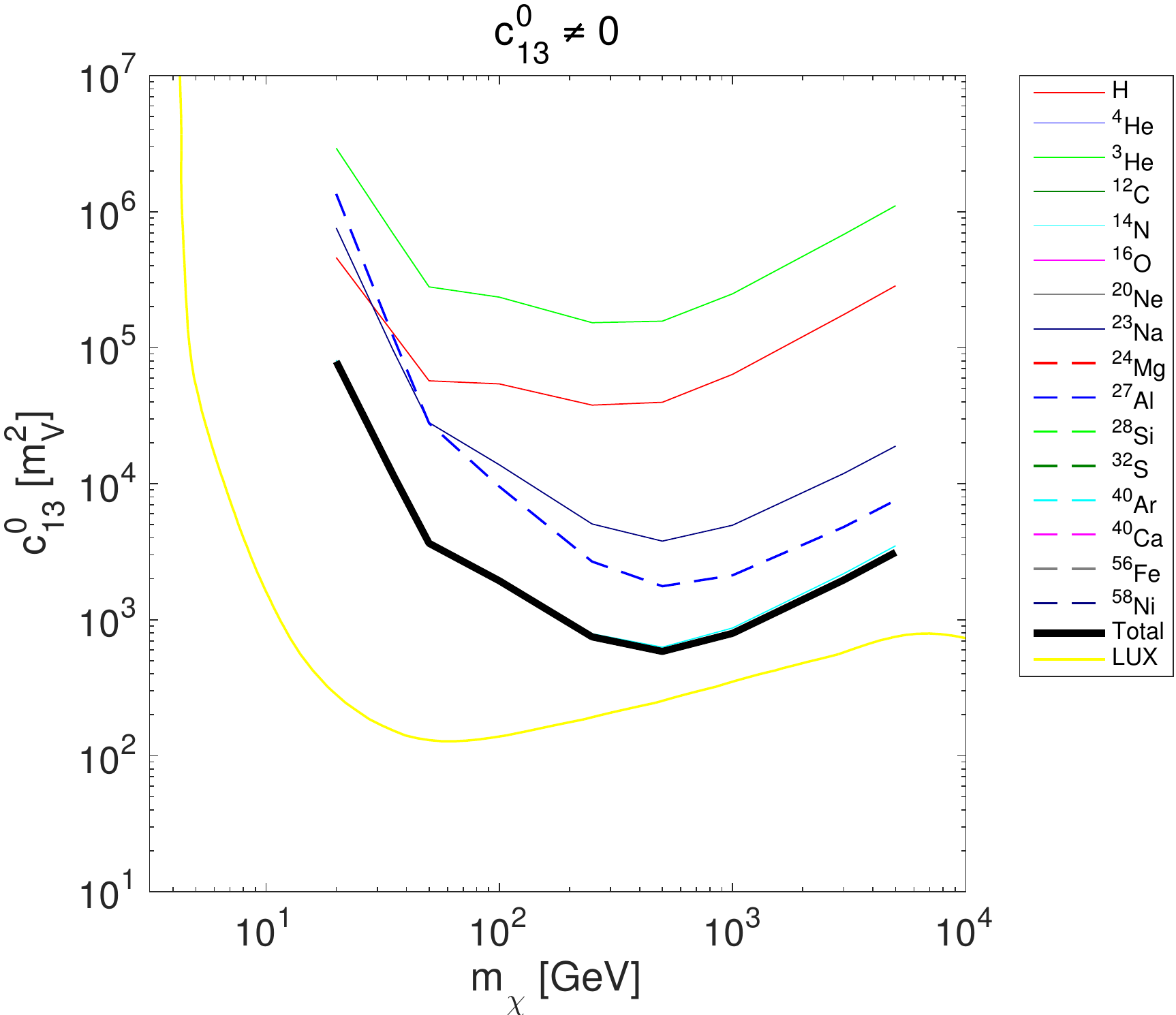}
\end{minipage}
\begin{minipage}[t]{0.49\linewidth}
\centering
\includegraphics[width=\textwidth]{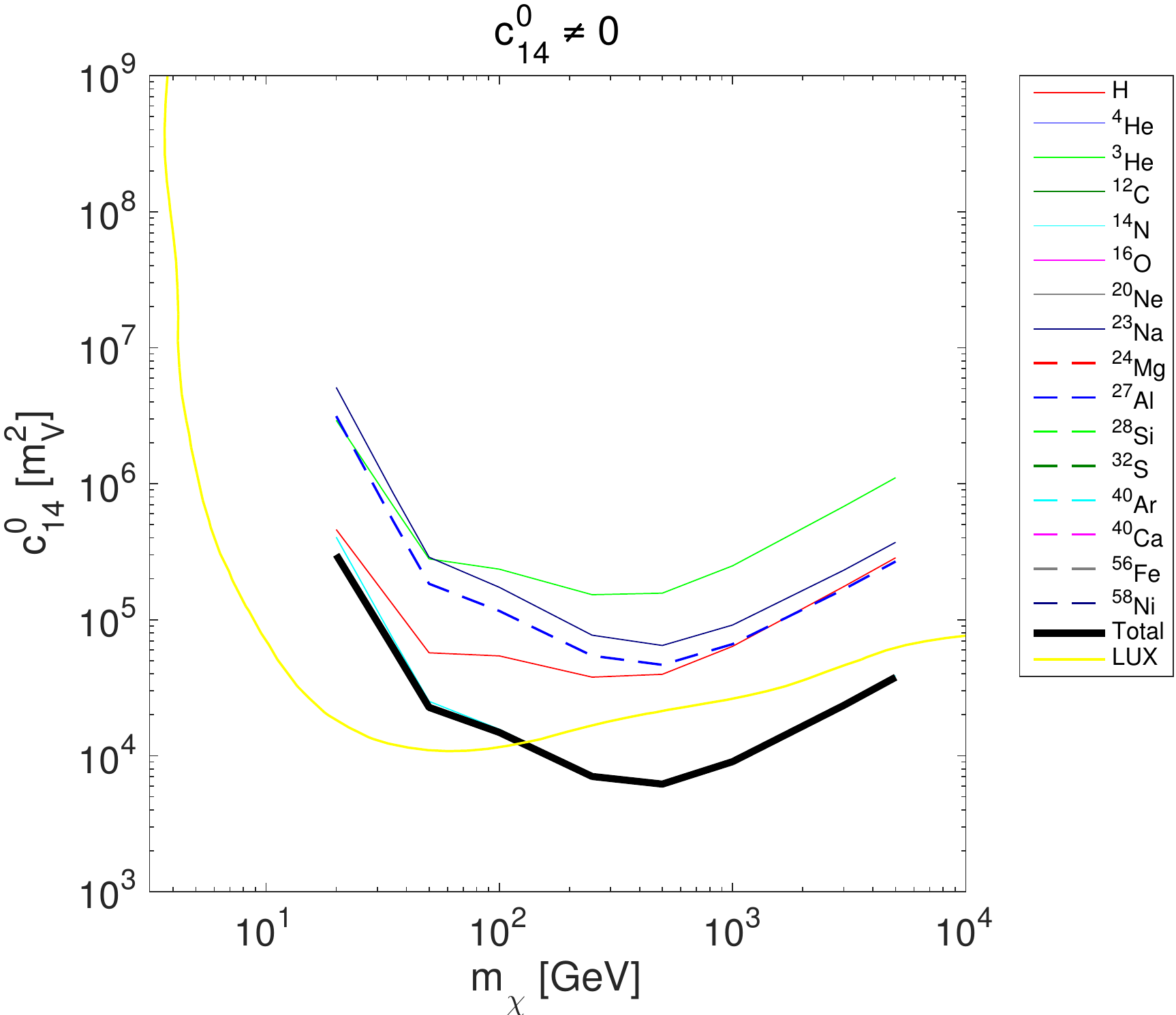}
\end{minipage}
\begin{minipage}[t]{0.49\linewidth}
\centering
\includegraphics[width=\textwidth]{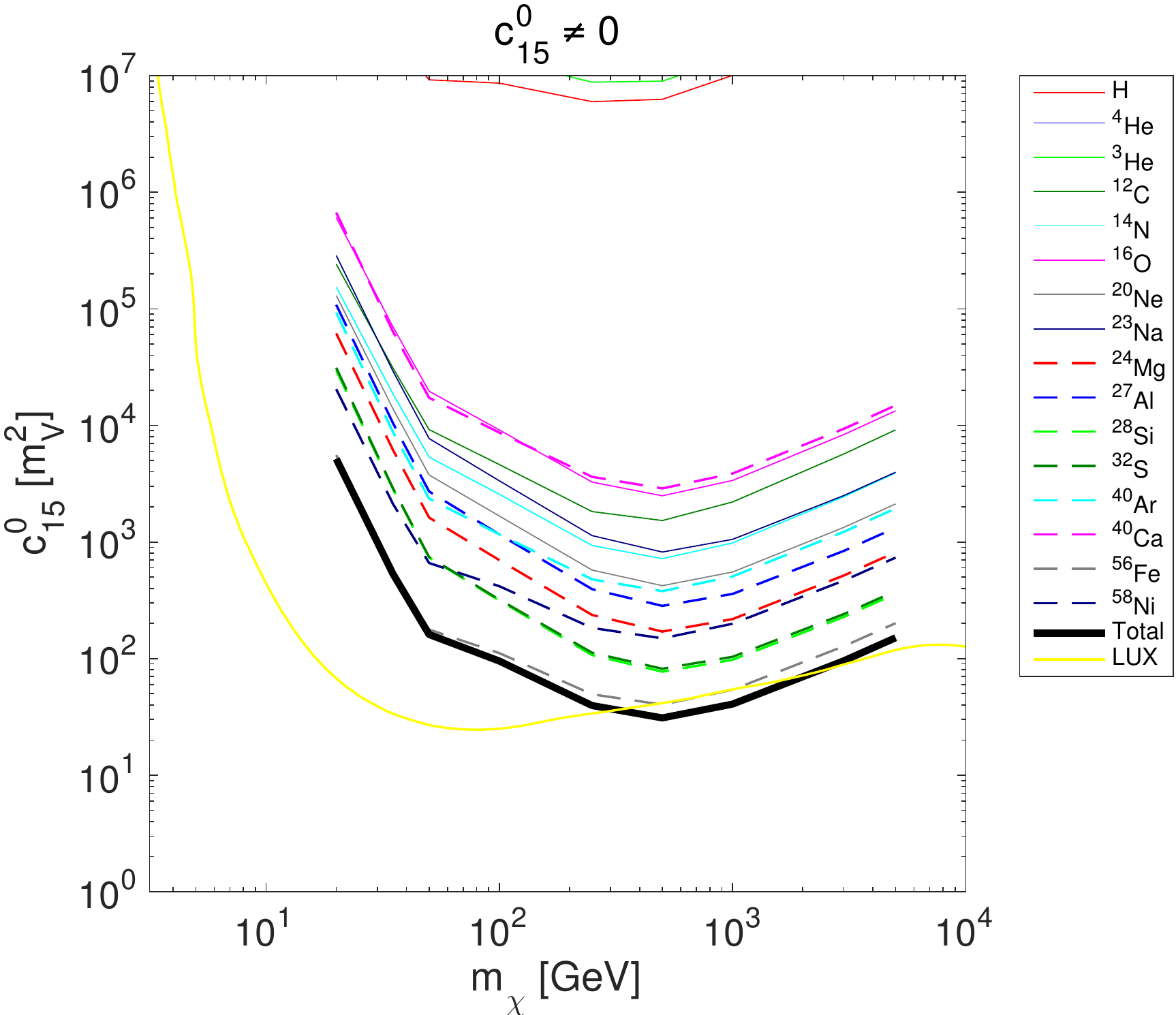}
\end{minipage}
\end{center}
\caption{Same as for Fig.~\ref{fig:i1}, but now for the coupling constant $c^0_{12}$,  $c^0_{13}$, $c^0_{14}$, and $c^0_{15}$.}
\label{fig:i3}
\end{figure}

\section{Model independent analysis}
\label{sec:analysis}
In this section we compare the general effective theory of one-body dark matter-nucleon interactions mediated by a heavy spin-1 or spin-0 particle with current neutrino telescope observations. We focus on data collected by the 79-string configuration of the IceCube/DeepCore observatory~\cite{Aartsen:2012kia}, and during 3109.6 days of SUPER-K dark matter searches~\cite{Tanaka:2011uf}. Both experiments report limits on the neutrino-induced muon flux from dark matter annihilation above a given energy threshold, and below a given angle from the direction of the centre of the Sun. In our analysis, we use Tab.~1 from~\cite{Aartsen:2012kia} as IceCube/DeepCore data, and Tabs.~1 and~2 from~\cite{Tanaka:2011uf} as SUPER-K data. 

We use {\sffamily darksusy} to compute the muon yields at the detector, and our routines and nuclear response functions (see Figs.~\ref{fig:W1}, \ref{fig:W2} and \ref{fig:W3}) to calculate the rate of dark matter capture by the Sun. 
For the dark matter primary annihilation channel, we consider two extreme scenarios corresponding to dark matter pair annihilation into $W^{+}W ^{-}$, as for models with hard annihilation spectra, and to dark matter pair annihilation into $b\bar{b}$, as for models with soft annihilation spectra. 

For $m_\chi < 80.3$~GeV, the $W^{+}W ^{-}$ channel is not kinematically allowed. Following~\cite{Tanaka:2011uf,Aartsen:2012kia}, in our IceCube/DeepCore hard spectrum analysis we replace the $W^{+}W ^{-}$ channel with the final state $\tau^+\tau^-$. We instead neglect the $W^{+}W ^{-}$ channel in the hard spectrum analysis of the SUPER-K data. 

Demanding that the predicted neutrino-induced muon flux does not exceed its experimental 90\% confidence level upper limit, we derive exclusion limits at the 90\% confidence level on the isoscalar and isovector coupling constants corresponding to the interaction operators in Tab.~\ref{tab:operators}.

For each interaction operator in Tab.~\ref{tab:operators}, we consider the corresponding isoscalar and isovector coupling constants separately. In the figures, we report our 90\% confidence level exclusion limits as a function of the dark matter particle mass, varying $m_\chi$ in the range 10 GeV - 10 TeV. 

We also compare our limits from neutrino telescope observations with the 2D 90\% credible regions that we obtain from a Bayesian analysis of the LUX and COUPP direct detection experiments. The details of our COUPP and LUX analysis are discussed in Appendix~\ref{sec:dd}.

\subsection{Momentum/velocity independent operators}
We start with an analysis of the momentum and velocity independent interaction operators $\hat{\mathcal{O}}_1$ and $\hat{\mathcal{O}}_4$. The operator $\hat{\mathcal{O}}_1$ generates a constant spin-independent cross-section $\sigma_{p}^{\rm SI}$ ($\sigma_{n}^{\rm SI}$) for dark matter interactions with protons (neutrons) given by
\begin{equation}
\label{eq:sigmaSI}
\sigma_{p}^{\rm SI} = \frac{\mu_{N}^2}{\pi} \, \frac{ |c_1^0 + c_1^1|^2}{4}\,; \qquad\qquad \sigma_{n}^{\rm SI} = \frac{\mu_{N}^2}{\pi} \, \frac{ |c_1^0 - c_1^1|^2}{4} \,.
\end{equation}
Analogously, the operator $\hat{\mathcal{O}}_4$ generates a constant spin-dependent cross-section $\sigma_{p}^{\rm SD}$ ($\sigma_{n}^{\rm SD}$) for dark matter-proton (dark matter-neutron) interactions given by
\begin{equation}
\label{eq:sigmaSD}
\sigma_{p}^{\rm SD} = \frac{\mu_N^2 j_\chi (j_\chi+1) }{4\pi}\, \frac{ |c_4^0 + c_4^1|^2}{4}\,; \qquad\qquad \sigma_{n}^{\rm SD} = \frac{\mu_N^2 j_\chi (j_\chi+1) }{4\pi}\, \frac{ |c_4^0 - c_4^1|^2}{4}  \,.
\end{equation}
In the expressions above, $\mu_{N}=m_\chi m_{N}/(m_\chi+m_{N})$ is the reduced dark matter-nucleon mass, and $j_\chi$ is the dark matter particle spin. For definitiveness, in the calculations we assume $j_\chi=1/2$. 

Fig.~\ref{fig:c1c4} shows the 90\% confidence level exclusion limits that we obtain on the coupling constants $c_1^0$ (top left panel), $c_1^1$ (top right panel), $c_4^0$ (bottom left panel), and $c_4^1$ (bottom right panel). Solid red (blue) lines correspond to the hard annihilation spectrum analysis of the IceCube/DeepCore (SUPER-K) data, whereas dashed red (blue) lines represent our soft annihilation spectrum analysis of the IceCube/DeepCore (SUPER-K) data. For comparison, we also report 2D 90\% credible regions obtained from LUX and COUPP as explained in Appendix~\ref{sec:dd}. 

Our exclusion limits on the coupling constants $c_1^0$ and $c_4^0$ (top left and bottom left panels in Fig.~\ref{fig:c1c4}) agree with those in~\cite{Tanaka:2011uf,Aartsen:2012kia} at the few \% level for small values of $m_\chi$. For $m_\chi \sim10$~TeV, our findings differ from those of~\cite{Tanaka:2011uf,Aartsen:2012kia}  by up to 1 order of magnitude in the cross-sections, as a consequence of using different nuclear response functions. 
Here we adopt the nuclear response functions recently derived in~\cite{Catena:2015uha} through numerical nuclear structure calculations. In contrast, in previous analyses approximate exponential form factors have often been used. 

\subsection{Momentum/velocity dependent operators}
\label{sec:qvint}
We now move on to describe our analysis of the momentum/velocity dependent dark matter-nucleon interactions operators. All interaction operators in Tab.~\ref{tab:operators} but $\hat{\mathcal{O}}_1$ and $\hat{\mathcal{O}}_4$ depend on ${\bf \hat{q}}$, on ${\bf \hat{v}}^\perp$, or on both.

For momentum or velocity dependent operators exclusion limits are weaker than for $\hat{\mathcal{O}}_1$ and $\hat{\mathcal{O}}_4$. The reason is that for operators $\hat{\mathcal{O}}_k$, $k\neq1,4$, the rate of dark matter capture by the Sun is suppressed by powers of $q^2$ and $v_T^{\perp2}=w^2-q^2/{(4 \mu_T^2})$ in Eq.~(\ref{eq:R}). For instance, the operator $\hat{\mathcal{O}}_6$ generates a term proportional to $q^4$ in $R_{\Sigma''}^{\tau\tau'}$, and the operator $\hat{\mathcal{O}}_{11}$ generates a term proportional to $q^2$ in $R_{M}^{\tau\tau'}$. Nevertheless, dark matter can scatter in the Sun with a larger relative velocity $w$ (and momentum transfer) than in terrestrial detectors, where the scattering velocity is less then about 800 km~s$^{-1}$. Indeed, being $w^2=u^2+v^2(R)$ and 620~km~s$^{-1}\lesssim v(R)\lesssim1400$ km~s$^{-1}$~\cite{Gondolo:2004sc}, then $w\gtrsim$ 620 km~s$^{-1}$. Therefore, since the scattering cross-section~(\ref{eq:sigma}) is evaluated at $w$, and not at $u$, velocity dependent scattering cross-sections weaken the exclusion limits from neutrino telescopes less dramatically than those from direct detection experiments. 

For momentum or velocity dependent operators, exclusion limits also depend on nuclear response functions different from  $W_{M}^{\tau\tau'}$, $W_{\Sigma'}^{\tau\tau'}$ and $W_{\Sigma''}^{\tau\tau'}$. For reference, in Figs.~\ref{fig:W1}, \ref{fig:W2}, and \ref{fig:W3} we plot $W_{M}^{\tau\tau'}$, $W_{\Phi''}^{\tau\tau'}$, $W_{\Sigma'}^{\tau\tau'}$, $W_{\Sigma''}^{\tau\tau'}$, $W_{\Delta}^{\tau\tau'}$ and $W_{\tilde{\Phi}'}^{\tau\tau'}$, with $\tau=\tau'$, for the 16 isotopes in Tab.~\ref{tab:massfrac}. We do not report here $W_{\Phi'' M}^{\tau\tau'}$, and $W_{\Delta \Sigma'}^{\tau\tau'}$, and the $\tau\neq\tau'$ components of the other nuclear response functions, since they are not relevant when the interference between different operators, or between isoscalar and isovector components of the same operator is neglected, as in the present analysis. In~\cite{Catena:2015uha} we list all nuclear response functions in Eq.~(\ref{eq:W}) for the 16 elements in Tab.~\ref{tab:massfrac} in analytic form.

In general, exclusion limits depend on the solar composition. For reference,  in Tab.~\ref{tab:massfrac} we report the average mass fractions of the 16 most abundant elements in the Sun as implemented in {\sffamily darksusy}.

Figs.~\ref{fig:c3c5}, \ref{fig:c6c7}, \ref{fig:c8c9}, \ref{fig:c10c11}, \ref{fig:c12c13}, and \ref{fig:c14c15} show the exclusion limits on the isoscalar and isovector coupling constants corresponding to the operators $\hat{\mathcal{O}}_k$, $k=3,5,6\dots,15$ that we find from IceCube/DeepCore and SUPER-K data. These figures reveal that the operators most severely constrained by current neutrino telescope observations are $\hat{\mathcal{O}}_{11}$, $\hat{\mathcal{O}} _{12}$, $\hat{\mathcal{O}}_8$ and $\hat{\mathcal{O}}_3$ (besides the operators $\hat{\mathcal{O}}_1$ and $\hat{\mathcal{O}}_4$ discussed in the previous section). Notably, our current exclusion limits on the interaction $\hat{\mathcal{O}}_{11}= i{\hat{\bf{S}}}_\chi\cdot({\hat{\bf{q}}}/m_N)$, are more stringent that those set on the familiar $\hat{\mathcal{O}}_4=\hat{{\bf{S}}}_{\chi}\cdot \hat{{\bf{S}}}_{N}$ interaction. We focus here on isoscalar interactions only. This result is expected~\cite{Catena:2015uha}, in that $\hat{\mathcal{O}}_{11}$ is independent of the transverse relative velocity operator, and of the nucleon spin, which implies a large capture rate proportional to $A^2$.

\subsection{Comparison with LUX and COUPP}
We now compare our limits on the constants $c_k^\tau$ from neutrino telescopes with the limits that we obtain from LUX and COUPP. 

For dark matter-nucleon interaction operators mainly coupling to the nuclear response operators $M_{LM;\tau}$ and $\Phi''_{LM;\tau}$, dark matter direct detection experiments tend to set stronger limits on the coupling constants $c_k^\tau$. The reason is that $M_{LM;\tau}$, and with some restriction also $\Phi''_{LM;\tau}$, favor heavy elements. Indeed, in the $q\rightarrow 0$ limit, $M_{LM;\tau}$ measures the nuclear mass number, and $\Phi''_{LM;\tau}$ the content of nucleon spin-orbit coupling in the nucleus, which is large for heavy elements with orbits of large angular momentum not fully occupied~\cite{Fitzpatrick:2012ix}. Elements in the Sun are hence disfavored compared to Xe or I, which respectively compose the LUX and (partially) COUPP experiments. Interaction operators of this type are $\hat{\mathcal{O}}_{1}$, $\hat{\mathcal{O}}_{3}$, $\hat{\mathcal{O}}_{5}$, $\hat{\mathcal{O}}_{8}$, $\hat{\mathcal{O}}_{11}$, $\hat{\mathcal{O}}_{12}$ and $\hat{\mathcal{O}}_{15}$.

The operators $\hat{\mathcal{O}}_{5}$ and $\hat{\mathcal{O}}_{8}$ also couple to the nuclear response operator $\Delta_{LM;\tau}$, which in the low momentum transfer limit measures the nucleon angular momentum content of the nucleus. The nuclear response operator $\Delta_{LM;\tau}$ is generated by the nuclear convection current in Eq.~(\ref{eq:Hx}). It favors nuclei with an unpaired nucleon in a non s-shell orbit, like for instance ${}^{14}$N and ${}^{27}$Al in the Sun's interior. The coupling of the operators $\hat{\mathcal{O}}_{5}$ and $\hat{\mathcal{O}}_{8}$ to the nuclear response operator $\Delta_{LM;\tau}$ strengthens our limits on $c_5^\tau$ and $c_8^\tau$ from neutrino telescope observations.

The remaining operators, namely $\hat{\mathcal{O}}_{4}$, $\hat{\mathcal{O}}_{6}$, $\hat{\mathcal{O}}_{7}$, $\hat{\mathcal{O}}_{9}$, $\hat{\mathcal{O}}_{10}$, $\hat{\mathcal{O}}_{13}$, and $\hat{\mathcal{O}}_{14}$ mainly couple to the nuclear response operators $\Sigma'_{LM;\tau}$ and $\Sigma''_{LM;\tau}$, which in the low momentum transfer limit measure the nucleon spin content of the nucleus. Within this group of interaction operators, of particular interests are the operators $\hat{\mathcal{O}}_{6}$ and $\hat{\mathcal{O}}_{7}$. For the operators $\hat{\mathcal{O}}_{6}$ and $\hat{\mathcal{O}}_{7}$, we find that neutrino telescopes can set stronger limits on $c_6^{\tau}$ and $c_7^{\tau}$ than dark matter direct detection. Notably, for the operator $\hat{\mathcal{O}}_{7}$ our exclusion limits on the coupling constant $c_7^0$ are stronger than those from COUPP and LUX in a large $m_\chi$ range, even within the soft annihilation spectrum analysis. The competitive limit on $c_7^0$ from neutrino telescopes depends on the $v_T^{\perp 2}$ term in $R_{\Sigma'}^{\tau\tau'}$, which is less penalizing for neutrino telescopes than for direct detection experiments. In addition, it also depends on the abundance of Hydrogen in the Sun, which for $\hat{\mathcal{O}}_{7}$ determines rate of dark matter capture, and on the large value of $W_{\Sigma'}^{\tau\tau'}$ for this element. 

\subsection{Spin-dependent operators and the heavy elements in the Sun}
We conclude this section exploring the spin-dependent dark matter-nucleon interaction operators more in depth. In particular we show that only for the spin-dependent operators $\hat{\mathcal{O}}_4$ and $\hat{\mathcal{O}}_7$ Hydrogen is the most important element in the exclusion limit calculation. For all other spin-dependent operators, i.e. $\hat{\mathcal{O}}_{3}$, $\hat{\mathcal{O}}_{6}$, $\hat{\mathcal{O}}_{9}$, $\hat{\mathcal{O}}_{10}$, $\hat{\mathcal{O}}_{12}$, $\hat{\mathcal{O}}_{13}$, $\hat{\mathcal{O}}_{14}$, and $\hat{\mathcal{O}}_{15}$, heavier elements are  significantly more important. 
This conclusion is illustrated in Figs.~\ref{fig:i1}, \ref{fig:i2}, and \ref{fig:i3}, where we report exclusion limits obtained assuming that dark matter scatters from single elements in the Sun. Conventions for colors and lines are those in the legends. For simplicity, we focus here on the isoscalar components only.

For the operators $\hat{\mathcal{O}}_{3}$, $\hat{\mathcal{O}}_{12}$ and $\hat{\mathcal{O}}_{15}$ the most important element is $^{56}$Fe, since they mainly couple to the nuclear response operator $M_{ML;\tau}$, which favors elements with large $A$ (and in the Sun $^{56}$Fe is more abundant than $^{58}$Ni).

For the operators $\hat{\mathcal{O}}_{6}$, $\hat{\mathcal{O}}_{9}$, $\hat{\mathcal{O}}_{10}$, $\hat{\mathcal{O}}_{13}$ and $\hat{\mathcal{O}}_{14}$, exclusion limits mostly depend on $^{14}$N, since in this case dark matter scattering from Hydrogen is suppressed for $m_\chi\gtrsim 20$~GeV. Indeed, for $m_i$ equal to the Hydrogen mass, the integral  in Eq.~(\ref{eq:omega}) is largely dominated by small recoil energies, where the differential cross-section for these operators is particularly small. Notice also that the operators $\hat{\mathcal{O}}_{6}$, $\hat{\mathcal{O}}_{9}$, $\hat{\mathcal{O}}_{10}$, $\hat{\mathcal{O}}_{13}$ and $\hat{\mathcal{O}}_{14}$ mainly couple to the nuclear response operators $\Sigma'_{ML;\tau}$ and $\Sigma''_{ML;\tau}$, and that $^{14}$N is characterized by fairly large $W_{\Sigma'}^{\tau\tau'}$ and $W_{\Sigma''}^{\tau\tau'}$ nuclear response functions. 

\section{Conclusions}
\label{sec:conclusions}
We derived exclusion limits on the coupling constants of the general effective theory of one-body dark matter-nucleon interactions using current IceCube/DeepCore and SUPER-K observations. In this study, we exploited new nuclear response functions computed in~\cite{Catena:2015uha} through nuclear structure calculations for all dark matter-nucleon isoscalar and isovector interaction operators in Tab.~\ref{tab:operators}, and for the 16 most abundant elements in the Sun. Exclusion limits were presented at the 90\% confidence level, and separately assuming dark matter pair annihilation into $W^+W^-$ and $b\bar{b}$.

We found that the most severely constrained interaction operators are $\hat{\mathcal{O}}_{11} =i{\hat{\bf{S}}}_\chi\cdot ({\hat{\bf{q}}}/m_N)$, $\hat{\mathcal{O}}_{12} = \hat{{\bf{S}}}_{\chi}\cdot (\hat{{\bf{S}}}_{N} \times{\hat{\bf{v}}}^{\perp})$,  $\hat{\mathcal{O}}_8 = \hat{{\bf{S}}}_{\chi}\cdot {\hat{\bf{v}}}^{\perp}$ and $\hat{\mathcal{O}}_3 = i\hat{{\bf{S}}}_N\cdot[({\hat{\bf{q}}}/m_N)\times{\hat{\bf{v}}}^{\perp}]$, besides the familiar spin-independent and spin-dependent operators $\hat{\mathcal{O}}_1 = \mathbb{1}_{\chi N}$ and $\hat{\mathcal{O}}_4 = \hat{{\bf{S}}}_{\chi}\cdot \hat{{\bf{S}}}_{N}$, respectively. For each operator we found a physical interpretation for the relative strength of the corresponding limit. 

We compared our limits on the coupling constants from neutrino telescopes with benchmark dark matter direct detection exclusion limits. We found that the operator $\hat{\mathcal{O}}_7={\hat{\bf{S}}}_N \cdot {\bf{v}}^\perp$ is significantly more constrained by neutrino telescopes than by LUX and COUPP. This conclusion is mainly related to the $v_T^{\perp 2}$ term in $R_{\Sigma'}^{\tau\tau'}$, which is less penalizing for neutrino telescopes than for direct detection experiments, to the abundance of Hydrogen in the Sun, and to the corresponding large value of $W_{\Sigma'}^{\tau\tau'}$. 

Another important result of this work is to show that Hydrogen is not the most important element in the exclusion limit calculation for the majority of the spin-dependent operators in Tab.~\ref{tab:operators}. As a consequence, nuclear structure calculations as those initiated in~\cite{Catena:2015uha} appear to be key tools for model independent analyses of dark matter signals at neutrino telescopes.

\acknowledgments It is a pleasure to thank Alejandro Ibarra, Chris Kouvaris, Paolo Panci, Bodo Schwabe, and Sebastian Wild, for many inspiring conversations on subjects related to this work. This work has partially been funded through a start-up grant of the University of G\"ottingen. I acknowledges partial support from the European Union FP7 ITN INVISIBLES (Marie Curie Actions, PITN-GA-2011-289442).

\appendix

\section{LUX and COUPP analysis}
\label{sec:dd}

We compare the general effective theory of one-body dark matter-nucleon interactions mediated by a heavy spin-1 or spin-0 particle with the LUX and COUPP experiments in a Bayesian analysis based on the Likelihood function
\begin{eqnarray}
-\ln \mathcal{L}(\mathbf{d}|m_\chi,\mathbf{c}) 
&\simeq& \mu_S(m_\chi,\mathbf{c}) + (2-k) \ln \Big[\mu_S(m_\chi,\mathbf{c})+\hat{\mu}_B\Big]  \nonumber\\
&-&\ln \left\{ \frac{(k^2-k)}{2}\sigma_B^2 + \left[ \mu_S(m_\chi,\mathbf{c})+\hat{\mu}_B - \frac{k}{2}\sigma_B^2 \right]^2 \right\} \nonumber\\
&+& {\rm constant} \,.
\label{Like_eff}
\end{eqnarray}    
In Eq.~(\ref{Like_eff}) $k$ is the number of observed recoils in a given dataset ${\bf d}$, $\mu_S(m_\chi,\mathbf{c})$ is the number of predicted scattering events at a given mass $m_\chi$, and coupling constant $\mathbf{c}$, and $\hat{\mu}_B$ is the corresponding number of expected background events (with error $\sigma_B$). The constant term in Eq.~(\ref{Like_eff}) corresponds to the arbitrary normalization of $\mathcal{L}$, which we fix by imposing $\ln \mathcal{L}_{\rm}=0$, for $\mu_S+\hat{\mu}_B=k$.

For the COUPP Likelihood function we assume $k=2$,  $k=3$ and $k=8$ bubble nucleations above a threshold energy $E_{\rm th}$ of 7.8 keV$_{\rm nr}$, 11 keV$_{\rm nr}$ and 15.5 keV$_{\rm nr}$, respectively \cite{Behnke:2012ys}. For the 3 threshold configurations the expected number of background events is, respectively, $\mu_B=0.8$, $\mu_B=0.7$ and $\mu_B=3$. In the analysis we use an energy dependent experimental exposure $\epsilon$ characterized by: $\epsilon(7.8~{\rm keV})=55.8$ kg-days, $\epsilon(11~{\rm keV})=70$ kg-days and $\epsilon(15.5~{\rm keV})=311.7$ kg-days~\cite{Behnke:2012ys}. For COUPP, we compute $\mu_{S}$ as follows
\begin{equation}
\mu_{S}(m_\chi,\mathbf{c}) =\epsilon(E_{\rm th}) \sum_{T={\rm C,F,I}} \int_{E_{\rm th}}^{\infty}{\rm d}E_{R} \, \mathcal{P}_T(E_{R},E_{\rm th}) \frac{{\rm d}\mathcal{R}_{T}}{{\rm d} E_{R}}
\label{coupp}
\end{equation}
where ${\rm d}\mathcal{R}_{T}/{\rm d} E_{R}$ is the differential rate of scattering from the target material $T$. The probability $\mathcal{P}_{T}(E_{R},E_{\rm th})$ that an energy $E_{R}$ nucleates a bubble above $E_{\rm th}$ is given by \cite{Behnke:2012ys}:
\begin{equation}
\mathcal{P}_{T}(E_{R},E_{\rm th}) = 1 - \exp\left[ -\alpha_{T} \frac{E_{R}-E_{\rm th}}{E_{\rm th}}\right] \,.
\end{equation}
We assume perfect efficiency for bubble nucleation for Iodine ($\alpha_{\rm I}\rightarrow +\infty$), we neglect scattering from Carbon ($\alpha_{\rm C}=0$), and marginalize over $\alpha_{\rm F}\equiv a_{\rm COUPP}$, assuming for the latter a log-prior with mean 0.15. 

We construct the LUX Likelihood function as in~\cite{Catena:2014uqa}, assuming $k=1$ and $\hat{\mu}_B=0.64\pm0.16$. We consider a Gaussian resolution for photoelectron detection with standard deviation $\sigma_{\rm PMT}=0.37$, an exposure of 250$\times$85.3 kg-days, and the experimental efficiency in Fig.~1 of Ref.~\cite{Akerib:2013tjd} times 0.5, in order to account for the 50\% nuclear recoil acceptance quoted by the LUX collaboration.

Results are presented in terms of x\% credible regions, i.e. portions of the parameter space containing x\% of the total posterior probability, and where the posterior probability density function $\mathcal{P}(\mathbf{\Theta}|\mathbf{d}) \propto  \mathcal{L}(\mathbf{d}|\mathbf{\Theta}) \pi(\mathbf{\Theta})$ at any point $\mathbf{\Theta}$ in parameter space inside the credible region is larger than at any external point. We assume a uniform prior probability density function $\pi(\mathbf{\Theta})$ for the logarithm of $m_\chi$ and of $\mathbf{c}$. Prior and volume effects in dark matter direct detection are discussed in~\cite{Catena:2014uqa,Conrad:2014nna,Strege:2012kv,Arina:2011si}.

\section{Dark matter response functions}
\label{sec:appDM}
In this appendix we list the dark matter response functions that appear in Eq.~(\ref{eq:sigma}). The notation is the same used in the body of the paper.
\begin{eqnarray}
 R_{M}^{\tau \tau^\prime}\left(v_T^{\perp 2}, {q^2 \over m_N^2}\right) &=& c_1^\tau c_1^{\tau^\prime } + {j_\chi (j_\chi+1) \over 3} \left[ {q^2 \over m_N^2} v_T^{\perp 2} c_5^\tau c_5^{\tau^\prime }+v_T^{\perp 2}c_8^\tau c_8^{\tau^\prime }
+ {q^2 \over m_N^2} c_{11}^\tau c_{11}^{\tau^\prime } \right] \nonumber \\
 R_{\Phi^{\prime \prime}}^{\tau \tau^\prime}\left(v_T^{\perp 2}, {q^2 \over m_N^2}\right) &=& {q^2 \over 4 m_N^2} c_3^\tau c_3^{\tau^\prime } + {j_\chi (j_\chi+1) \over 12} \left( c_{12}^\tau-{q^2 \over m_N^2} c_{15}^\tau\right) \left( c_{12}^{\tau^\prime }-{q^2 \over m_N^2}c_{15}^{\tau^\prime} \right)  \nonumber \\
 R_{\Phi^{\prime \prime} M}^{\tau \tau^\prime}\left(v_T^{\perp 2}, {q^2 \over m_N^2}\right) &=&  c_3^\tau c_1^{\tau^\prime } + {j_\chi (j_\chi+1) \over 3} \left( c_{12}^\tau -{q^2 \over m_N^2} c_{15}^\tau \right) c_{11}^{\tau^\prime } \nonumber \\
  R_{\tilde{\Phi}^\prime}^{\tau \tau^\prime}\left(v_T^{\perp 2}, {q^2 \over m_N^2}\right) &=&{j_\chi (j_\chi+1) \over 12} \left[ c_{12}^\tau c_{12}^{\tau^\prime }+{q^2 \over m_N^2}  c_{13}^\tau c_{13}^{\tau^\prime}  \right] \nonumber \\
   R_{\Sigma^{\prime \prime}}^{\tau \tau^\prime}\left(v_T^{\perp 2}, {q^2 \over m_N^2}\right)  &=&{q^2 \over 4 m_N^2} c_{10}^\tau  c_{10}^{\tau^\prime } +
  {j_\chi (j_\chi+1) \over 12} \left[ c_4^\tau c_4^{\tau^\prime} + \right.  \nonumber \\
 && \left. {q^2 \over m_N^2} ( c_4^\tau c_6^{\tau^\prime }+c_6^\tau c_4^{\tau^\prime })+
 {q^4 \over m_N^4} c_{6}^\tau c_{6}^{\tau^\prime } +v_T^{\perp 2} c_{12}^\tau c_{12}^{\tau^\prime }+{q^2 \over m_N^2} v_T^{\perp 2} c_{13}^\tau c_{13}^{\tau^\prime } \right] \nonumber \\
    R_{\Sigma^\prime}^{\tau \tau^\prime}\left(v_T^{\perp 2}, {q^2 \over m_N^2}\right)  &=&{1 \over 8} \left[ {q^2 \over  m_N^2}  v_T^{\perp 2} c_{3}^\tau  c_{3}^{\tau^\prime } + v_T^{\perp 2}  c_{7}^\tau  c_{7}^{\tau^\prime }  \right]
       + {j_\chi (j_\chi+1) \over 12} \left[ c_4^\tau c_4^{\tau^\prime} +  \right.\nonumber \\
       &&\left. {q^2 \over m_N^2} c_9^\tau c_9^{\tau^\prime }+{v_T^{\perp 2} \over 2} \left(c_{12}^\tau-{q^2 \over m_N^2}c_{15}^\tau \right) \left( c_{12}^{\tau^\prime }-{q^2 \over m_N^2}c_{15}^{\tau \prime} \right) +{q^2 \over 2 m_N^2} v_T^{\perp 2}  c_{14}^\tau c_{14}^{\tau^\prime } \right] \nonumber \\
     R_{\Delta}^{\tau \tau^\prime}\left(v_T^{\perp 2}, {q^2 \over m_N^2}\right)&=&  {j_\chi (j_\chi+1) \over 3} \left[ {q^2 \over m_N^2} c_{5}^\tau c_{5}^{\tau^\prime }+ c_{8}^\tau c_{8}^{\tau^\prime } \right] \nonumber \\
 R_{\Delta \Sigma^\prime}^{\tau \tau^\prime}\left(v_T^{\perp 2}, {q^2 \over m_N^2}\right)&=& {j_\chi (j_\chi+1) \over 3} \left[c_{5}^\tau c_{4}^{\tau^\prime }-c_8^\tau c_9^{\tau^\prime} \right].
 \label{eq:R}
\end{eqnarray}


\providecommand{\href}[2]{#2}\begingroup\raggedright\endgroup

\end{document}